\documentclass[fleqn,usenatbib]{mnras}
\pdfoutput=1
\usepackage{newtxtext,newtxmath}

\usepackage[T1]{fontenc}
\usepackage{ae,aecompl}

\usepackage{graphicx}	
\usepackage{amsmath}	
\usepackage{amssymb}	
\usepackage{ulem}
\usepackage{float}
\usepackage[first=0,last=9]{lcg}

\title[SZ scaling relations]{
Weak-lensing mass calibration of the Sunyaev-Zel'dovich effect using APEX-SZ galaxy clusters}

\author[A. Nagarajan et al.]{A. Nagarajan,$^{1}$\thanks{Member of the International Max Planck Research School (IMPRS) for Astronomy and Astrophysics at the Universities of Bonn and
Cologne, E-mail: aarti$@$astro.uni-bonn.de}
F. Pacaud,$^{1}$
M. Sommer,$^{1}$
M. Klein,$^{2,3}$
K. Basu,$^{1}$
F. Bertoldi,$^{1}$
\newauthor A. T. Lee,$^{4,5}$
P. A. R. Ade,$^{6}$
A. N. Bender,$^{7,8}$
D. Ferrusca,$^{9}$
N. W. Halverson,$^{8,10}$
\newauthor C. Horellou,$^{11}$
 B. R. Johnson,$^{12}$
J. Kennedy,$^{7}$
R. Kneissl,$^{13,14}$
K. M. Menten,$^{15}$
\newauthor C. L. Reichardt,$^{16}$
C. Tucker,$^{6}$
B. Westbrook,$^{4}$\\
$^{1}$ Argelander Institut fuer Astronomie, University of Bonn, Germany\\
$^{2}$ Faculty of Physics, Ludwig-Maximilians-Universit\"at, Scheinerstr. 1, D-81679 Munich, Germany\\
$^{3}$ Max Planck Institute for Extraterrestrial Physics, Giessenbachstr. 1, D-85748 Garching, Germany\\
$^{4}$ Department of Physics, University of California, Berkeley, CA 94720, USA\\
$^{5}$ Physics Division, Lawrence Berkeley National Laboratory, Berkeley, CA 94720, USA\\
$^{6}$ School of Physics and Astronomy, Cardiff University, Queens Buildings, The Parade, Cardiff, CF24 3AA, UK\\
$^{7}$ Department of Physics, McGill University, Montreal H3A 2T8, Canada\\
$^{8}$ Center for Astrophysics and Space Astronomy, Department of Astrophysical and Planetary Sciences,\\ University of Colorado, Boulder, CO 80309, USA\\
$^{9}$ Instituto Nacional de Astrof\'{i}sica, \'{O}ptica y Electr\'{o}nica, Luis Enrique Erro 1, Tonantzintla, Puebla C.P. 72840, M\'{e}xico\\ 
$^{10}$ Department of Physics, University of Colorado, Boulder,CO, 80309\\
$^{11}$ Chalmers University of Technology, Dept. of Space, Earth and Environment, Onsala Space Observatory, SE-43992 Onsala, Sweden\\
 $^{12}$ Columbia University, Department of Physics, New York, NY 10027, USA\\
 $^{13}$ European Southern Observatory, Alonso de Cordova 3107,
Vitacura, Santiago, Chile\\
$^{14}$ Atacama Large Millimeter/submillimeter Array, Joint
ALMA Observatory, Alonso de Cordova 3107, Vitacura, Santiago, Chile\\
$^{15}$ Max Planck Institute for Radio Astronomy, 53121 Bonn, Germany\\
$^{16}$ School of Physics, University of Melbourne, Parkville, 3010, VIC, Australia\\
}

\date{Accepted XXX. Received YYY; in original form ZZZ}

\pubyear{2018}

\begin{document}
\label{firstpage}
\pagerange{\pageref{firstpage}--\pageref{lastpage}}
\maketitle


\begin{abstract}
The use of galaxy clusters as precision cosmological probes relies on an accurate 
determination of their masses. However, inferring the
relationship between cluster mass and observables from direct
observations is difficult and prone to sample selection biases.
In this work, we use weak lensing as the best possible
proxy for cluster mass to calibrate the Sunyaev-Zel'dovich
(SZ) effect measurements from the APEX-SZ experiment. For a
well-defined (ROSAT) X-ray complete cluster sample, we calibrate the integrated
Comptonization parameter, $Y_{\rm SZ}$, to the weak-lensing derived
total cluster mass, $M_{500}$.
We employ a novel Bayesian approach to account for the selection
effects by jointly fitting both the SZ Comptonization, $Y_{\rm
SZ}\text{--}M_{500}$, and the X-ray luminosity, $L_{\rm x}\text{--}M_{500}$, scaling
relations.
We also account for a possible correlation between the intrinsic (log-normal)
scatter of $L_{\rm x}$ and $Y_{\rm SZ}$ at fixed mass. We find the
corresponding correlation coefficient to be $r= 0.47_{-0.35}^{+0.24}$, and
at the current precision level our constraints on the scaling relations
are consistent with previous works.
For our APEX-SZ sample, we find that ignoring the covariance between the SZ and X-ray
observables biases the normalization of the $Y_{\rm SZ}\text{--}M_{500}$
scaling high by $1\text{--}2\sigma$ and the slope low by $\sim 1\sigma$, even
when the SZ effect plays no role in the sample selection.
We conclude that for higher-precision data and larger cluster samples,
as anticipated from on-going and near-future cluster cosmology
experiments, similar biases (due to intrinsic covariances of cluster observables) in the scaling relations will dominate the cosmological error budget if not accounted for correctly. 
\end{abstract}




\section{Introduction}

The $\Lambda$CDM model of Big Bang cosmology predicts a hierarchical, gravity-driven scenario of structure formation in which galaxy clusters are the largest and most recently assembled quasi-virialized structures. The abundance of galaxy clusters in mass-redshift space depends on cosmological parameters, making it a sensitive probe of cosmology. To derive cosmological constraints from the cluster abundance evolution, it is crucial to have an accurate mass calibration of cluster observables. 
  The baryonic components in galaxy clusters, such as stars, galaxies, and the intra-cluster medium (ICM), are visible in a wide range of the electromagnetic spectrum. In contrast, the cold dark matter component can only be measured indirectly, e.g., through the gravitational distortion of background light, which becomes increasingly challenging to measure for galaxy clusters at higher redshifts.
 To relate direct observables to cluster mass, it is of great advantage that, for a scale-free initial matter power spectrum, cosmic structures evolve in a self-similar way \citep{1986Kaiser}. Under the simplifying assumption that the cluster ICM is in isothermal, hydrostatic equilibrium with an isothermal dark matter distribution, cluster observable global properties and the total mass of the cluster are related by simple scaling relations. These scaling relations are an essential link between cluster observables and cosmology, but also show considerable advantage in probing the thermodynamic history of the ICM \citep{2013Giodini}.
 There have been major efforts to empirically study the scaling relations from cluster observations, in order to constrain cosmological parameters from cluster number count measurements (e.g., \citealt{2009Vikhlinin, 2010Mantz}). The observational selection of representative cluster samples always relies on some directly observable quantity, such as X-ray luminosity, Sunyaev-Zel'dovich Comptonization observables or optical richness (e.g., \citealt{ 2013Boehringer,2015Bleem,2014Oguri}). Despite the limited size of current cluster samples, the calibration of scaling relations already emerges as the limiting factor in the error budget of number count studies of galaxy clusters \citep{2011Allen}. Mass calibration will become critical with on-going and next generation galaxy cluster surveys (SPT-3G: \citealt{2014Benson}, eROSITA: \citealt{eROSITA}, Euclid: \citealt{2011Euclid}, LSST: \citealt{2009LSST}), which are expected to increase sample sizes by two orders of magnitude.

Any sample of galaxy clusters is generally affected by a number of biases that depend on the underlying mass distribution, 
the intrinsic covariance of the cluster observables, additional measurement uncertainties and the selection method 
\citep[e.g.,][]{2006Stanek,2007Pacaud,2009Vikhlinin,2010Mantz}. Mass functions predicted by simulations 
\citep[e.g.,][]{2008Tinker} and determined from cluster surveys \citep[e.g.,][]{2002Reiprich,2009Vikhlininb}
have shown the number density of clusters to be an exponentially decreasing function of cluster mass, 
including a trend with redshift. In the presence of scatter (intrinsic as well as that arising from 
measurement uncertainties), this will cause more low-mass clusters to up-scatter to a given observed mass 
than high mass clusters to down-scatter to that same level, thus distorting the distribution of sources in the space of observables - an effect known as Eddington bias \citep{1913Eddington}. These distortions are further exacerbated in the presence of sample selection thresholds that truncate the scattered distributions. In addition, depending on their distances, the selected clusters are not drawn from the same mass distribution due to the combined effect of the cosmological growth of structures, the surveyed volume and source selection thresholds - the well known Malmquist bias \citep{1920Malmquist}. As a consequence, samples selected by luminosity would typically be biased towards low masses and intrinsically bright sources. In the presence of a (positive) correlation in the intrinsic scatters of the selecting mass observable and a follow-up mass observable, the follow-up observable would also be, on average, biased towards intrinsically bright sources (e.g. \citealt{2011Allen}).  An accurate calibration of cluster scaling relations requires that these biases are controlled and corrected for.

In the work presented here, we focus on scaling relations involving the Sunyaev-Zel'dovich (SZ) effect \citep{1970SunyaevZel}, a spectral distortion of the blackbody cosmic microwave background (CMB) radiation caused by inverse Compton scattering by the hot electrons. While the recent availability of SZ-selected galaxy clusters for cosmological analysis has resulted in several precise constraints on cosmology (e.g. \citealt{2016deHaanupdate,2016Planckcosmo,2016Planckxxiv}), these studies largely rely on prior information on the SZ-mass calibration obtained from X-ray derived masses and/or weak-lensing masses.
Thus, directly calibrating the integrated SZ Comptonization ($Y_{\rm SZ}$) with cluster mass ($M_{500}$) has generated much interest. Weak-lensing mass estimates are best suited for calibrating cluster masses as they directly measure the line-of-sight matter distribution and do not rely on further assumptions about the physical state of matter inside clusters (like hydrostatic equilibrium or thermal pressure support). 
Simulations indicate that lensing masses are biased by at most a few percent \citep{2011Becker,2010Meneghetti, 2012Rasia}. 

Early studies of the scaling between weak-lensing mass and SZ Comptonization either suffered from not statistically complete samples (e.g., \citealt{2012Hoekstra, 2015Hoekstra, 2015Sereno}), 
or were limited by the availability of lensing and SZ observations (e.g., \citealt{2012Marrone}). 
Additionally, in cases where the sample selection was based on X-ray luminosities, the effects of possible correlations in the intrinsic scatters of SZ Comptonization and X-ray luminosity at fixed mass were assumed to be negligible or approximated with fixed values (e.g., \citealt{2012Marrone,2016Mantzwtg, 2017Sereno}). 
Numerical simulations have predicted that at a given cluster mass, the dispersion of global thermodynamic properties are correlated (e.g., \citealt{2016Truong, 2012Angulo, 2010Stanek}). In particular, these authors find a correlation in the intrinsic scatter of X-ray luminosity and integrated Comptonization in the range of 0.5--0.8. If this correlation is unaccounted for, it can bias the inferred $ Y_{\rm SZ}\text{--}M_{500}$ scaling relation for a sample that is selected on X-ray luminosities (an illustration of the impact of this correlation is given in Appendix \ref{ch3:sec:samplebias}). Observationally, this correlation remains largely unconstrained. 

In this work, we employ a sample of 39 galaxy clusters observed with the SZ effect using the APEX telescope \citep{2011Schwan}. To provide an accurate mass calibration of the integrated Comptonization, we measure the scaling relation of the Comptonization with weak-lensing derived masses of an X-ray selected sample with a well-defined selection function.
The sample, henceforth eDXL, is a sub-sample of galaxy clusters observed by the APEX-SZ experiment. We present a Bayesian method to account for the
sample selection while placing an emphasis on controlling
the bias in the scaling relations due to the correlated intrinsic
scatter of the selection observable (X-ray luminosity) and scaling observable
(integrated Comptonization) at fixed mass. 

This paper is organized as follows: In Section \ref{sec:sample} we describe the APEX-SZ sample and our complete X-ray selected sub-sample. 
The cluster follow-up observations in the SZ and weak-lensing are described in Section \ref{sec:obs}. In Section \ref{sec:massproxy} we discuss our mass proxy measurements in detail. In Section \ref{sec:method} we present a Bayesian method for fitting scaling relations while accounting for selection effects. Our results are presented in Section \ref{sec:results} and their robustness, systematics and limitations are discussed in Section \ref{sec:robust}. We discuss the significance of our results in Section \ref{sec:discussion}. We offer our conclusions in Section \ref{sec:conclusions}.
Unless otherwise noted, we assume a $\Lambda$CDM cosmology with $\Omega_{\rm m}=0.3$, $\Omega_{\Lambda}=0.7$ and $H_{0}=70 \;\rm km \;s^{-1}\; Mpc^{-1}$. 
 
\begin{figure*}
\includegraphics[width=15cm]{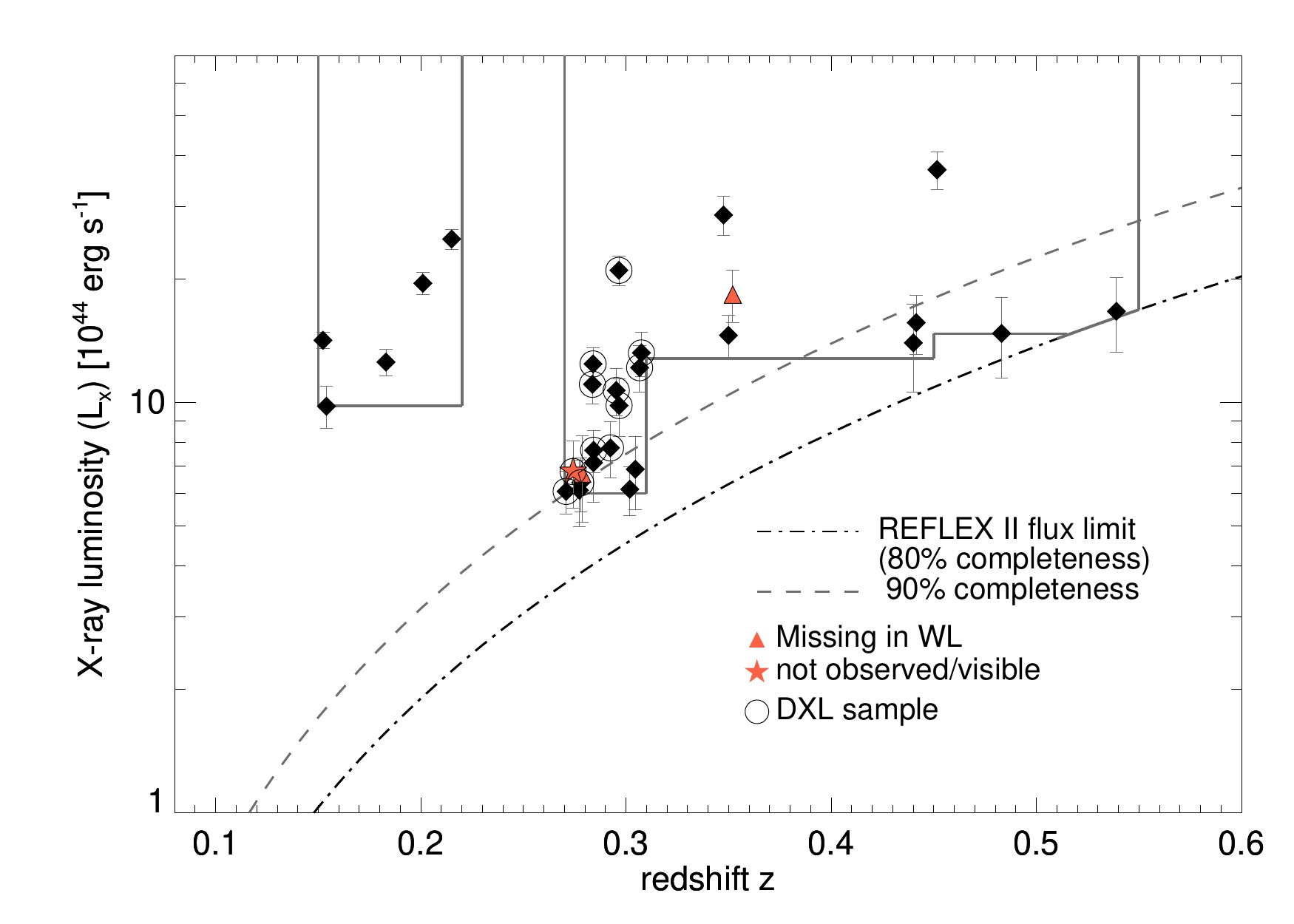} 
    \caption{The extended DXL (eDXL) sample selection. Above $z=0.27$, the sample is selected in the luminosity-redshift plane of the REFLEX II catalogue. The low redshift sample is selected from NORAS and REFLEX.  In total, 30 galaxy clusters are selected. The grey rectangular outlines enclose the sample selection plane. 
    The details on multi-wavelength follow-up observations are given in section \ref{sec:obs}. As indicated in this figure, three galaxy clusters are missing in the follow-up program.
    The dotted-dashed curve represent the nominal flux limit ($1.8 \times 10^{-12} \,\rm erg \,s^{-1}$) of the REFLEX II catalogue in the luminosity-redshift plane. This curve corresponds to approximately 80\% completeness. The dashed curve is the luminosity curve for higher flux limit ($3 \times 10^{-12} \, \rm erg\, per\, sec$), which represent the 90\% completeness of the REFLEX II catalogue. 
   The APEX-SZ cluster targets include the eDXL sample and additional clusters that were selected in an {\it ad hoc} manner. 
      }\label{fig:sample}
    \end{figure*}
\section{APEX-SZ cluster sample}\label{sec:sample}


The APEX-SZ (Section \ref{sec:apexdata}) cluster targets were initially selected in an {\it ad hoc} manner, focusing on well-studied or seemingly interesting clusters with bright X-ray emission and hot ($T_{\rm e} \gtrsim 5~ \rm KeV$) temperatures to ensure highly significant detections.
To make a robust scaling relation analysis possible, later APEX-SZ observations were dedicated to follow-up a complete sample of 30 clusters, selected from the ROSAT All-Sky Survey (RASS) catalogues by applying well-defined cutoffs in the ROSAT luminosity-redshift plane. This sub-sample is essentially an extension of the REFLEX-DXL sample (\citealt{2006Zhang}), and will, henceforth, be referred to as the extended Distant X-ray luminous galaxy clusters (eDXL) sample. 
In the following we describe the selection and characteristics of the eDXL sample. For completeness, a summary of the APEX-SZ clusters not belonging to the eDXL sample are given in Section \ref{sec:other}. 


\subsection{The eDXL cluster sample}\label{sec:edxl}

The sample was constructed as an extension of the volume complete DXL sample \citep{2006Zhang}, which consisted of the 13 clusters in the southern hemisphere with $0.27<z<0.31$ and ROSAT luminosities in the [0.1-2.4] keV band $L_X>10^{45}$ erg/s. Taking advantage of the updated and deeper REFLEX-II catalogue enabled us to lower the luminosity cutoffs in the DXL redshift range to increase the mass coverage, and include some higher redshift clusters (up to $z=0.54$). The precise luminosity cuts for each redshift range were set to maximize the overlap with earlier APEX-SZ observations, while staying above the nominal flux limit of the parent REFLEX-II catalogue \citep{2013Boehringer}. 

The REFLEX II nominal flux limit, transposed onto the luminosity-redshift plane, is indicated in Figure \ref{fig:sample}. At this limit, the completeness of the parent sample is approximately 80\% (as inferred from Figure 11 of the paper by \citealt{2013Boehringer}). 
We also show in the same figure the location of the 90\% completeness curve. Most of the clusters
falls above this curve, ensuring a high completeness. As explained in more details in Section \ref{sec:completeness}, our own scaling relation model
permits to estimate a global completeness $\gtrsim$ 90\% over our luminosity
- redshift selection.

In the low redshift range ($0.15<z<0.22$), all of the X-ray brightest clusters from REFLEX \citep{2004Boehringer} and NORAS \citep{2000Boehringer} catalogues are part of the APEX-SZ target list. 
This enables us to extend our sample selection to lower redshifts {\it a posteriori}, but 
requires the inclusion of NORAS to reach a meaningful number addition of five 
clusters. The high luminosity and redshift cuts were set to exclude other bright 
sources not observed with APEX-SZ. This luminosity cut is well above the nominal 
flux limit of both REFLEX and NORAS catalogues ensuring an effectively volume 
complete selection.

With 30 galaxy clusters in total, the extended DXL selection more than doubles the number of clusters from the initial DXL sample. It was designed to provide both a good leverage on the slope of scaling relations at $z\sim0.3$ and a large redshift coverage. This should permit breaking the degeneracy between the inferred slope and redshift evolution of scaling relations (e.g., \citealt{2014Andreon}). 

The exact, redshift-dependent, luminosity thresholds used for the selection are given in Table \ref{tab:sample}. A graphical representation of the corresponding parameter space and the selected clusters is provided in Figure \ref{fig:sample}. 
The sample is complete within the selected luminosity and redshift ranges.

\begin{table*}\caption{Luminosity cuts of the eDXL sample in different redshift bins. The selection is represented graphically in Figure \ref{fig:sample}. The luminosities quoted here are computed in the energy range $[\text{0.1-2.4}]$ keV. 
In the final column, we give the number of clusters in each redshift range. In bracket, we mention the actual number of clusters that were completely followed-up in our multi-wavelength observations.}
\label{tab:sample}
\renewcommand{\arraystretch}{1.2}
 \begin{tabular}{l l l c}
 \hline
 Redshift bin & Luminosity cut & Parent  &  Number\\
 & $L^{\rm min}$&sample& of\\
 & $ [10^{44}$ erg s$^{-1}$] && clusters\\
 \hline
  $0.15$ $< z <$ $0.22$ & $9.78$& REFLEX \& &  5 (5)\\
  &&NORAS&\\
  $0.27$ $< z <$ $0.31$ & $6.0$ &REFLEX II & 17 (15)\\
  $0.31$ $< z <$ $0.45$ & $12.8$ &REFLEX II & 5 (4)\\
  $0.45$ $< z <$ $0.55$ & $14.7$ & REFLEX II &3 (3)\\
  \hline
\end{tabular}
\end{table*}

XMM observations are available for all 30 galaxy clusters in the eDXL sample. 
However, one of them could not be observed from the APEX site due to its very low declination. 
For two others, the lensing data were not of sufficient quality to provide any 
mass information due to bad weather conditions and poor seeing.

In the rest of the paper, only those 27 cluster with complete follow-up data are included in the complete eDXL sample. 
Since the exclusion of the two clusters in this down-selection was random, i.e. does not depend on the cluster physical properties, we assume that the selection function of the sample remains unaffected.

\subsection{Other APEX-SZ clusters}\label{sec:other}
The full APEX-SZ sample does not have a well-defined selection. In addition to the eDXL sub-sample, it contains a number of high redshift clusters and a few massive local clusters whose inclusion in our complete selection would have required the observation of many more targets to reach a complete sample.
In total, 12 additional APEX-SZ clusters have complete multi-wavelength follow-up in X-rays (either by the {\it XMM-Newton} or the {\it Chandra} satellite) and  were followed up with optical observations. The latter follow-up is summarized in section \ref{sec:opt}. For completeness, we provide the global observable measurements of these 12 additional clusters along with our eDXL clusters.

\section{Observations and Data Analysis}
\label{sec:obs} 
\subsection{APEX-SZ instrument and observations}
\label{sec:apexdata}

The APEX-SZ (\citealt{2006Dobbs,2011Schwan,2012Dobbs}) instrument was a  bolometer array which operated from 2005 to 2010 on the 12-meter Atacama Pathfinder Experiment (APEX) telescope (\citealt{2006Guesten}). It consisted of 280 transition-edge-sensor (TES) bolometers spread over an instantaneous 23 arcminute field-of-view (FoV).  The camera was designed for observations of the Sunyaev-Zel'dovich (SZ) effect at 150 GHz where the SZE is a decrement. 
APEX-SZ had a resolution of one arcmin and was used to observe 47 known massive galaxy clusters, with a total observation time of over 800 hours.
\subsection{APEX-SZ data analysis}\label{sec:datareduction}

The APEX-SZ data were flagged and filtered using the bolometer array
data analysis software
BoA\footnote{\url{http://www.apex-telescope.org/bolometer/laboca/boa/}}. A
series of linear filtering steps was carried out on the time-stream data of each
target, using universal settings to ensure a uniform analysis. We begin this Section with summaries of the calibration and time-stream filtering steps, and proceed to discuss our analysis in terms of the {\it point source transfer function} (described in section \ref{sec:pst}), constructed to take into account both the filtering steps and the instrument beam when modelling the sky signal.  

\subsubsection{Calibration}

The beam position and shape of each bolometer flux in the focal plane were
measured from daily scans of a calibrator (Mars, Uranus, Saturn or
Neptune). Side lobes were characterized by combining the individual
detector beams into a composite beam. Absolute flux calibration was
performed based on the response of each detector using scans of Mars
and Uranus. Depending on visibility of the primary calibrators, 
bootstrapped observations of secondary calibrators were also used. To account for
differences in atmospheric opacity between the data and calibration
scans, a correction was applied based on radiometer readings. A
further correction was applied to account for gain fluctuations due to
bolometers being biased near the upper edge of the superconducting
transition. The total calibration uncertainty for APEX-SZ is $\pm$10\%. The details of all these steps were discussed by \cite{2016Bender}, and are thus only summarized here.

\subsubsection{Time stream processing}
The time stream processing of the APEX-SZ data is similar but not
identical to that performed by \cite{2016Bender}. Thus, we give a relatively
detailed account of this process here.
The observations with APEX-SZ were carried out using circular drift scan
patterns centred on a constant horizontal coordinate, allowing the
target to drift through the pattern and the FoV. Circle radii were
chosen such as to maximize the integrated signal-to-noise ratio of
each target, based upon considerations of filtering effects (see
Section \ref{sec:pst}). The details of the APEX-SZ drift scan
pattern were discussed by \cite{2016Bender}. 
As a first step, the data were parsed into separate, full circles on
the sky, and re-grouped based on a common centre in horizontal
coordinates, resulting in what we shall call subscans. Data not
belonging to circle sets were discarded. Optically unresponsive
channels (bolometers) were rejected. Spikes were cut using sigma
clipping, and jumps (in DC level) were identified and corrected for using a wavelet-based algorithm. An
additional data cut was performed by analysing the correlation
between channels; channels found to correlate poorly with their
neighbours were rejected along with channels exhibiting levels of noise
significantly higher than the median noise level. Typically, 140-170 live channels were used for further analysis. After these initial
steps, an optical time constant (time delay in bolometer response) was de-convolved from each channel,
using the approach of \cite{2016Bender}.

The polysecant (a polynomial $+$ secant model) fitting employed by \cite{2016Bender} was also
used here. To the time stream of each channel and subscan, we fit a
6th order polynomial plus a normalization of the expected variation of
signal along a circle due to air mass load, and subtracted this
baseline from the data. Following this step, we removed a signal
correlated across all channels, constructed by taking the mean signal
adjusted for individual channel normalisations. Finally, a polynomial
of order 3 was fit to each set of two circles on the sky, before the
data were again de-spiked using sigma clipping.

\subsubsection{Point source transfer function}\label{sec:pst}

APEX-SZ observations were generally carried out at relatively high
(for the site) levels of precipitable water vapour due to significant amount of observation run concurring with the \textit{Bolivian Winter}. For this reason,
the APEX-SZ data suffer from excess low-frequency noise correlated on
scales much smaller than the FoV, requiring high-pass filtering of
individual bolometer time streams to be applied after removing the
correlated atmospheric signal. While this step enhances the
signal-to-noise ratio of detections, it also significantly attenuates
astrophysical signals. To account for this, we make use of a point
source transfer function (as described by \citealt{2009Halverson} and
\citealt{2009Nord}) to model the systematic signal loss. The point source transfer
function is unique for each target. It is constructed from a noiseless
simulation of a perfect point source at the position of the target,
convolved with the instrument point-spread function, de-gridded to the
bolometer time streams and processed in parallel with the data,
applying identical filtering to both the data and the
simulation. After filtering, the point source transfer function
represents the impulse response of the filter, and can be used, under
the assumptions of directional independence and linearity of the
filter, to model the response of any model that one may wish to
compare to the data.
Images of the data and the transfer functions were made using the
methods outlined by \cite{2016Bender}. For each target we also created 100
noise images by randomly inverting half the data (randomly chosen subscans), to characterize
instrument noise properties. 

\subsection{Optical follow-up observations and data}\label{sec:opt}

To obtain weak-lensing mass estimates for as many clusters in the full APEX-SZ sample as possible, we used a combination
of archival data and dedicated follow-up observations of clusters lacking sufficient amounts of high-quality weak-lensing data.
Follow-up observations were carried out between January 2010 and February 2012, using the Wide Field Imager (WFI) at the 2.2 MPG/ESO telescope at La Silla, Chile.

The observations were done in the B, V and R bands, with exposure times dependent on cluster redshift, reaching 12, 4.5 and 15 kilo-seconds per band, respectively, at z=0.3.
In combination with archival data, we were able to obtain quality weak-lensing data with the WFI for 21 clusters. 16 clusters had Suprime-Cam data from the 
Subaru telescope with imaging in at least three bands and sufficient quality for a weak-lensing analysis. For an additional 6 clusters we used a combination of WFI and Suprime Cam data, 
with at least one photometric band supplied by
the other instrument. For three clusters, sufficient amounts of data were available to perform independent weak-lensing analysis using both instruments separately.

For all clusters we used three band-photometry for background selection and required sub-arcsecond seeing for the shape measurement band. 
All colours were matched to the nearest colours available in COSMOS photo-z catalogues \citep{2009ApJ...690.1236I}, which were used as reference catalogues for background galaxy selection for all targets.

The optical follow-up and the weak-lensing analysis are described in detail by Klein et. al (in preparation). We present a summary of the weak-lensing analysis in Section \ref{sec:wlmass}.

\section{Measurements of mass proxies}\label{sec:massproxy}
We are interested in measuring the integrated Compton parameter, which probes the total thermal energy of the intra-cluster medium, using the filtered APEX-SZ images to fit parametric models for the pressure distribution of the intracluster medium (ICM). This process is described in section \ref{sec:inty}. For the absolute mass calibration we need corresponding mass estimates from the weak-lensing data, which are obtained from the optical data and for which the procedure is briefly described in section \ref{sec:wlmass}. For the eDXL sample, we re-measure the X-ray luminosities in a homogeneous manner in order to account for the sample selection (Section \ref{sec:method}). The associated steps are outlined in section \ref{sec:lum}. 
The mass measurement within a spherical radius $R_{500}$ is such that \begin{equation}\label{eq:m500}
M_{500}=500\rho_{c}(z)\frac{4\pi}{3} R_{500}^3\, ,
\end{equation}
where $\rho_{c}$ is the critical density of the Universe at a given redshift.

\subsection{Weak-lensing (WL) masses}\label{sec:wlmass}

We summarize the analysis of the weak-lensing data here. The lensing analysis adopted by Klein et al. (in preparation) implements a multicolour background selection in two colour and magnitude space. 
It differs from standard background selection methods (e.g., \citealt{2010Israel, 2012Israel, 2010Medezinski}) by including some of the efficiency of detailed photo-$z$ methods (e.g., \citealt{2017Serenolensing}) 
into typical colour-cut methods (e.g., \citealt{2016Okabe}) by calibrating the colour selection on a reference photo-$z$ catalogue from $z$-COSMOS \citep{2009ApJ...690.1236I}. 
%
The complete analysis and details are fully described in Klein et al. (in preparation) and \cite{2013Kleinthesis}. Below is a summary of the details on the background selection and further analysis investigating the contamination due to cluster members.

Both WFI and Suprime Cam instruments have demonstrated their suitability for weak-lensing measurements \citep[e.g.,][]{2002A&A...395..385C, 2002PASJ...54..833M}. 
For the optical follow-up data described in section \ref{sec:obs}, we use the \citet{2007A&A...468..823S} implementation of the KSB+ algorithm \citep{1995ApJ...449..460K, 2001A&A...366..717E} to measure the shapes of individual galaxies. Distortions of the point spread function (PSF) could be well modelled and corrected for, using polynomials of orders up to five. 
We used the data reduction process, PSF anisotropy correction and shape measurements similar to the analysis done by \cite{2010Israel}, \cite{2012Israel}.

The image distortion in the weak-lensing limit by a radially symmetric lens at an angular diameter distance $D_d$ from the observer can be measured as an average tangential ellipticity about the lens centre. The average tangential ellipticity $\varepsilon_t$ of source images is a direct measurement of the reduced shear $\langle g\rangle =\langle \varepsilon_t \rangle$. The reduced tangential shear is related to the shear, $\gamma$, and convergence or surface mass density, $\kappa$ as
\begin{equation} \label{eq:redshear}
g(\theta, \beta)=\frac{\gamma(\theta, \beta)}{1-\kappa(\theta, \beta)}\, ,
\end{equation}
where $\theta$ is the angular projected radial distance from lens centre and $\beta$ is a scale factor for the strength of the lensing effect.
It is defined as the angular distance ratios such that $\beta=\frac{D_{\rm ds}}{D_{\rm s}}$, where $D_{\mathrm{s}}$, $D_{\mathrm{ds}}$ are the angular diameter distances between observer and source and between deflector and source.  
 For redshifts lower than or equal to the cluster redshift, $\beta$ is equal to zero by construction. 
For higher redshift sources, $\beta(z_s)$ is a strictly monotonously increasing function of the source redshift $z_{\rm s}$. As such, it can be seen as a distance measurement that is proportional to the lensing signal and could be used for the selection of background sources. We select our background sources by determining a cut in $\beta$ that ensures to some degree the exclusion of cluster members and foreground. 
For this, we calibrate the colours against a reference photo-$z$ catalogue, COSMOS \citep{2009ApJ...690.1236I}. 

We estimate $\beta_i$ for each galaxy $i$ in the observation field as the weighted mean of $\beta_{k}$ of all sources $k$ in the COSMOS photo-z catalogue \citep{2009ApJ...690.1236I} within a region in colour-colour-magnitude space defined by the size of the photometric errors in colour and magnitude,

\begin{equation}\label{eq:beta}
 \beta_{i}=\!\frac{\sum_{k=1}^{N}w_k(\Delta c_1,\Delta c_2) \beta_k}
{\sum_{k=1}^{N}w_k(\Delta c_1,\Delta c_2)}\, .
\end{equation}
Each reference source ($k$) is weighted by a two dimensional Gaussian function, $w_{k} (\Delta c_1, \Delta c_2)$, where $\Delta c_1$, $\Delta c_2$ are the distance coordinates in colour-colour region of reference galaxy $k$, from the observed colour of the source galaxy, $i$. The dispersion of the Gaussian function is given by the actual measurement uncertainty on the observed colour of the galaxy, $i$. 
Due to a limited precision of the estimated $\beta_i$, a cut $\beta_{\mathrm{cut}}>0$  was applied to exclude cluster members and foreground galaxies. The first step in obtaining a meaningful background selection is finding the $\beta_{\mathrm{cut,max}}$ that maximizes the signal to noise of the lensing signal. Klein et al. (in preparation) show that this cut results in a bias of $\sim 1\%$ in $R_{200}$ due to noise fluctuations. This bias is avoided in the final mass analysis by increasing the applied cut, $\beta_{\mathrm{cut,fin}}=\beta(z_\mathrm{cut,max} + 0.05)$, the value that is obtained for a redshift 0.05 higher than that of $\beta_{\mathrm{cut,max}}$.

The mass estimate for each cluster was obtained by fitting a reduced tangential shear profile predicted by a projected Navarro-Frenk-White (NFW) profile \citep[e.g.,][]{1996A&A...313..697B} to the observed ellipticities . 
We derive the best fitting profile parameters $R_{200}$ and $c_{200}$ by minimizing the merit function 
\begin{equation} \label{eq:meritWL}
\chi^{2}\!=\!\sum_{i=1}^{N}{\frac{\left|g_{i}(\theta_{i},\beta_i;R_{200},c_{\mathrm{NFW}})\!-\!
\tilde{\varepsilon}_{\mathrm{t},i}(\theta_{i})\right|^{2}}
{\tilde{\sigma}_{\!i}^{2}\left(1\!-\!\left|
g_{i}(\theta_{i},\beta_i;R_{200},c_{\mathrm{NFW}})\right|^{2}\right)^{2}}}\quad . 
\end{equation} Here $g_{i}(\theta_{i},\Sigma_{\mathrm{crit},i};R_{200},c_{\mathrm{NFW}})$ is the model prediction for galaxy $i$ and $\tilde{\varepsilon}_{\mathrm{t},i}$ the observed ellipticity times 1.08 for the same galaxy. The factor 1.08 is the multiplicative shear calibration bias of the used KSB+ pipeline (\citealt{1995ApJ...449..460K, 2001A&A...366..717E}) to convert from measured to true ellipticity. This calibration bias has an uncertainty of $\sim 5$\%. This uncertainty is a dominant source of systematic uncertainty in the mass measurements.
Each shear profile was centred on the BCG, using distances in the range of 0.2 to 4.2 Mpc for the fitting procedure. We minimised the $\chi^2$ on a grid of $R_{200}$ and $c_{200}$. Finally, we used the mass-concentration relation described by \cite{2013ApJ...766...32B} to put priors on the concentration parameter to break the degeneracies in the profile models. 
The initial mass estimates from Equation \eqref{eq:meritWL} are biased. In evaluating the NFW shear profile we make use of the ratio in Equation \eqref{eq:redshear} when averaging the value of $\beta$ over the reference catalogue sources. However, $\frac{\gamma(\langle \beta \rangle)}{1-\kappa(\langle \beta \rangle)}\neq \left\langle \frac{\gamma(\beta)}{1-\kappa(\beta)}\right\rangle$. Given the finite width of the $\beta$ distribution that are averaged over when calculating $\beta_i$ from a reference catalogue (Equation \ref{eq:beta}), we find a biased point estimate for $\beta_i$. Especially in the inner regions of the cluster, this would model the shear profile incorrectly.
We estimate the final masses by correcting for the averaging over $\beta$ in two subsequent iterations.
We utilize the best-fit mass estimate from the zeroth iteration to predict the reduced shear, $g$, 
at the projected distance $\theta$ from the cluster centre, and $\beta_{k}$. 
We then introduce $\beta_i'$, which satisfies the equation:
\begin{equation}\label{eq:gbeta}
 g(\beta_{i}')=\!\frac{\sum_{k=1}^{N}w_kg(\theta_{i},\beta_k)}
{\sum_{k=1}^{N}w_k}\frac{1}{v_{b}(c_{1},c_{2})}\, .
\end{equation}

Here, the first term is the weighted average of the reduced shear given the projected distance $\theta_i$ of galaxy $i$ to the cluster centre and angular diameter distance ratios $\beta_k$ of references sources. The weights $w_k$ are identical to those used to derive $\beta_i$ and solely depend on the distance between reference and observed source in colour-colour space.  

The second term in equation \eqref{eq:gbeta} contains the map $\nu_{b}(c_{1},c_{2})$, an estimator of the overdensity of galaxies in colour-colour space with respect to a background estimate. This term addresses the different redshift distributions in the reference and cluster fields, assuming that they are caused by the addition of cluster galaxies. This can be seen as a radial and colour dependent contamination correction. We divide the cluster field into annuli. The background annulus is chosen to be beyond $R_{200}$ (using the $R_{200}$ estimate from the first iteration, equation \ref{eq:meritWL}). The region inward of $R_{200}$ is split into three overlapping annuli of width $0.3-0.5 R_{200}$. For each galaxy $i$, we compute $\nu_{b}(c_1,c_2)$ for one specific bin $b$ (depending on its angular separation $\theta_i$ from the cluster centre), with respect to the background annulus.

Under the assumption that the redshift distribution in the outskirts of the observed fields is close to the reference distribution, the density ratio maps $\nu_b$ in colour-colour space reflect the difference of the two distributions at a given position in colour-colour space.
Ignoring the impact of lensing magnification, the cluster always causes an excess of galaxies compared to the average distribution. To avoid correcting to insignificant noise fluctuations, $v_{b}(c_{1},c_{2})$ is set to 1 for all colour-colour regions with an excess smaller than two sigma above the mean value. 
Visual inspection of the $v_{b}$ images was performed to ensure that overdensities caused by additional clusters in the observed fields with redshifts higher than those of the targeted clusters are not considered in the correction.

Equation \eqref{eq:gbeta} describes the expected reduced shear at position $\theta_i$ given the expected redshift distribution of reference sources and the expected contamination caused by cluster galaxies given the colours of the observed galaxy $i$. As such $\beta_{i}'$ is a less biased estimator than $\beta_{i}$. 
The $\chi^2$ is re-computed in a grid of $c_{200}$ and $R_{200}$ with the updated $\beta'$. We re-iterate once more and compute final mass estimates minimising Equation \eqref{eq:meritWL} and applying a procedure identical to the first iterations of mass estimates. 
In Table \ref{tab:39clusters}, we give the spherical masses within $R_{500}$ defined in Equation \eqref{eq:m500} of the final iteration.

%
 
\subsection{Integrated Comptonization}\label{sec:inty}
The SZE distortion, $\Delta T_{\rm SZE}$, of the cosmic microwave background temperature $T_{\rm CMB}$, is given by, 
\begin{equation}\label{eq:tsz}
         \frac{\Delta T_{\rm SZE}}{T_{\rm CMB}}= f(x) \int \sigma_{\rm T}n_{\rm e}\frac{ k_{\rm B} T_{\rm e}}{(m_{\rm e}c^2)} dl     =f(x) y \; ,    
\end{equation} 
where $l$ is the line of sight variable, $\sigma_{\rm T}$ is the Thomson scattering cross-section for electrons, $m_{\rm e}$ is the electron mass, $k_{\rm B}$ is the Boltzmann constant, and $c$ is the speed of light. $T_{\rm e}$ is the electron temperature of the X-ray emitting plasma and $f(x)$ gives the spectral shape of the effect, given by
\begin{equation}
       f(x) \equiv \left(x\frac{e^{x}+1}{e^{x}-1}-4\right)(1+\delta_{\rm SZE}(x,T_{e})) \;,
\end{equation}
where $x$ is the dimensionless frequency related to the frequency by $x=h\nu/k_{\rm B}T_{\rm CMB}$. $\delta_{\rm SZE}(x,T_{\rm e})$ is a correction due to relativistic effects (e.g., \citealt{1998Itoh, 1998Challinor}).
The frequency independent measure $y$ is the line-of-sight Compton parameter, proportional to the electron pressure integrated along the line of sight as 
\begin{equation}
 y= \sigma_{\rm T}/m_{\rm e}c^2 \int P_{\rm e}(l) \mathrm{d}l \;,
\end{equation}
where $P_{\rm e}=n_{\rm e} k_{\rm B} T_{\rm e}$ is the electron pressure.
The integrated Compton parameter, denoted $Y$, is defined by
\begin{equation}
 Y =\int y d\Omega\;,
\end{equation}
where the integration is over solid angle $\Omega$ in a given aperture, resulting in a cylindrically integrated quantity which we shall refer to as $Y_{\rm cyl}$. Given an azimuthally symmetric radial model, $Y_{\rm cyl}$ can be converted to the spherical counterpart $Y_{\rm sph}$, representing the integrated Comptonization in a sphere of corresponding radius. The SZ Comptonization in terms of its physical units (or extent) is given by
$Y_{\rm SZ} =D_{\rm A}^{2}\; Y_{\rm sph}$, where $D_{\rm A}$ is the angular diameter distance of the cluster determined by cosmology and redshift.

We next describe the parametric model used in our analysis and the procedure by which we extract the best-fit parameters. The correlation in the measurement of the weak-lensing mass and the SZ integrated Comptonization (due to the choice of $r_{500}$) is discussed in section \ref{sec:covar}.
%
\subsubsection{Generalized Navarro-Frenk-White profile }\label{sec:model}
To model the pressure of the ICM we use the generalized Navarro-Frenk-White (gNFW) profile as motivated by dark matter halo profiles found from simulations \citep{2007Nagai}. In this framework, the pressure profile $P(R)$ as a function of radius $R$ is given by
\begin{equation}
P(R) = \frac{p_{0}}{\left(\frac{R}{R_{\rm s}}\right)^{\gamma} \left(1+\left(\frac{R}{R_{\rm s}}\right)^{\alpha}\right)^{\left(\frac{\beta - \gamma}{\alpha}\right)}},
\end{equation}
where $p_0$ is a normalization, the logarithmic slope parameters $\alpha$, $\beta$ and $\gamma$ describe the intermediate, outer and inner part of the pressure profile, respectively, and the scale radius ($R_{\rm s}$) is related to $R_{500}$ by the concentration parameter $c_{\rm gnfw}$ as 
\begin{equation}\label{eq:rs}
  R_{\rm s} = \frac{R_{500}}{c_{\rm gnfw}} \;.
\end{equation}

The peak signal-to-noise ratios of the APEX-SZ detections in beam smoothed maps range from $\sim16$ down to non-detections. Due to scan pattern and high pass filtering, scales larger than $\sim 10^{\prime}$ are not recovered in the filtered data. 
Due to the degeneracy in the  $c_{\rm gnfw}$ and $R_{500}$ of the gNFW model and the limitations set by the data, we use the weak-lensing estimate of the spherical over-density radius. Furthermore, we assume the slope parameters of \cite{2010arnaud}, with \{$c_{\rm gnfw}$, $\alpha$, $\beta$, $\gamma$\}=\{1.177, 1.0501, 5.4905, 0.3081\}. Thus, the normalization of the gNFW profile of each cluster is the only free parameter. 

In order to be consistent with the weak-lensing analysis, the centroids were fixed to the BCG centres used for the former in Section \ref{sec:wlmass}. 
 Based on the assumed parametrisation of the gNFW profile, $Y_{\rm  sph,500}$ is obtained by dividing the $Y_{\rm cyl,500}$ by a factor of 1.203. 
 To check for statistical compatibility between the integrated Comptonizations reported in \cite{2016Bender} and this work, we use the centroids and apertures quoted in the former and re-measure the integrated Comptonizations for 41 clusters. We find a statistical agreement between the two pipelines across this sample. Further details are given in Appendix \ref{app:bender}.
\subsubsection{Model Fitting}\label{sec:modelling}
\begin{figure}
 \includegraphics[width=\columnwidth]{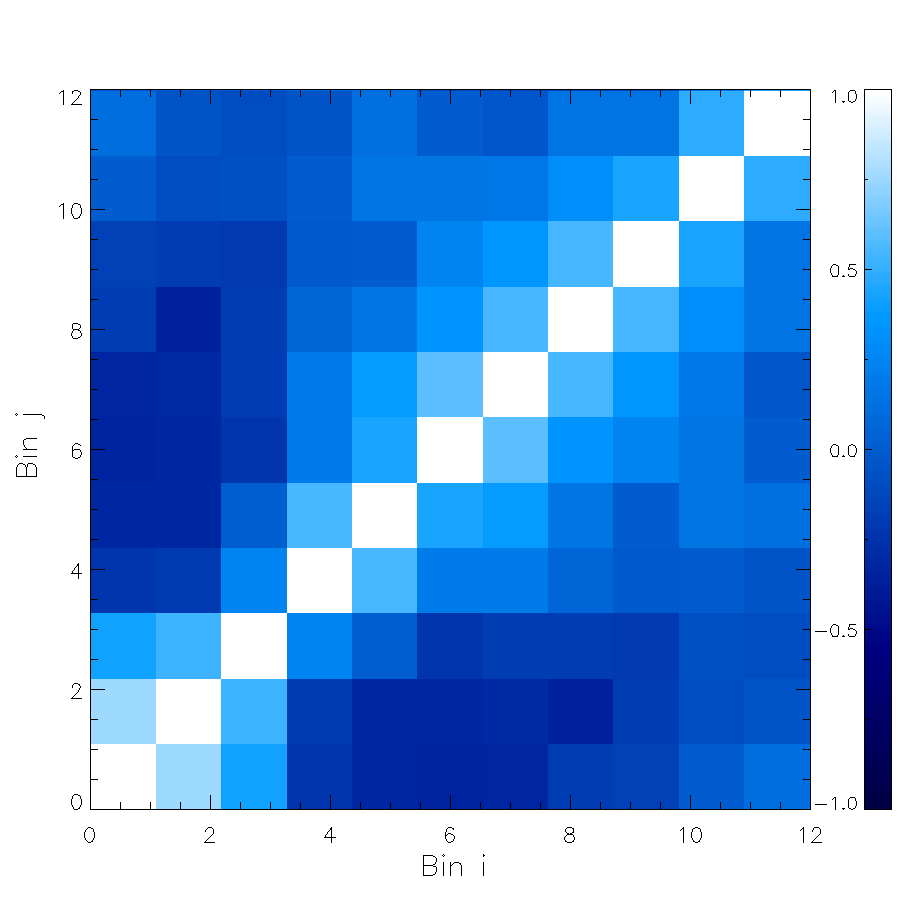}\caption{Bin-to-bin correlation matrix for the Bullet cluster field. The bin-width is 1$'$. The neighbouring bins are correlated due to the beam size of the telescope. The anti-correlations between intermediate and central bins are due to the low order polynomial baseline filtering of the time streams. }\label{fig:Bulletcm}
\end{figure}
To fit the gNFW model to the data, we bin the data about the BCG centroids used in the weak-lensing analysis, considering all data within a radius of $12^{\prime}$, using a bin width of $1^{\prime}$ (corresponding to the FWHM of the APEX-SZ beam) and taking pixel weights into account in the averaging. The model image is convolved with the point source transfer function to account for the finite resolution and filtering effects, as discussed in section \ref{sec:pst}, and binned in the same way as the data. 

For each target, the bin-to-bin noise covariance matrix was computed from 100 noise realizations, produced by randomly inverting half the data. Because noise realizations produced in this way do not account for noise covariance components produced by astronomical signals, we generate random realizations of primary CMB anisotropies using the \cite{2015PlanckXI} best fit CMB power spectrum, convolve these with the transfer function and add the filtered CMB realizations to the instrument noise realisations. The noise contributions from unresolved point sources emitting synchrotron and dust emission at $150\; \rm GHz$ can be neglected for the APEX-SZ noise levels \citep{2009Reichardt}.
We radially bin the final noise images using the scheme described above. The ensemble of noise
realisations in each radial bin is used to compute the full bin-to-bin covariance matrix. An example correlation matrix is illustrated for the Bullet cluster in Figure \ref{fig:Bulletcm}. Neighbouring bins are strongly correlated due to the telescope resolution, while intermediately separated radial bins are anti-correlated due to the low-order polynomial filtering applied to attenuate low-frequency noise. 

We use the $\chi^2$ statistic to define our likelihood $\mathcal{L} \propto \mathrm{exp}(-\chi^2/2)$, with
\begin{equation}\label{eq:chi}
\mathbf{
 \chi^{2} = (d - m)^{T} C^{-1} (d - m ) },
\end{equation}
where  $\mathbf{d,~ m}$ are vectors of the radially binned data and the filtered model, and $\mathbf{C}$ is the bin-to-bin covariance. The fitting was done using an Markov chain Monte Carlo analysis (MCMC) to estimate the confidence levels of the normalisation parameter. The $Y_{\rm SZ,500}$ for each cluster is computed using the formulation given in previous section for the recovered models.  In the following section we use the MCMC approach to estimate the correlation in the measurements of $Y_{\rm SZ,500}$ and $M_{\rm WL,500}$.   

\begin{figure*}
	\includegraphics[width=15cm]{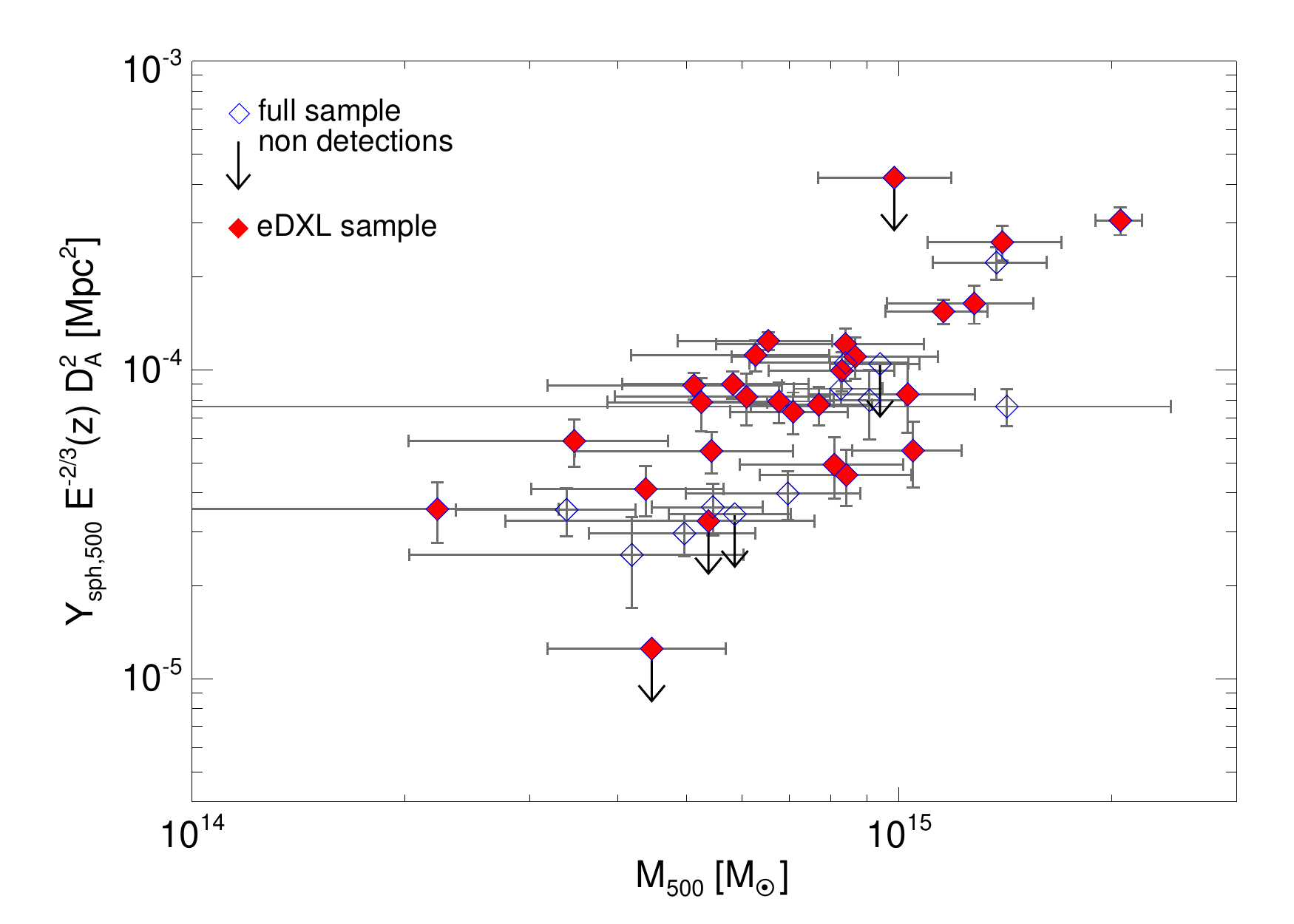} 
    \caption{ $Y_{\rm sph,500}$ measurements vs. the weak-lensing spherical mass estimates within $R_{500,\rm WL}$. The non-detections in the integrated Compton parameter are indicated by upper limits at a 2$\sigma$ level (downward arrows).}
    \label{fig:yszmass}
\end{figure*}

The measured $Y_{\rm SZ,500}$ for the full set of 39 APEX-SZ clusters are given in Table \ref{tab:39clusters}. The measured $Y_{\rm SZ}$ and the lensing masses $M_{\rm WL, 500}$ for the full sample of 39 clusters is shown in Figure \ref{fig:yszmass}. In total, five clusters are non-detections in the measured Comptonizations, whose measured Compton-Y's is within $3\sigma$ of the noise level. Three of them are part of the eDXL sample.

\begin{table*}\caption{The 39 targets of the full APEX-SZ sample. Each cluster position is taken to be the position of the BCG in the optical data. We report the integrated Compton-y parameters ($Y_{\rm SZ,500}$), the weak-lensing derived masses ($M_{\rm WL,500}$), and the re-measured ROSAT luminosities ($L_{\rm x}$). The clusters in the sample that are non-detections in Compton-Y's are indicated by asterisk ($*$).}
\label{tab:39clusters}
\renewcommand{\arraystretch}{1.4}
\begin{tabular}{l c c c c r c }
\hline
\multicolumn{1}{c}{Name} & RA & Dec & redshift&\multicolumn{1}{c}{ $Y_{\rm SZ,500}$}  &\multicolumn{1}{c}{ $M_{\rm WL,500}$} &\multicolumn{1}{c}{$L_{\rm x}[0.1-2.4 \rm \, keV]$}\\
& & &$z$& $[10^{-5}\, \rm Mpc^{2}]$ & $[ 10^{14}\rm M_{\sun}]$&$[ 10^{44} \, \rm erg \, s^{-1}]$\\
\hline
\multicolumn{7}{c}{eDXL clusters}\\
\hline
 A$2204$           &$16$:$32$:$46.9$	&$+05$:$34$:$32.3$ & 0.152 &$13.05   \pm      0.86 $   &      $6.53 \, _{-1.67 }^{+1.51} $  &  $14.2 \pm 0.6$     \\
 RXCJ$2014.8-2430^{*}$ &$20$:$14$:$51.7$	&$-24$:$30$:$22.3$ &  0.160  &   $ 0.97 \pm   1.22 $   &       $5.37\, _{-2.60 }^{+2.22 } $   &  $\; 9.8 \pm 1.1$ \\
A$1689$         	& $ 13$:$11$:$29.5$	&$-01$:$20$:$27.9$ &0.183  &$  32.4   \pm  3.36    $ &      $20.56\, _{-1.59}^{+ 1.51}$ & $12.5 \pm 0.9$   \\
A$2163$          	& $16$:$15$:$49.0$	&$-06$:$08$:$41.5$ &0.203 &$ 17.56 \pm  2.47 $  &            $12.78\, _{-3.17}^{+ 2.72  }  $  & $19.5 \pm 1.2$  \\ 
RXJ$1504$         & $15$:$04$:$07.5$	&$-02$:$48$:$16.5$ & 0.215 & $  8.45  \pm 1.66  $     &      $5.25\, _{-1.38}^{+ 1.25} $  &    $25.0 \pm 1.4$   \\
RXCJ$0532.9-3701$ &$05$:$32$:$55.7$	&$-37$:$01$:$36.0$ & 0.275  &     $8.68 \pm 1.31 $     &     $6.76\, _{-1.53}^{+ 1.33  } $     &   $\; 6.1 \pm 0.7$   \\
RXCJ$0019.0-2026$ &   $00$:$19$:$08.0$&$-20$:$26$:$28.0$ & 0.277&     $ 8.47 \pm 1.22   $   &$7.70\, _{-1.53}^{+ 1.45} $  &  $\; 6.1 \pm 1.1$\\      
RXCJ$2337.6+0016$ &$23$:$37$:$39.7$	&$+00$:$16$:$17.2$ &  0.278   &  $ 8.04  \pm 1.25  $   & $     7.08\, _{-1.32 }^{+1.37}$ &$\; 6.4 \pm 1.0$\\
RXCJ$0232.2-4420$ &$02$:$32$:$18.6$	&$-44$:$20$:$48.0$ & 0.284 &      $ 9.82\pm 0.98  $     &     $5.13\, _{-1.94 }^{+1.69}$        & $11.1 \pm 1.2$  \\  
RXCJ$0437.1+0043$ &$04$:$37$:$09.5$	&$+00$:$43$:$52.1$ & 0.284 &      $  5.44 \pm  1.22 $   &      $8.10\, _{-2.15 }^{+2.03 } $     &  $\; 7.6 \pm 0.9 $\\
 RXCJ$0528.9-3927$ &$05$:$28$:$53.0$	&$-39$:$28$:$17.8$ & 0.284  &    $4.53   \pm   0.84 $  &    $  4.38\, _{-1.36}^{+ 1.26 }  $    & $ 12.4 \pm 1.2$   \\
RXCJ$2151.0-0736$ &$21$:$51$:$00.8$	&$-07$:$36$:$31.0$ & 0.284   &  $ 3.90 \pm 0.86     $  &       $2.22\, _{-1.31 }^{+1.10}   $ &  $\; 7.1 \pm 1.4$   \\
A$2813$         	&$00$:$43$:$25.1$	&$-20$:$37$:$01.2$ & 0.292& $  10.99  \pm 1.60    $    &$8.30\, _{-1.75 }^{+1.54} $ & $\; 7.8 \pm 1.2$  \\
   RXCJ$0516.6-5430$ &$05$:$16$:$37.6$	&$-54$:$30$:$38.1$ & 0.295  &    $ 5.05\pm 1.03      $ &      $8.42\, _{-2.06}^{+ 1.99}$       & $10.7 \pm 1.4$   \\ 
  Bullet        	&$06$:$58$:$36.4$	&$-55$:$57$:$19.2$ &0.297  &$  12.35   \pm 1.44       $ &      $6.30\, _{-2.09}^{+ 1.71}  $  &  $21.0 \pm 1.7$ \\
   A$2537$       	&$23$:$08$:$22.2$	&$-02$:$11$:$31.6$ & 0.297 &$     6.06    \pm 1.44 $   &      $10.46\, _{-1.87 }^{+1.80} $ & $\; 9.8 \pm 1.6$\\  
   RXCJ$0245.4-5302$ &$02$:$45$:$31.3$	&$-53$:$02$:$07.8$ & 0.302 &      $ 6.53 \pm 1.14$     &      $3.47\, _{-1.45}^{+ 1.24 }$        &  $\; 6.1 \pm 0.8  $ \\ 
   RXCJ$1135.6-2019^{*}$ &$11$:$35$:$21.4$	&$-20$:$19$:$56.6$ & 0.305  &  $  -0.64\pm 1.02 $      &       $4.47\, _{-1.29}^{+ 1.22 }  $       &  $\; 6.9 \pm 1.4$ \\
   A$2744$       	& $00$:$14$:$18.5$	&$-30$:$22$:$51.2$ &	 0.307&   $ 17.19    \pm 1.59 $    &$11.55\, _{-2.00}^{+ 1.79}$  & $12.2 \pm 1.6  $ \\
  A$1300$          	&$11$:$31$:$54.2$	&$-19$:$55$:$39.8$ &0.308 & $    9.99  \pm 1.02   $     &      $5.82\, _{-1.76 }^{+1.62 } $   & $13.2 \pm 1.6$    \\
  MACSJ$1115.8+0129$& $11$:$15$:$52.0$	&$+01$:$29$:$55.0$ &  0.348 &   $ 6.18  \pm 0.94  $    &       $5.43\, _{-1.95}^{+ 1.65 }    $      & $14.6 \pm 1.7$ \\
RXCJ$2248.7-4431^{*}$ &$22$:$48$:$44.0$	&$-44$:$31$:$51.0$ &  0.348    &     $6.44\pm  20.42$  &     $9.84\, _{-2.16}^{+ 2.01}  $&$28.6 \pm 3.1$   \\
RXCJ$1206.2-0848$ &  $12$:$06$:$12.1$	&$-08$:$48$:$03.4$ & 0.441   &   $ 12.91\pm 1.97 $      &      $8.67\, _{-2.87}^{+ 2.68 }  $     &  $15.6 \pm 2.6$  \\
    RXCJ$2243.3-0935$ &$22$:$43$:$22.8$	&$-09$:$35$:$22.0$ &  0.447    & $9.78  \pm 2.46 $    &      $10.29\, _{-2.84}^{+ 2.52}  $  & $14.0 \pm 3.4$ \\
RXJ$1347-1145$    &  $13$:$47$:$30.6$	&$-11$:$45$:$09.5$ & 0.451  & $ 30.4 \pm   3.92   $    &     $14.00\, _{-3.02 }^{+2.96} $ &  $ 36.9 \pm 3.8$     \\  
RXCJ$2214.9-1359$ &$22$:$14$:$57.2$	&$-14$:$00$:$12.3$ &  0.483    & $ 14.53\pm  1.78   $   &      $8.39\, _{-2.88 }^{+2.45}   $ & $14.7 \pm 3.3$ \\
MS$0451.6-0305$   &$04$:$54$:$10.8$	&$-03$:$00$:$51.4$ & 0.539 &   $  9.98 \pm 1.91   $    &     $6.08\, _{-2.12 }^{+1.90 } $  &  $16.7 \pm 3.4$   \\  
\hline
\multicolumn{7}{c}{Other clusters}\\
\hline
A$907$          	&$09$:$58$:$22.0$	&$-11$:$03$:$50.2$ & 0.153 & $   3.71   \pm 0.66 $    &      $ 3.38\, _{-1.02}^{+0.85 } $   &   -\\ 
A$3404$         	&$06$:$45$:$29.5$	&$-54$:$13$:$37.1 $& 0.167 & $   11.12 \pm  1,40 $    &      $8.40\, _{-2.26 }^{+1.91}    $  & -   \\ 
A$383^{*}$        	&$02$:$48$:$03.4$	&$-03$:$31$:$45.1$ & 0.187 & $ 1.80    \pm 0.91  $   &      $5.86\, _{-1.13}^{+ 1.17}$    &  - \\   
A$520$         	&$04$:$54$:$13.7$	&$+02$:$56$:$10.2 $&0.199 & $   3.84   \pm  0.73 $   &       $5.45\, _{-0.98 }^{+0.96} $    &  - \\ 
   A$209$         	&$01$:$31$:$52.5$	&$-13$:$36$:$40.7$ &0.206 &$  8.55  \pm  2.16    $   &$9.08\, _{-1.22 }^{+1.12} $  &   - \\       
   A$2390^{*}$          	&$21$:$53$:$36.8$	&$+17$:$41$:$43.7$ & 0.228& $   4.91  \pm  3.18 $    &$       9.40\, _{-1.41 }^{+1.28} $&  -   \\
   A$1835$         	&  $14$:$01$:$02.1$	&$+02$:$52$:$42.6$ & 0.253& $   24.25  \pm 2.92  $     &     $13.74\, _ {-2.58 }^{+2.45}  $ &  -   \\
RXCJ$1023.6+0411$ &$10$:$23$:$39.2$	&$+04$:$10$:$58.0$ & 0.280  &   $  9.59 \pm 1.16  $   &     $8.28\, _{-1.19}^{+ 1.20}  $      & - \\ 
XLSSC-$006$       &$02$:$21$:$45.2$	&$-03$:$46$:$02.7$ &	0.429&  $  3.44 \pm   0.54  $    &$     4.97\, _{-1.33}^{+ 1.28}$   & - \\
MACSJ$1359.2-1929$&  $13$:$59$:$10.3$	&$-19$:$29$:$24.7$ &  0.447  & $ 2.95  \pm 0.96 $     &       $4.19\, _{-2.16 }^{+1.84 }   $ & -   \\
MACSJ$1311.0-0311$&  $13$:$11$:$01.8$	&$-03$:$10$:$39.7$ & 0.494   & $  4.75 \pm 0.86$     &         $6.96\, _{-1.97}^{+ 1.86 }   $     & -\\
MS$1054.4-0321$   &$10$:$56$:$60.0$	&$-03$:$37$:$36.2$ & 0.831  & $   10.39 \pm 1.42  $     &    $14.20\, _{-14.57 }^{+10.06 } $  &  -    \\

\hline
  \end{tabular}                              
		
\end{table*}

\subsubsection{Propagation of uncertainties in $R_{500}$ into the SZ modelling}\label{sec:covar}

Since by definition $M_{\rm WL, 500}$ is proportional to $R_{500}^{3}$ (Equation \ref{eq:m500}), the obtained $Y_{\rm SZ,500}$ values are correlated with the weak-lensing masses because the same apertures were used for measuring both quantities. 
      
We estimate this correlation by using the MCMC and re-fitting the gNFW profile with $R_{\rm s}$ and $p_{0}$ as free parameters. We use a prior on $R_{\rm s}$ from the  weak-lensing estimates of the $R_{500}$ distribution via the relation given by Equation \eqref{eq:rs}. We propagate the $R_{500}$ uncertainties modelled as a two-sided Gaussian distribution into our modelling of the SZ signal. The correlation in the measurements is determined using a Pearson correlation coefficient from the recovered distribution of $Y_{\rm SZ,500}, ~M_{\rm WL,500}$.

 \subsection{X-ray observables and parameter estimation}\label{sec:lum}
 
Our procedure to consistently recompute the ROSAT X-ray luminosities for all the eDXL clusters derives from the REFLEX-II recipes described in \cite{2013Boehringer}. 
The measurements rely on ROSAT PSPC photon and exposure maps in the [$0.5\text{--}2$] keV, where the signal-to-noise ratio is highest. 
However, the final luminosities quoted and used in this work correspond to the full [$0.1\text{--}2.4$] keV band, as is customary for ROSAT sources. 
The conversion between the two bands make use of the redshift and temperature dependent K-correction tables provided by \cite{2004Boehringer} which 
show little variation over a wide temperature range. 

\begin{figure}
\includegraphics[keepaspectratio=1,width=\columnwidth]{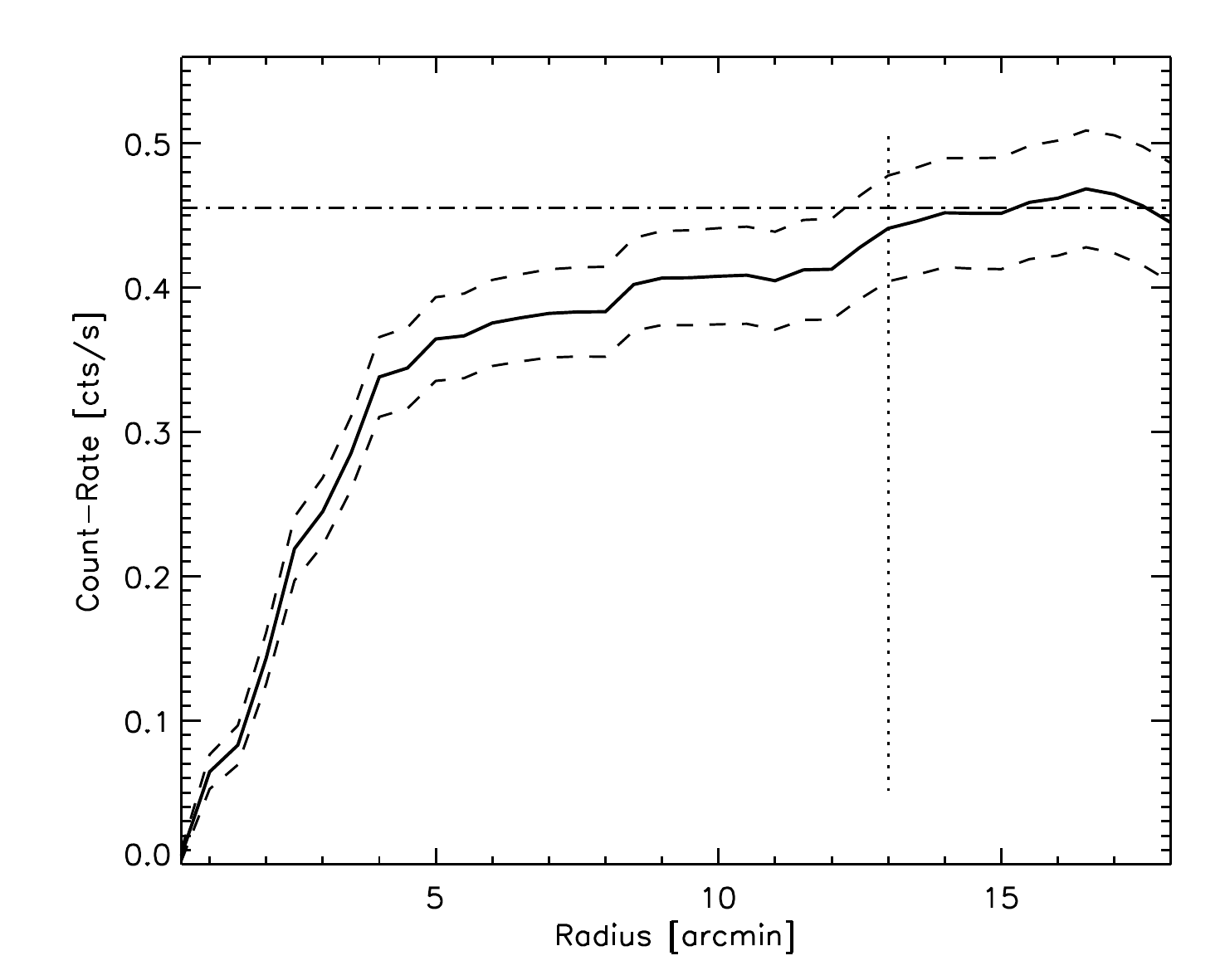}
\caption{Result of the growth curve analysis for the Bullet cluster. 
The net aperture count-rate of the cluster is plotted against the radial distance from the centre (solid line) 
together with its $1\sigma$ uncertainty (dashed lines).
The integrated count-rate flattens, i.e. shows fluctuations lower than the $1\sigma$ error range, after the 
radius $R_{\rm x}$, indicated by the vertical dotted line. 
A constant count-rate (CR), indicated by the horizontal dot-dashed line, is fitted to this plateau region 
and serves as the main source photometry indicator.}\label{fig:CR}
\end{figure}

\begin{figure}
 \includegraphics[width=\columnwidth,angle=0]{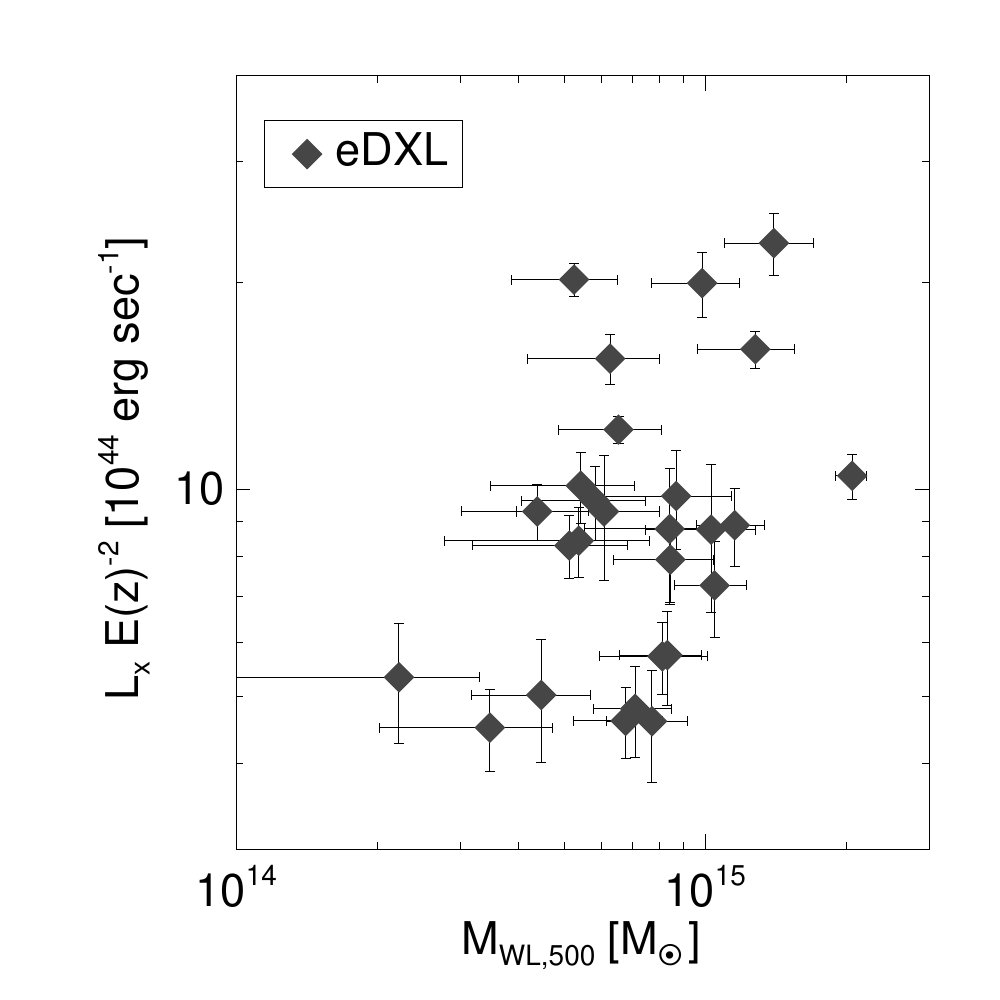} \caption{ROSAT luminosities $L_{\rm x}[0.1\text{--}2.4\rm \, keV]$ and weak-lensing spherical masses. The spherical weak-lensing masses are measured within $R_{500}$. The luminosities are measured within a $R_{500}$ that is independent of the weak-lensing analysis (See section \ref{sec:lum}).
 }\label{fig:mlscat}
\end{figure}
\begin{figure}
      \includegraphics[scale=1,width=\columnwidth,keepaspectratio=true, angle=0]{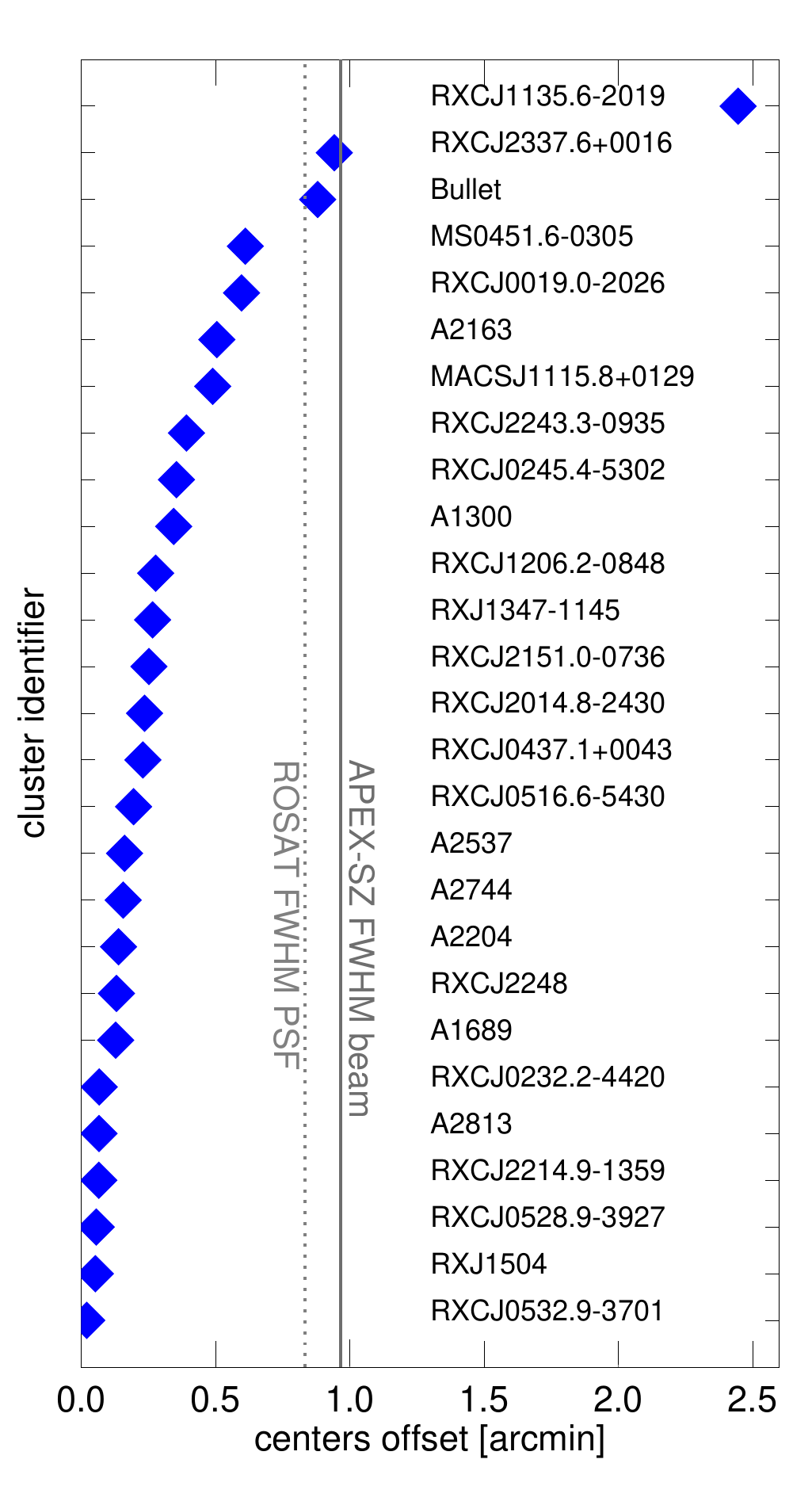}
\caption{The angular offset between the centres from ROSAT X-ray and the centres adopted for weak-lensing mass estimate are shown for the 27 eDXL clusters. 80\% of the clusters have an offset between  optical and X-ray centres that is smaller than half of the APEX-SZ beam. RXCJ1135 shows by far the largest offset of 2.5 arcmin.  }\label{fig:centeroffset}
     \end{figure}
The process can be split into the following main steps:
\begin{enumerate}
  \item The X-ray centroid for each cluster in the sample was calculated from the ROSAT photon map within a $3^{\prime}$ aperture, iteratively 
  updating the centre of the aperture until convergence. 
  \item The local background for each cluster was computed inside an annulus covering the radial range $20^{\prime}$-41.3$^{\prime}$. 
  To account for the possible contamination by surrounding AGNs, this annulus was split into 12 sectors azimuthally. 
  The background count-rate in each sector was estimated and contaminated areas were rejected using an iterative $2.3\sigma$ clipping. 
  The mean background was finally computed from the remaining sectors. Such a procedure is justified by the low AGN density in the ROSAT maps.
  \item A growth curve analysis as prescribed in \cite{2013Boehringer} was used to estimate the integrated net aperture count-rate 
  of the source in a suitable radius. The integration radius, $R_{\rm x}$, is first defined as the radius above which all changes in the 
  integrated flux stay within the 1-sigma error range at that radius. 
  The corresponding integrated source count-rate,  $\rm CR$, is then estimated by fitting a straight line to the plateau at larger radii, as shown in Figure \ref{fig:CR}.
  \item Finally, we estimated the value of $L_{\rm x,500}$ in the [0.1--2.4] keV band corresponding to the measured $\mathrm{CR}(<R_{\rm x})$. 
  For this, we first use the $L_{\rm x,500}\text{--}T_{\rm x}$  relation of \cite{2009Pratt}, 
  \begin{equation}        
       \frac{T_{\rm x}}{1 \rm \, keV} = 3.31  \left(\frac{L_{\rm x,500}}{10^{44} \rm erg ~s^{-1}}\right)^{0.332} h_{70}^{0.666}\;, 
  \end{equation}
  to estimate the temperature dependent K-correction suitable for any given $L_{\rm x,500}$, and convert it to the [$0.5\text{--}2$] $\rm keV$ ROSAT 
  count-rate in $R_{500}$, $\mathrm{CR}_{500}$.
 $L_{\rm x,500}$ is the X-ray luminosity within $R_{500}$. 
 $T_{\rm x}$ is the X-ray temperature. 
  Following the results of \cite{2011Reichert}, we assumed the redshift dependence of the $L_{\rm x,500}-T_{\rm x}$ relation to be negligible.
  Then, we use the \cite{2011Reichert} Mass-Luminosity relation expressed as
 \begin{equation}
    \frac{R_{500}}{1 \rm ~Mpc} = 0.957 \left(\frac{L_{\rm x}}{10^{44} \rm erg ~ s^{-1}}\right)^{0.207} E(z)^{-1}h_{70}^{0.586}\;,
 \end{equation} 
 to estimate the radius $R_{500}$ within which $\mathrm{CR}_{500}$ should be measured. 
 Lastly, we assumed a fixed beta-model with $\beta=2/3$ and $R_{\rm c} = R_{500}/7$ to estimate the extrapolation factor from $\mathrm{CR}_{500}$ 
 to $\mathrm{CR}(<R_{\rm x})$. The full conversion process is performed for a grid of $L_{\rm x,500}$ and the correct value is obtained 
 after interpolation over the estimated $\mathrm{CR}$.
\end{enumerate}

 The X-ray luminosities obtained from ROSAT vs. the lensing masses for the eDXL sample are shown in Figure \ref{fig:mlscat}.
The above procedure provides us with complementary and independent information
on cluster centroids for the baryonic component emitting X-rays, however, these estimates are less precise due to the low resolution of the ROSAT PSF, which makes these estimates sub-optimal. 
The $Y_{\rm SZ}$ measured in section \ref{sec:inty} used optical centres (i.e., BCG). %
The centroid offset between BCG and the ICM gas profile can bias the measured $Y_{\rm SZ}$, however, the optical centroids determined for the lensing analysis offers the best possible option for centering the gNFW profiles, given that the X-ray centres are determined from the low resolution X-ray observations. In Figure \ref{fig:centeroffset}, we compare the X-ray centres of the eDXL clusters identified in ROSAT survey and the optical centres that were used to measure the weak-lensing masses. 
We find that most of the clusters including merging systems like Bullet have centroid offsets in optical and X-ray at a level lower than the APEX-SZ FWHM beam. The single most extreme outlier is RXCJ$1135$, which is a double cluster system that appears to be in a pre-merger state. 
It has two dark matter peaks and a third diffuse one in between the two. The ROSAT X-ray centre lies between the two DM peaks. Measuring the $Y_{\rm SZ}$ signal measured at the optical centre yields a non-detection, whereas at the X-ray centre we obtain a 3$\sigma$ level detection. 
 Therefore, we analyze the scaling relation also for the re-measured $Y_{\rm SZ}$ at X-ray centres as a robustness check of the scaling parameters constraints to mis-centering of the gNFW profiles and discuss this later in Section \ref{sec:xraycenters}.  
      
\section{Method}\label{sec:method}

We present a Bayesian method to account for sample selection biases in the scaling relations for the eDXL sample in which the sample selection is well-defined. 
 Several authors have discussed using Bayesian techniques for measuring cluster scaling relations \citep[e.g.,][]{2007Kelly, 2013Andreon, 2014Maughan, 2010Mantz, 2015Sereno}.
In this work, we apply a Bayesian formalism for measuring jointly multiple mass-observable scaling relations by accounting for a truncated selection in a measured cluster property.  
We differ from some of the other work by not requiring for a model to predict the number counts of the underlying or missing population \citep[e.g.,][]{2010Mantz} but still accounting for the shape of the underlying cluster mass function, the sample selection, the measurement uncertainties in cluster properties and masses, and the intrinsic covariances of cluster properties. 
The completeness of the sample allows us to compute a semi-analytical approximation for accounting for non-ignorable sample selection effects. In particular, we deal with this impact for measuring the mass scaling relations of cluster observables that do not play a role in the selection of cluster members of a sample.  
Our likelihood presented here bears the most similarity to the XXL likelihood used by \cite{2016Giles}, however, they use temperature function to model the underlying cluster population, include a scaling relation between two cluster properties with only one intrinsic scatter, their selection depends on two observables rather than measured properties, and they measure the scaling relation between cluster properties that play some role in the sample selection. 

In Section \ref{sec:statmodel}, we outline the general framework of the method while Section \ref{sec:method:application} discusses the application of that statistical model to the eDXL sample. In Section \ref{sec:method:tests}, we validate our application of the statistical model for our eDXL measurements through analyses of mock data. 


\subsection{Statistical model}\label{sec:statmodel}

 The key ingredients for the statistical model to determine the posterior distribution of the parameters of interest are described below: 
\begin{enumerate}
\item 
The mass variable, $m$, is the fundamental variable that describes a cluster and relates to all other observables, arranged in a vector $\boldsymbol{\xi}$, through a scaling model $P(\boldsymbol{\xi}|m,\theta)$ that is fully described by parameters $\theta$, which needs to be determined. 
  The full probability distribution in mass-observable plane is obtained from the
conditional probability rule:
\begin{equation}\label{eq:pxim}
 P(\boldsymbol{\xi},m|\theta)=P(\boldsymbol{\xi}|m,\theta) P(m)\, .
\end{equation}
\item To model the conditional probability $P(m)$, some authors leave a large freedom for this function by introducing flexible parametric model \citep[such as the 
multiple Gaussians of][]{2007Kelly}, to be constrained simultaneously in the
fit. However, there is some knowledge of the cluster mass function
and using it reduces degeneracies in the fit. Therefore, we use the cluster mass function as the $P(m)$.
In practice, it is evaluated using the Tinker mass function \citep{2008Tinker} in our reference
cosmology, where $\Omega_{m}=0.3$, $\Omega_{\Lambda}=0.7$, $H_{0}=70 \rm ~ km/s/Mpc$, $\sigma_{8}=0.82$,  $\Omega_{b} = 0.045$, $n_s=-1.0$, for a density contrast of 500$
\times\rho_{\rm c}$.
$P(m)$ is then proportional to the mass function $\frac{dn}{dm}(m, z)$.  A proper estimation of the normalisation
constant would, in general, require to set a
lower limit to the cluster mass, but since we
do not vary $P(m)$ in our model that estimation
is not required in practice.
 \item 
 Numerical simulations and observations both demonstrate that the 
average mass observable scalings have power-law shapes (possibly 
broken power-laws when including groups) and support a Gaussian 
scatter in log space around these average power laws \citep[e.g.,][]{2013Giodini, 2010Stanek, 2012Angulo}.
The true ensemble average (over a large volume) of a global observable, $\hat{\xi}^i$, is related to mass of clusters $m$ and redshift $z$ as 

\begin{equation}
\hat{\xi}^{i}(m, z)  = \alpha^{i} (z) [m]^{\beta^{i}}\, ,
\end{equation}
where $\alpha^i$ and $\beta^i$ are the logarithmic normalisation and slope respectively and $m$ is the independent variable. 

Deviations from a perfect power-law scaling relation are expected due to the diversity of dynamical states in galaxy cluster population, non-gravitational physics, projection effects, etc, affecting the cluster observables. 
The random variables $\boldsymbol{\xi}$ can be modelled as originating from a multi-variate log-normal probability density function $P(\boldsymbol{\xi}|m,\boldsymbol{\Sigma}, \boldsymbol{\alpha},\boldsymbol{\beta})$, where $\boldsymbol{\Sigma}$ is the log-normal intrinsic covariance matrix of cluster observables at fixed mass. 
The diagonal elements of $\boldsymbol{\Sigma}$ give the log-normal intrinsic scatter for a corresponding cluster observable at fixed mass, which we denote as $\sigma_{\ln \xi^{i}}$. The off-diagonal terms quantify the covariance of different cluster observables at fixed mass. 
For $i \neq j$, the covariance between the cross-terms is related by the correlation coefficient:
\begin{equation}\label{eq:intrinsic_corr}
 r_{ij} \equiv \frac{\Sigma_{ij}}{\sqrt{\Sigma_{ii} \Sigma_{jj}}}\, .
\end{equation}


\item Typically, we access the cluster observables through a set of observations. The measured cluster properties and mass are denoted with a tilde as $\bf \tilde{\boldsymbol{\xi}}$, $\tilde{m}$ respectively. 
The link between the true observable and its
noisy estimate is provided by a
measurement model $P(\tilde{\boldsymbol{\xi}}, \tilde{m}|\boldsymbol{\xi}, m)$.


 The probability of measured cluster observables, $\tilde{\boldsymbol{\xi}}$ and mass, $\tilde{m}$ for a single cluster is 
 \begin{equation}\label{Eq:meas_prob}
  P(\tilde{\boldsymbol{\xi}},\tilde{m}|\theta)= \int_{\Xi}d\boldsymbol{\xi} \int_{0}^{+\infty} dm \, P(\tilde{\boldsymbol{\xi}}, \tilde{m}|\boldsymbol{\xi}, m) P(\boldsymbol{\xi}|m, \theta) P(m)\, ,
 \end{equation}
where $\Xi$ is the domain in which $\boldsymbol{\xi}$ is defined. 
\item The mass distribution and scaling relation model used to derive equation (\ref{Eq:meas_prob}) 
refer to the whole cluster population. In practice, one can never access a pure mass selected 
sample and always has to deal with a censored population, were a sub-sample has been selected 
based on some of the observables. We here describe this selection process through a detection 
probability $P(\mathcal{I}=1\,|\,\boldsymbol{\xi}_k,\phi)$, where $\mathcal{I}$ is a boolean random 
variable specifying whether a $k^{\rm th}$ cluster was detected or not and $\phi$ are a number of additional 
model parameters that describe the selection process.
The generative probability model for a cluster that passed the selection
is now conditional on $\mathcal{I}=1$ and can be expressed using Bayes theorem as:
\begin{equation}
   \label{Eq:CensoredLik}
   P(\boldsymbol{\tilde{\xi}}_k, \tilde{m}_k\,|\,\mathcal{I}=1,\theta,\phi) = 
         \frac{ P(\mathcal{I}=1\,|\,\boldsymbol{\tilde{\xi}}_k,\tilde{m}_k,\phi)\,P(\boldsymbol{\tilde{\xi}}_k,\tilde{m}_k\,|\,\theta) }{ P(\mathcal{I}=1\,|\,\theta,\phi) }\,.
\end{equation}

The overall probability for clusters to be selected, which appears in the denominator,
can be estimated by averaging the observable dependent selection probability over the 
global distribution of cluster observables provided by equation (\ref{Eq:meas_prob}), i.e.:
\begin{multline}
   P(\mathcal{I}=1|\theta,\phi) = \int_{\tilde{\Xi}} \mathrm{d}\boldsymbol{\tilde{\xi}}_k \int_{\tilde{\mathcal{M}}}\mathrm{d}\tilde{m}_k \,P(\mathcal{I}=1|\boldsymbol{\tilde{\xi}}_k, \tilde{m}_k, \phi)\,\\ \times P(\boldsymbol{\tilde{\xi}}_k, \tilde{m}_k\,|\,\theta)\, .
\end{multline}

\item The likelihood of the scaling relation parameters given a complete set of $N_{\rm det}$ detected clusters 
follows from equation (\ref{Eq:CensoredLik}):
\begin{equation}
\mathcal{L}(\theta|\,\boldsymbol{\tilde{\xi}}_{\rm obs}, \tilde{m}_{\rm obs},\phi) = \prod_\mathrm{k=1}^\mathrm{N_{\rm det}}\ P(\boldsymbol{\tilde{\xi}}_{k}, \tilde{m}_k\,|\,\mathcal{I}=1,\theta,\phi)\, ,
\end{equation}
where $\tilde{\boldsymbol{\xi}}_{\rm obs}$ is used to denote the full matrix of cluster observables measurements of all the detected clusters, $\tilde{m}_{\rm obs}$ denotes the full set of mass measurements for the detected sample of clusters and the posterior reads:
\begin{equation}
\mathcal{P}(\theta|\,\boldsymbol{\tilde{\xi}}_{\rm obs}, \tilde{m}_{\rm obs},\phi) = \pi(\theta)\,\times\,\mathcal{L}(\theta|\,\boldsymbol{\tilde{\xi}}_{\rm obs}, \tilde{m}_{\rm obs},\phi)\, ,
\end{equation}
where $\pi(\theta)$ is the prior on the model parameters.
\end{enumerate}

\subsection{Application to the eDXL sample}\label{sec:method:application}
          
We apply the method discussed in the previous subsection to our X-ray selected sample (eDXL).
For this sample the class of cluster properties on which the selection function depends on is the measured X-ray luminosity ($\tilde{L}_{\rm x}$) of clusters in the energy band $0.1\text{--}2.4$ keV  and redshift. 
Our primary goal is to measure the $Y_{\rm SZ}\text{--}M_{500}$ scaling relation using the eDXL sample. We remind here that the statistical model for the likelihood described above takes into account the impact of the sample selection function, the measurement uncertainties of cluster observables and mass, intrinsic covariances of cluster observables at fixed mass, and the underlying cluster mass function. 
We describe below briefly the essential components required for the likelihood in Equation \eqref{Eq:CensoredLik} to determine the posterior of the scaling relation parameters. 
%
%

We use the mass as the fundamental variable and cluster properties (such as $Y_{\rm SZ}$ and $L_{\rm x}$) as the response variables. 
Unbiased weak-lensing masses provide an absolute mass calibration for scaling relations and as already mentioned earlier, we expect the bias in lensing masses to be negligible as predicted by numerical simulations \citep[e.g.,][]{2010Meneghetti, 2011Becker, 2012Rasia}. Thus, our weak-lensing masses are a natural choice for anchoring the cluster masses. 
We note here that intrinsic scatter in the lensing masses can, however, occur due to elongation and projection effects along the line-of-sight \citep[e.g.,][]{2011Becker, 2015Gruen, 2016Shirasaki}, which can, in turn, produce biases in measuring scaling relations if not modelled correctly \citep[e.g.,][]{2015Sereno}. 
%

In order to properly account for several sources of uncertainties and systematic effects simultaneously, we consider two ways of modelling the scaling relations. In the first model, we assume no intrinsic scatter  
in the weak-lensing mass, essentially making the true lensing mass same as the spherical overdensity halo mass ($M_{\rm HM}$ or $M_{500}$). We describe the corresponding set of scaling models in Section \ref{sec:method:application:nointscatter}.
 In the second model, we assume a fixed intrinsic scatter in the true weak-lensing masses (section \ref{sec:method:application:withintscatter}).       
In both cases, the underlying cluster mass function in the redshift-mass space is described by the Tinker halo mass function \citep{2008Tinker}. 
 The inverse situation with either luminosity or $Y_{\rm SZ}$ being the independent variable would require knowledge of their number density which in turn depends on the scaling law with the total mass of the cluster. 
 We note that the use of the mass function depends on the cosmological parameters, most prominently on $\sigma_{8}$, $\Omega_{\rm m}$, $\Omega_{\Lambda}$. 
 By fixing these parameters, we are assuming an \textit{a priori} perfect knowledge of the mass function. 
 The impact of this somewhat strong
assumption is mitigated by the fact that the
number density of galaxy clusters is not
included in our likelihood model. We only rely
on the distribution in the measurements $(\tilde{M}_{\rm WL},\, \tilde{Y}_{\rm SZ},\, \tilde{L}_{\rm x})$ space, which only depends mildly
on the shape of the cluster mass function.
 
 The measurements $\tilde{Y}_{\rm SZ}$ and $\tilde{M}_{\rm WL}$ are drawn from a bi-variate Gaussian distribution. 
 Incorporating this probability density as such, naturally takes into account the non-detections in the SZ and does not require any special correction to the probability density. 
 The measured values of X-ray luminosities are treated as coming from a log-normal distribution with the log-normal uncertainty $\sigma_{\ln \tilde{L}_{\rm x}}$ and is independent of other measured properties.
 The explicit expression of the probability densities are given in Appendix \ref{app:prob}. In our implementation of the likelihood, we marginalise over the true variables (X-ray luminosities, integrated Comptonizations, masses) through an MCMC.

The selection function for the eDXL sample is a Heaviside step function that depends on the observed luminosities and the applied minimum luminosity threshold, i.e., $P(\mathcal{I}=1|\tilde{L}_{\rm x}, \tilde{L}^{\rm min})=1$ only when $\tilde{L}_{\rm x} \geq \tilde{L}^{\rm min}$, where the thresholds correspond to the defined values in Section \ref{sec:edxl}. 
The normalisation of the likelihood (Equation \ref{Eq:CensoredLik}) is computed for each redshift of the eDXL sample and is dependent on the scaling parameters of the $L_{\rm x}\text{--}M$ relation. This necessitates the joint modelling of multi-observable to mass scaling relations. Moreover, this joint modelling also has the advantage of considering a possible covariance between $Y_{\rm SZ}$ and $L_{\rm x}$ at fixed mass.
 The log-normal measurement uncertainty and the log-normal intrinsic scatter in X-ray luminosities allows us to analytically integrate the normalisation in Equation \eqref{Eq:CensoredLik} over the variables $L_{\rm x}$ and $\tilde{L}_{\rm x}$. Furthermore, the nature of the threshold cut selection gives an expression with an error function and this modulates the mass function for the sample especially at the low mass end.   
The explicit expression of the normalised likelihood for the eDXL sample is given in Appendix \ref{app:lik}. 
This expression of the normalisation of the likelihood remains the same for both set of scaling models discussed in Section \ref{sec:method:application:nointscatter} and \ref{sec:method:application:withintscatter}.
%

%

 In the subsections below, we describe the two different scaling models.%
\begin{figure*}%

 \includegraphics[width=\columnwidth]{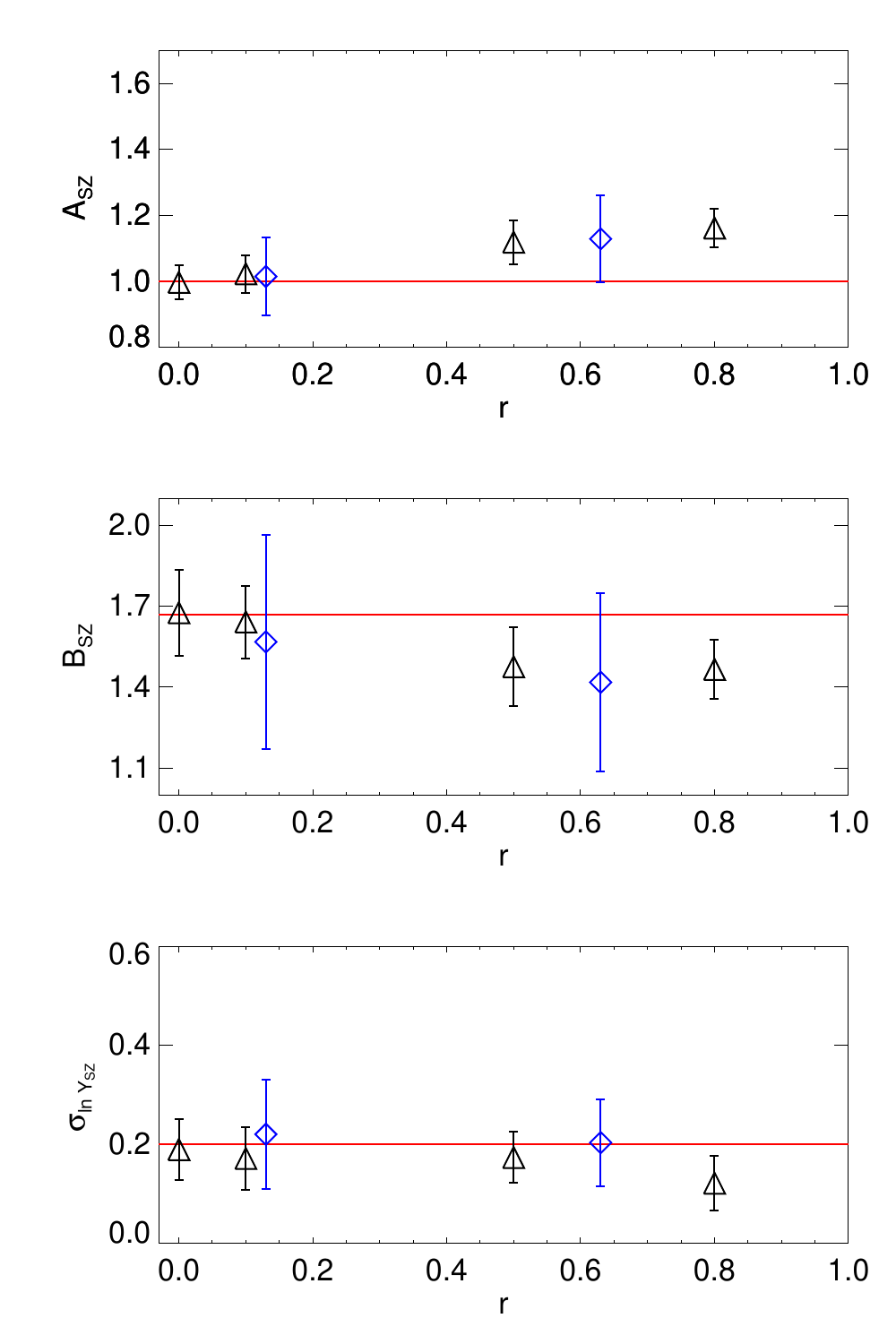}
 \includegraphics[width=\columnwidth]{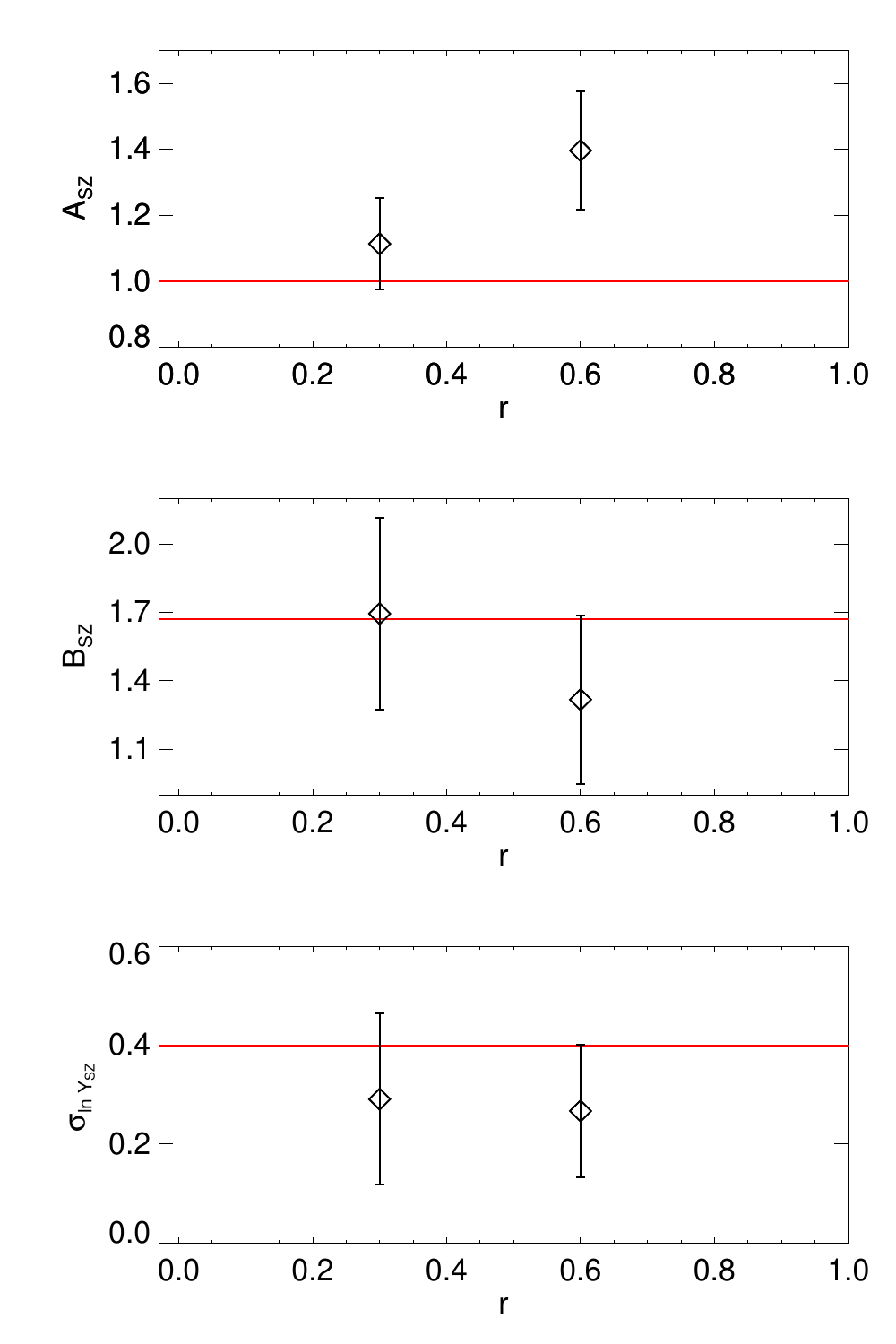}\caption{The recovered mean parameters from mock data realisations are shown here for the $Y_{\rm SZ}\text{--}M_{500}$ relation. 
 The red lines mark the input values of the scaling parameters. The error bar on the recovered mean value represents the uncertainty level from single set of mock sample analysis.
 \textit{Left:} The black triangular symbols correspond to results from mock data with 10\% measurement uncertainties. 
 The blue diamond symbols correspond to results for mock data with realistic uncertainties (See text). 
 \textit{Right:} The recovered mean parameters for different input scaling relations, where $\sigma_{\ln L_{\rm x}}$ and $\sigma_{\ln Y_{\rm SZ}}$ were increased to 0.6 and 0.4 respectively. 
 We simulated the mock data with realistic uncertainties. The bias we see in the normalisation for realistic mock samples indicates that even for cluster sample size of 30 and with realistic measurements, ignoring the correlation show significant bias.}\label{fig:rbiasdiff}
\end{figure*}

\begin{figure}
  \includegraphics[width=\columnwidth,keepaspectratio=true]{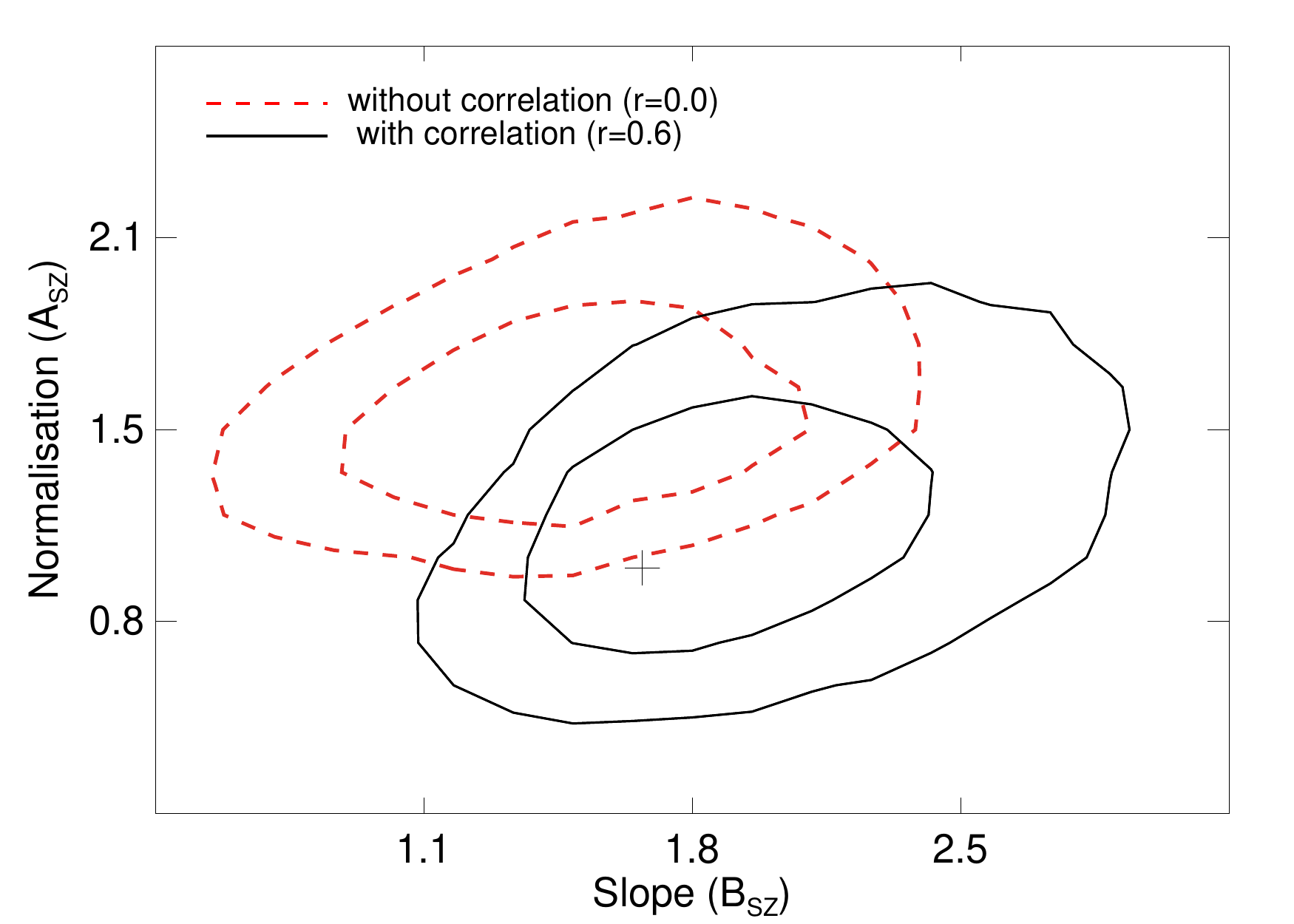}
  \caption{An example of 68\% and 95\% confidence levels of the normalisation and slope parameters of the $Y_{\rm SZ}\text{--}M_{500}$ scaling relation fit to a mock sample. Red dashed contours: $r=0$, and the black solid contours: $r=0.6$. The mock sample was generated using realistic measurement errors and with a correlation ($r=0.6$) in intrinsic scatters of $L_{\rm x}$ and $Y_{\rm SZ}$. The input value of the $Y_{\rm SZ}\text{--}M_{500}$ relation is shown as cross.}\label{fig:rbiasfit}
\end{figure}   
       \subsubsection{Without intrinsic scatter in lensing mass}\label{sec:method:application:nointscatter}
       
       \paragraph*{Scaling model:}{ The prescription for the scaling laws of the observables with the mass of a cluster is defined as 
       
 \begin{align}\label{eq:lm}
 \frac{\hat{L}_{\rm x}}{L_{\rm pvt}}\, E(z)^{-2} = A_{\rm LM}\, \left(\frac{M_{\rm HM, 500}}{M_{\rm pvt}}\right)^{B_{\rm LM}}\; ; \\
 \label{eq:sz-mwl}
 \frac{\hat{Y}_{\rm SZ,500}}{Y_{\rm pvt}}\, E(z)^{-2/3} = A_{\rm SZ}\, \left(\frac{M_{\rm HM, 500}}{M_{\rm pvt}}\right)^{B_{\rm SZ}}\; ,
  \end{align}
 where $M_{\rm HM, 500}$ is the spherical halo mass or the true total mass of a galaxy cluster and, where the pivot values for luminosities, masses, and SZ Compton parameters are 
 $L_{\rm pvt}$ $=$ $8.77 \times 10^{44} \, \rm erg \, s^{-1}$, $M_{\rm pvt}$ $=$ $7.084 \times 10^{14}\, M_{\sun}$, $Y_{\rm pvt}$ $=$ $7.93  \times 10^{-5} \, \rm Mpc^2$ respectively. 
  The pivot values reflect the median values of the measurements $\tilde{L}_{\rm x}E(z)^{-2}$, $\tilde{M}_{\rm WL, 500}$, and $\tilde{Y}_{\rm SZ, 500}E(z)^{-2/3}$ across the eDXL sample. We choose these values to minimise the degeneracy in measuring the normalisation and slope of
the scaling relations.  The above scaling power-law are modelled with log-normal intrinsic scatter in $L_{\rm x}$ and $Y_{\rm SZ}$ at fixed mass with correlation parameter $r$. The intrinsic covariance matrix is given as follows:
      }
      \begin{equation} \label{eq:matrix}
\begin{pmatrix}
   \sigma_{\ln L_{\rm x}}^2       & r\sigma_{\ln L_{\rm x}}\sigma_{\ln Y_{\rm SZ}}  \\
    r\sigma_{\ln L_{\rm x}}\sigma_{\ln Y_{\rm SZ}}       & \sigma_{\ln Y_{\rm SZ}}^2  \\
   \end{pmatrix}
   \; ,
\end{equation} where $\sigma_{\ln L_{\rm x}}$, $\sigma_{\ln Y_{\rm SZ}}$ are the log-normal intrinsic scatters in $L_{\rm x}$ and $Y_{\rm SZ}$ at fixed mass, respectively, and $r$ is the correlation coefficient. 
%

  In this model, we anchor the halo masses to the lensing masses by a one-to-one scaling of true lensing mass, $M_{\rm WL}$, to halo mass, $M_{\rm HM}$, by setting $M_{\rm HM}=M_{\rm WL}$.

       The redshift evolutions of the scaling relations are power-law of $E(z)$, the time evolution of the Hubble parameter. We use the logarithmic self-similar slope for the evolution in the $Y_{\rm SZ}\text{--}M_{500}$ and $L_{\rm x}\text{--}M_{500}$ relations. Throughout the analysis, we keep them fixed. We fix the logarithmic slope of the redshift evolution of the $L_{\rm x}\text{--}M_{500}$ relation to the self-similar evolution value for soft-band luminosities \citep{2015Ettori}. 
       This slope is shallower than the self-similar slope of bolometric luminosities and is confirmed by other authors (\citealt{2009Vikhlinin, 2015Sereno}).  
  Additionally, we choose uniform priors in the interval (0, 5.0] for the parameter set, \{$A_{\rm LM}$, $B_{\rm LM}$, $A_{\rm SZ}$, $B_{\rm SZ}$\}. The priors for the intrinsic scatters \{$\sigma_{\ln Y_{\rm SZ}}$, $\sigma_{\ln L_{\rm x}}$\} are uniform in the interval [0.02, 5.0] and we place an uniform prior on the correlation parameter $r$ in the open interval (-1, 1).
   
      %
    \subsubsection{With intrinsic scatter in lensing mass}\label{sec:method:application:withintscatter}

    To take into account a possible scatter in lensing masses, we add a scaling law between the lensing mass and true spherical overdense mass and model the lensing mass observable to scatter from the halo mass with a dispersion. 
    This additional scaling is given below:
   \begin{align}
  \label{eq:wl-hm}\frac{\hat{M}_{\rm WL, 500}}{M_{\rm pvt}} = A_{\rm WL}\, \left(\frac{M_{\rm HM, 500}}{M_{\rm pvt}}\right)^{B_{\rm WL}}\; ,
\end{align}
   where the normalization ($A_{\rm WL}$) and the slope ($B_{\rm WL}$) of the relation are both fixed to unity. 

      The scatter in the lensing mass from the true halo mass is predicted to be log-normal and of the level of 20--23\% for the massive clusters of $M_{500}  \geq ~1.4 \times 10^{14} M_{\sun}$ in the redshift range of 0.25--0.50 \citep{2011Becker}. The constraints from observations are consistent with the predictions \citep[e.g.,][]{2015MantzwtgIV, 2015Sereno}.
          Since we lack the statistical power to constrain the dispersion in lensing mass observable, we use this prior to fix the intrinsic scatter. We model it to be a Gaussian dispersion of $0.20 \times M_{\rm HM, 500}$. 
         Introducing this lensing scatter in our modelling requires a marginalisation over the true lensing mass variable, $M_{\rm WL}$, for all clusters. Since the lensing mass measurement itself is from a bi-variate Gaussian distribution, 
         our intuitive choice of a Gaussian intrinsic scatter in true lensing mass observable simplifies the marginalisation over these additional variables. 
Therefore, the marginalisation over these variables to relate $M_{\rm HM,500}$ to $\tilde{M}_{\rm WL,500}$ is done analytically in our implementation fully taking into account the measurement covariances between lensing masses and integrated Comptonizations (the calculations are outlined in Appendix \ref{app:method:intscatter}).

It is understood that not accounting for an intrinsic scatter in lensing masses can bias the estimate of the scaling parameters \citep{2015Sereno, 2015comalitII, 2015Gruen}. 
But including such a scatter also requires a consideration of the correlations between the lensing scatter and the intrinsic scatters of other cluster observables at fixed mass. 
Due to limitations set by our measurements and sample size, we are forced to fix this scatter to 20\% (Gaussian) and do not marginalise over this scatter. 
We assume zero correlations in the intrinsic covariances of lensing mass observable with other observables at fixed mass. We discuss this further in Section \ref{sec:limitations}.
Therefore, we give this model here as a consideration of the impact of such a scatter in our lensing masses on the scaling parameters and in this work, we follow the model given in Section \ref{sec:method:application:nointscatter} as our fiducial model.

\subsection{Tests with simulations}\label{sec:method:tests}
 
Based on reported findings from simulations \citep[e.g.,][]{2016Truong, 2010Stanek} that thermodynamic observables are correlated at fixed masses, we study the impact of the correlated scatters and selection biases on scaling relations of these observables. It is important and crucial to understand this impact for inferring deviations in self-similar scaling. In the following sections, we present results focusing on mock datasets that mimic the behaviour of our eDXL sample and measurements. In addition, we test for more precise measurements and present a detailed description of the results in Appendix \ref{app:test}, which is useful for a look-up of the level of bias in possible future surveys and follow-up studies that may have more precise measurements. 
But focusing on a more realistic measurement uncertainties that are representative of the uncertainties in our cluster observable measurements, we address two key issues relevant for our analysis for the eDXL sample:
\begin{enumerate}
\item First, we consider the presence of a correlation between intrinsic scatters of luminosity and Compton-Y at fixed mass. We parametrize this correlation between the scatters with `$r$'. We present this discussion in Section \ref{sec:method:test:r}.
\item Second, we consider that the weak-lensing masses have an intrinsic scatter due to projection effects, etc. We study the impact of ignoring this information in our analysis. We present this discussion in Section \ref{sec:method:wlscat}.
\end{enumerate}

 \subsubsection{Correlation in the scatters of selecting observable and follow-up observable at a fixed mass}\label{sec:method:test:r}
 In order to understand the level of bias that can occur in the recovered scaling relations if the correlation in the scatters of selecting observable and follow-up observable is ignored, 
 we simulate sets of mock data with measurement uncertainties. 
 Our mock samples have 30 clusters, which is similar in size to the eDXL sample.  We generate mock samples with three observables: the independent variable (e.g., $M_{\rm 500}$), and two response variables (e.g., $L_{\rm x}$, $Y_{\rm SZ}$) modelled using a power law relation with the independent variable. The selection is on one of the response variables, namely, $L_{\rm x}$. The samples were generated at a median redshift of 0.3 using the Tinker mass function \citep{2008Tinker}. 

      \paragraph*{Realistic measurement uncertainties}{
     We assume different values of the correlation coefficient $r$ between 0.0 and 1.0 in the intrinsic scatters of the two response variables as input for generating mock samples. We test with simulated measurements of 30\% uncertainty in $M_{\rm 500}$ and 25\% on the $Y_{\rm SZ}$. These measurement uncertainties reflect the median relative uncertainties of our eDXL mass and mass proxy measurements. 
     Simulation studies report a positive correlation in the intrinsic scatters of $L_{\rm x}$ and $Y_{\rm SZ}$ at fixed mass in the range of 0.5-0.9 (\citealt{2010Stanek,2012Angulo,2016Truong}). We choose values of correlation on the lower end (ranging between 0.1 and 0.6) for our mock samples to test when the impact starts becoming significant.  We fit numerous realisations of mock data sets for each input relation using the method prescription in section \ref{sec:method:application:nointscatter} with $r$ set to zero.
   The measured average recovered scaling relation parameters for the $Y_{\rm SZ}\text{--}M_{500}$ relation are plotted in Figure \ref{fig:rbiasdiff} for different set of input scaling relations. 
   The bias in the normalisation of the $Y_{\rm SZ}\text{--}M_{500}$ relation for a set of input relations with 20\% and 40\% intrinsic scatters in $Y_{\rm SZ}$ and $L_{\rm x}$ is about $1\sigma$. 
     For an eDXL like sample with larger intrinsic scatters ($\sigma_{\ln L_{\rm x}}=0.6$, $\sigma_{\ln Y_{\rm SZ}}=0.4$ were used based on recovered values in Section \ref{sec:results}), we find that the normalisation of the $Y_{\rm SZ}\text{--}M_{500}$ relation is significantly biased at the level of $2\sigma$. 
     An example of recovered normalisation and slope of the $Y_{\rm SZ}\text{--}M_{500}$ relation from a mock eDXL-like data set is shown in Figure \ref{fig:rbiasfit}. The input scaling relations lies outside the 95\% confidence level of the recovered parameter space when ignoring the correlation in the scatters. On fitting with the assumed correlation, the recovered parameter space is consistent with the input values within 68\% confidence.  
     
      Besides the normalisation, for different input relations we observe the average slope is almost $1\sigma$ shallower than the input. We also observe an under-estimation of the intrinsic scatter $\sigma_{\ln Y_{\rm SZ}}$ for the eDXL-like data set. %
      The full table of results is summarized in the Appendix \ref{app:test} in the Table \ref{tab:rbias}. A summary of results from this Section and Appendix \ref{app:r} is plotted in Figure \ref{fig:rbiasdiff} showing the means and standard deviations of the recovered modes for each scaling parameter of the $Y_{\rm SZ}\text{--}M_{500}$ relation.
       
   }

 \begin{figure*} \includegraphics[width=18cm]{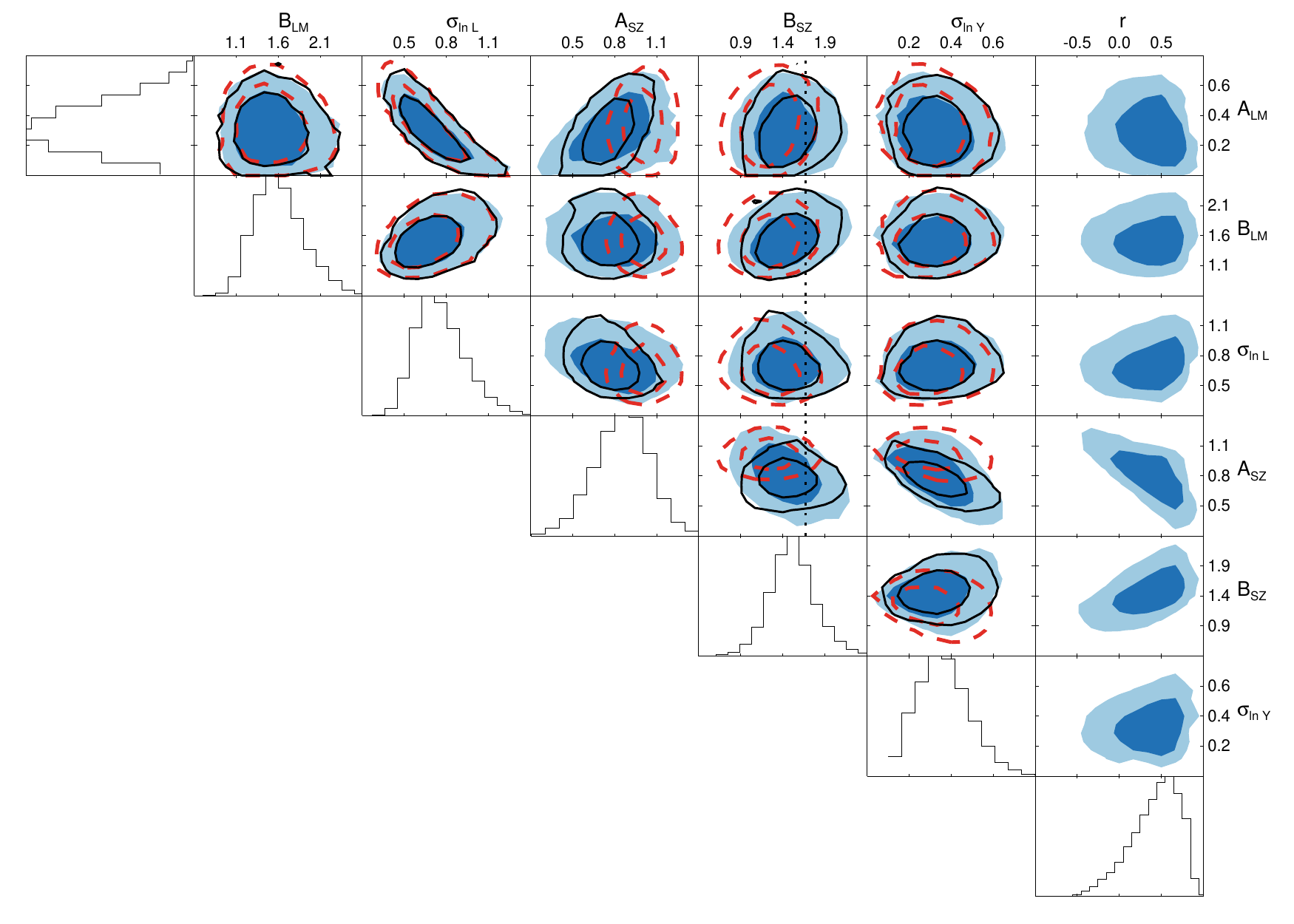}\caption{Marginalized posterior distributions of the parameters of the scaling models, shown as $68\%$ and $95\%$ confidence levels. Colour shaded contours are from the full joint likelihood fit including the correlation ($r$) in intrinsic scatter with a uniform prior on $r$ such that $-1 < r < 1$. Red contours indicate the marginalized confidence levels with $r=0$. The black contours indicate the marginalized confidence levels with $r=0.5$. The histograms show the marginalised distribution of each recovered scaling parameter when we allow $r$ to vary. The vertical dotted line corresponds to the self-similar expectation of the $Y_{\rm SZ}\text{--}M_{500}$ scaling slope.  
}\label{fig:ryml}\end{figure*}
 
\begin{figure*}
\includegraphics[width=15cm,keepaspectratio=true]{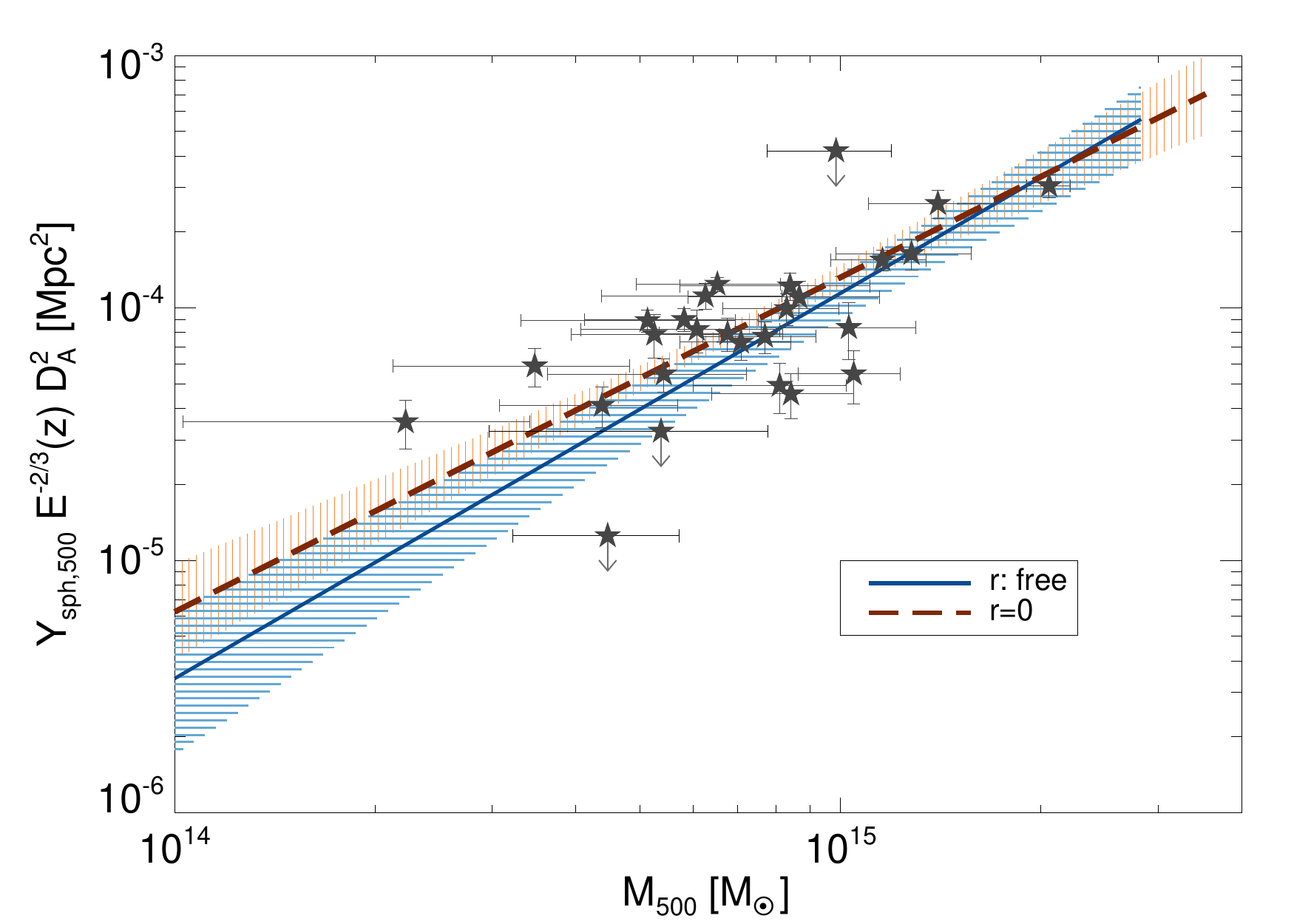}
\caption{The median $Y_{\rm SZ}\text{--}M_{500}$ scaling relation measured by including a possible correlation between intrinsic scatters is shown as the solid blue line. 
The red dashed line shows the median scaling relation assuming no correlation between the intrinsic scatters.
The red and blue line-filled regions denote 68\% confidence. 
Assuming un-correlated intrinsic scatters in luminosity and Comptonization finds a higher normalisation and shallower slope for the $Y_{\rm SZ}\text{--}M_{500}$ relation. The constraints shown here were obtained by ignoring the weak-lensing intrinsic scatter. The measured weak-lensing masses and integrated Comptonizations plotted here are same as for the eDXL sample shown in Figure \ref{fig:yszmass}.   }\label{fig:results:rboth}
\end{figure*}
\subsubsection{Intrinsic  scatter in lensing masses} \label{sec:method:wlscat}

     Here we present the analysis using mock data where we scatter the true lensing mass from the halo mass using a log-normal scatter of 20\%.
    \paragraph*{Realistic measurement uncertainties} As done in previous Section, we generate mock samples with the realistic measurement uncertainties and eDXL-like scaling parameters. 
   We consider the following two cases:
   \begin{enumerate}
    \item \textit{Without correlation in the intrinsic scatters of $L_{\rm x}$ and $Y_{\rm SZ}$:} We introduce un-correlated intrinsic scatters in the $L_{\rm x}$ and $Y_{\rm SZ}$ observables for the mock samples. We fit numerous realisations of mock samples with the model given in Section \ref{sec:method:application:nointscatter} (i.e., ignoring the lensing scatter) with $r=0$. From the analysis of mock data with realistic measurements, we find that the recovered values of the scaling parameters show bias values less than $0.7\sigma$. For the more precise measurements, we find a bias in the slope of the $Y_{\rm SZ}\text{--}M$ relation and the intrinsic scatter in $Y_{\rm SZ}$ at fixed mass to be $\sim 1.4\sigma$. The other scaling parameters show less than $1\sigma$ bias.
         
         \item \textit{With correlation in the intrinsic scatters of $L_{\rm x}$ and $Y_{\rm SZ}$:} Most importantly for our case, we test the impact of the presence of an intrinsic scatter in the lensing masses and simultaneously having a correlation in the intrinsic scatters of $L_{\rm x}$ and $Y_{\rm SZ}$ at fixed mass. Therefore, we inject a correlation of 0.6 in the intrinsic scatters of $L_{\rm x}$ and $Y_{\rm SZ}$ for our mock samples. We fit the scaling relations with the same procedure as done previously by fitting scaling models with $r=0$ and without intrinsic scatter in lensing mass. 
     We observe a total bias of $2.7\sigma$ in the normalisation ($A_{\rm SZ}$) parameter of $Y_{\rm SZ}\text{--}M_{500}$ relation. 
     The slope ($B_{\rm SZ}$) of this relation is found to be biased low by 0.34 ($\sim 1\sigma$) from the input value. 
     We fit the mock samples again but fixing the $r$ value to 0.6. 
     From the mean recovered scaling parameters of mock samples, we find that the bias in the normalisation reduces to $+0.5\sigma$. 
     The slope parameter is now lowered by $0.17$ ($0.9\sigma$) from the input value. This level of bias in the slope occurring due to the scatter in weak-lensing mass is consistent with the findings and discussion given in \cite{2015Sereno}. The results from this Section are summarised in Table \ref{tab:scat}.
    \end{enumerate}

   %
\section{Results }\label{sec:results}

We jointly fit the three observables ($L_{\rm x}$, $Y_{\rm SZ}$, $M_{\rm 500}$) of the eDXL sample to the $L_{\rm x}\text{--}M_{500}$ and $Y_{\rm SZ}\text{--}M_{500}$ scaling relations. 
The main purpose of fitting the $L_{\rm x}\text{--}M_{500}$ relation is to account for the sample selection. 

In this section, we present the results of these fits under progressively less conservative assumptions on intrinsic scatter of weak-lensing masses.
First, in Section \ref{sec:results:rfree}, we present fits with correlated scatter in $L_{\rm x}$ and $Y_{\rm SZ}$ (allowing the correlation parameter, $r$, to vary, and fixing it) while ignoring the intrinsic scatter in the lensing masses. In Section \ref{sec:results:wlscatter}, we also add the expected intrinsic scatter in weak-lensing masses. 
\begin{table*}\caption{
Results of the scaling relations analysis for the eDXL sample using the method described in Section \ref{sec:method}. The medians and 68\% confidence levels of the marginalised distributions are quoted. Centroids for gNFW model fits were fixed to the optical centres (BCG) or X-ray (X) values obtained from the ROSAT survey. 
}
\label{tab:selresults}
\renewcommand{\arraystretch}{1.6}
 \begin{tabular}{c c c| c  c  c  c  c  c  c}
 \hline
\multicolumn{3}{c|}{Priors} & \multicolumn{7}{c}{Recovered parameters} \\
\hline
{} & {} & {} & {} & \multicolumn{3}{c}{$L_{\rm x}\text{--}M_{500}$ scaling parameters} & \multicolumn{3}{c}{$Y_{\rm SZ}\text{--}M_{500}$ scaling parameters}\\

Centroid & $r$&$\sigma_{ \rm WL|HM}$& $r$ & $A_{\rm LM}$ & $B_{\rm LM}$ & $\sigma_{\ln L_{X}}$ & $A_{\rm SZ}$ & $B_{\rm SZ}$ & $\sigma_{\ln Y_{\rm SZ}}$ \\
\hline
 
 BCG & $\in (-1,1)$&-& $0.47 \;_{-0.35}^{+0.24}$ &$0.32\; _{-0.15}^{+0.17}$ & $1.59\; _{-0.27}^{+0.33}$  & $0.75\; _{-0.16}^{+0.19}$ &$0.86 _{-0.21}^{+0.18}$ & $1.51 \;_{-0.24}^{+0.28}$ &$ 0.36  \;_{-0.12}^{+0.13}$ \\
%
%
 BCG & fixed &-&($0.0$) &$0.36\;  _{-0.16}^{+0.18}$ & $1.58\; _{-0.25}^{+0.32}$  & $0.70\; _{-0.15}^{+0.19}$  & $1.06\; _{-0.10}^{+0.09}$  & $1.33\;  _{-0.22}^{+0.21}$ &$0.33\;  _{-0.10}^{+0.12}$ \\
 BCG & fixed &-&($0.5$)&$0.32\; _{-0.15}^{+0.17}$ & $1.60\; _{-0.27}^{+0.33}$& $0.74 \;_{-0.15}^{+0.19}$ & $0.81\; _{-0.13}^{+0.13}$ & $1.54\; _{-0.22}^{+0.24}$ &$0.37\; _{-0.10}^{+0.11}$\\
%
 BCG & $ \in (-1,1)$&$0.2M_{\rm HM}$&$> -0.51$ (at $16\%$)  & $0.41\; _{-0.17}^{+0.16}$ & $1.71 \;_{-0.29}^{+0.37}$ & $0.65\; _{-0.14}^{+0.19}$ & $0.97 \;_{-0.19}^{+0.16}$  & $1.67\; _{-0.27}^{+0.34}$  & $0.19\; _{-0.09}^{+0.14}$ \\%
%
%
%
 BCG & fixed&$0.2M_{\rm HM}$&($0.0$) & $0.43\; _{-0.18}^{+0.16}$ & $1.69\; _{-0.28}^{+0.40}$  & $0.63 \;_{-0.14}^{+0.19}$  & $1.00 \;_{-0.12}^{+0.11}$ & $1.64\; _{-0.27}^{+0.30}$  & $0.17\; _{-0.08}^{+0.14}$ \\
%
%
 BCG &fixed&$0.2M_{\rm HM}$&($0.5$) & $0.37 \;_{-0.15}^{+0.15}$  & $1.71\; _{-0.29}^{+0.37}$  & $0.68\; _{-0.13}^{+0.17}$  & $0.88\; _{-0.13}^{+0.13}$  & $1.78\; _{-0.26}^{+0.30}$  & $0.18\; _{-0.09}^{+0.16}$ \\
\hline
 X & $ \in (-1,1) $ & -& $0.49\; _{-0.33}^{+0.23}$ & $0.33\; _{-0.15}^{+0.16}$&$1.63 \;_{-0.26}^{+0.35}$ & $0.74 \;_{-0.15}^{+0.20}$  & $0.70 \;_{-0.21}^{+0.18}$& $1.73 \;_{-0.34}^{+0.36}$& $0.48\; _{-0.11}^{+0.14}$\\

\hline
\end{tabular}
\end{table*}

\subsection{Including correlated intrinsic scatters in $Y_{\rm SZ}$ and $L_{\rm x}$ at fixed mass}\label{sec:results:rfree}

We fit the $L_{\rm x}\text{--}M_{500}$ and $Y_{\rm SZ}\text{--}M_{500}$ relations using the model described in Section \ref{sec:method:application:nointscatter}. As discussed earlier, we include a correlation coefficient parameter in the intrinsic scatters of luminosity and Comptonization at fixed mass. We marginalise over the correlation parameter, $r$, allowing it to vary between $-1$ and $+1$. The result is summarised in Table \ref{tab:selresults}. Including correlated intrinsic scatters in Comptonization and luminosity at fixed mass results in a slope of $1.51 _{-0.22}^{+0.31}$ in the $Y_{\rm SZ}\text{--}M_{500}$ scaling relation, fully consistent with self-similarity.  
For the correlation between intrinsic  scatters of luminosity and Comptonization we find $r=0.47\, _{-0.35}^{+0.24}$. Approximately 90\% of the posterior distribution prefers a positive correlation. The marginalised posterior distributions are shown in Figure \ref{fig:ryml}. The correlation parameter, $r$,  correlates the strongest with the SZ normalisation $A_{\rm SZ}$ (anti-correlation) but also with the slope $B_{\rm SZ}$ (positive correlation). 
Ignoring the  correlation between intrinsic scatters of luminosity and Compton-Y at fixed mass (i.e., $r=0$) results in a $Y_{\rm SZ}\text{--}M_{500}$ scaling relation with a recovered slope of $1.33\, _{-0.22}^{+0.21}$, marginally shallower than what is expected from self-similarity (1.67) and the normalisation found is higher by 1$\sigma$. The uncertainties in the recovered scaling relation are lower when $r$ is set to a fixed value (either 0.0 or 0.5). If one indeed uses the prior of ignoring the correlation in scatter completely (as would be the case using a method similar to that of \citealt{2007Kelly}), the bias in the normalisation of the $Y_{\rm SZ}\text{--}M_{500}$ relation is on the order of $\sim 2\sigma$. A similar level of bias was found in our analysis of mock data sets in Section \ref{sec:method:test:r}.   
 By applying a method similar to \cite{2007Kelly} for measuring the $Y_{\rm SZ}\text{--}M_{500}$ relation, we note that the constraints are same as those obtained from ignoring the intrinsic covariance between $L_{\rm x}$ and $Y_{\rm SZ}$. 

 In Figure \ref{fig:ryml}, we compare the results of the analysis without correlation between the intrinsic scatters to the one obtained with leaving the correlation as a free parameter.
The marginal change in the $Y_{\rm SZ}\text{--}M_{500}$ scaling parameters is illustrated in Figure \ref{fig:results:rboth}, where it becomes evident that the bias from setting $r=0$ is more prominent at the low-mass end. Table \ref{tab:selresults} summarises the results for fitting with different assumptions.   

The normalization of the $L_{\rm x}\text{--}M_{500}$ relation, $A_{\rm LM}$, shows a strong anti-correlation with the intrinsic scatter $\sigma_{\ln L_{\rm x}}$ in luminosity at fixed mass, with a Pearson correlation coefficient of -0.81.   
Our recovered normalisation of the $L_{\rm x}\text{--}M_{500}$ relation is $0.32\, _{-0.15}^{+0.17}$, and the slope is $1.59\, _{-0.27}^{+0.33}$ for our fiducial analysis with varying $r$ parameter. From Figure \ref{fig:ryml} and Table \ref{tab:selresults}, we can observe that the $L_{\rm x}\text{--}M_{500}$ relation constraints are unaffected by the correlation parameter $r$. This relation with its 68\% confidence levels is shown in Figure \ref{fig:mllit}. 

The results summarised here with $r$ as a free parameter will be considered as our fiducial result. A further discussion on the constraints obtained on the $Y_{\rm SZ}\text{--}M_{500}$ and $L_{\rm x}\text{--}M_{500}$ relations is given in Section \ref{sec:compare}.

\subsection{Including uncorrelated intrinsic scatter in the weak-lensing masses} \label{sec:results:wlscatter}

We include an intrinsic scatter term in the model by adding the scaling relation between true halo mass and weak-lensing mass given in Section  \ref{sec:method:application:withintscatter}. 
We marginalise over the true weak-lensing masses analytically for a Gaussian scattered lensing masses with a 20\% dispersion.  
The implementation is detailed in Appendix \ref{app:method:intscatter}. We assume the bias in the lensing mass to be negligible. Due to the limited sample size, we forgo fitting and marginalisation of the percentage scatter of lensing masses w.r.t halo mass ($\sigma_{\rm WL|HM}$). \cite{2015Sereno} marginalised the scatter for a much larger sample and were able to constrain its value at approximately 20\% log-normal scatter. 

First, we consider the scenario with no correlation between the intrinsic scatters of $L_{\rm x}$ and $Y_{\rm SZ}$ (i.e., $r=0$). While the normalisation of the $Y_{\rm SZ}\text{--}M_{500}$ scaling is comparable to the case where the scatter in lensing mass was ignored, we find a steeper slope of $1.64\, _{-0.27}^{+0.30}$, which is a $1\sigma$ increase from $1.33$ found in the previous section for $r=0$.  Fixing $r$ to the mean value of $0.5$ recovered from the previous subsection, the $Y_{\rm SZ}\text{--}M_{500}$ slope increases marginally ($\sim 0.5\sigma$) from $1.64$ to $1.78_{-0.26}^{+0.30}$, while the normalization decreases by $\sim 1\sigma$. 

Finally, we carry out the analysis allowing all parameters, including $r$, to vary. The result is summarised in Table \ref{tab:selresults}. We note that the data do not have the leverage to constrain $r$ in this case, as is evident from Figure \ref{fig:rymlbecker}. We quote a lower limit of $-0.51$ for $r$ with 84\% of the distribution lying above this limit. 
 The marginalised posterior distributions are shown in Figure \ref{fig:rymlbecker} for all of the cases discussed here. 
 
Assuming a 20\% Gaussian intrinsic scatter in weak-lensing mass, the intrinsic scatters $\sigma_{\ln L_{\rm x}}$ and $\sigma_{\ln Y_{\rm SZ}}$ are both reduced, by 10\% and 17\% respectively. 
The normalisation, $A_{\rm LM}$, being anti-correlated with $\sigma_{\ln L_{\rm x}}$, increases by $0.5\sigma$. These differences in the $L_{\rm x}\text{--}M_{500}$ relation with respect to constraints obtained in Section \ref{sec:results:rfree} are marginal, however, this trend of increased normalisation and decrease in intrinsic scatter is consistent with the demonstrated effect due to scatter in weak-lensing mass \citep{2015Gruen}.

    \subsection{Correlated intrinsic scatters: interpretation from residuals}\label{sec:robust:residuals}

  We examine the distribution of residuals in $\log(L_{\rm x})$ and $\log(Y_{\rm SZ})$ obtained for our median scaling relations in Section \ref{sec:results:rfree}. 
 For each cluster in the sample that is a detection in APEX-SZ, the residual is computed at fixed lensing mass. 
 We predict the 68\% and 95\% confidence regions using Monte-Carlo realisations. 
 For this purpose, we generate population of masses from Tinker mass-function and scatter the masses with the measurement uncertainty in the lensing mass. Additionally, we generate other observables including the luminosities using our median scaling relations and covariance matrix (using the Equation \ref{eq:matrix} with $r=0.47$). The observables are scattered with their measurement uncertainties. The procedure for generating cluster observables is similar to the one described Appendix \ref{app:test}.
 For each cluster at a given redshift, we generate 6000 realisations of cluster observables that would make the selection of the eDXL sample. 
  The distribution of the generated residuals and their 68\% and 95\% confidence levels are shown in Figure \ref{fig:residuals} for individual clusters. The measured residual for each cluster is indicated in the same. 
  We combine all the measured residuals in the residual plane which is shown in the lower panel of Figure \ref{fig:residuals}. 
  The distribution of residuals show a positive alignment with a Pearson correlation coefficient of 0.73. The generated residuals from Monte-Carlo simulations are combined together in the residual plane for 24 clusters. %
  The model prediction of 68\% and 95\% confidence levels of the residuals in Figure \ref{fig:residuals} show that the distribution of residuals are consistent with our model prediction. %
 
 Next, we repeat the exercise for the $r=0$ recovered mean scaling relation. The residuals for the median scaling relations are plotted in Figure \ref{fig:residualsrzero}. The predicted 68\% and 95\% confidence levels for $r=0$ are represented as contours, where the 95\% confidence encompasses all of the residuals. We find a positive alignment in the residuals with a Pearson correlation coefficient of $0.66$. This being still positively aligned, we use 3000 mock random realisations for the $r=0$ model prediction of 24 cluster residuals and compute the Pearson correlation.  
 We iterate the process with $r=0.5$. Both distributions of Pearson coefficients are shown in Figure \ref{fig:residualsrzero}. Finding a strong correlation in the residuals appears to be less likely when there is no intrinsic correlation. However, it does not altogether rule out the $r=0$ value as 10\% of the distribution lies above $0.66$. This reflects our weak constraint on $r$. 
 
\begin{figure} 
  \includegraphics[width=\columnwidth,keepaspectratio=1]{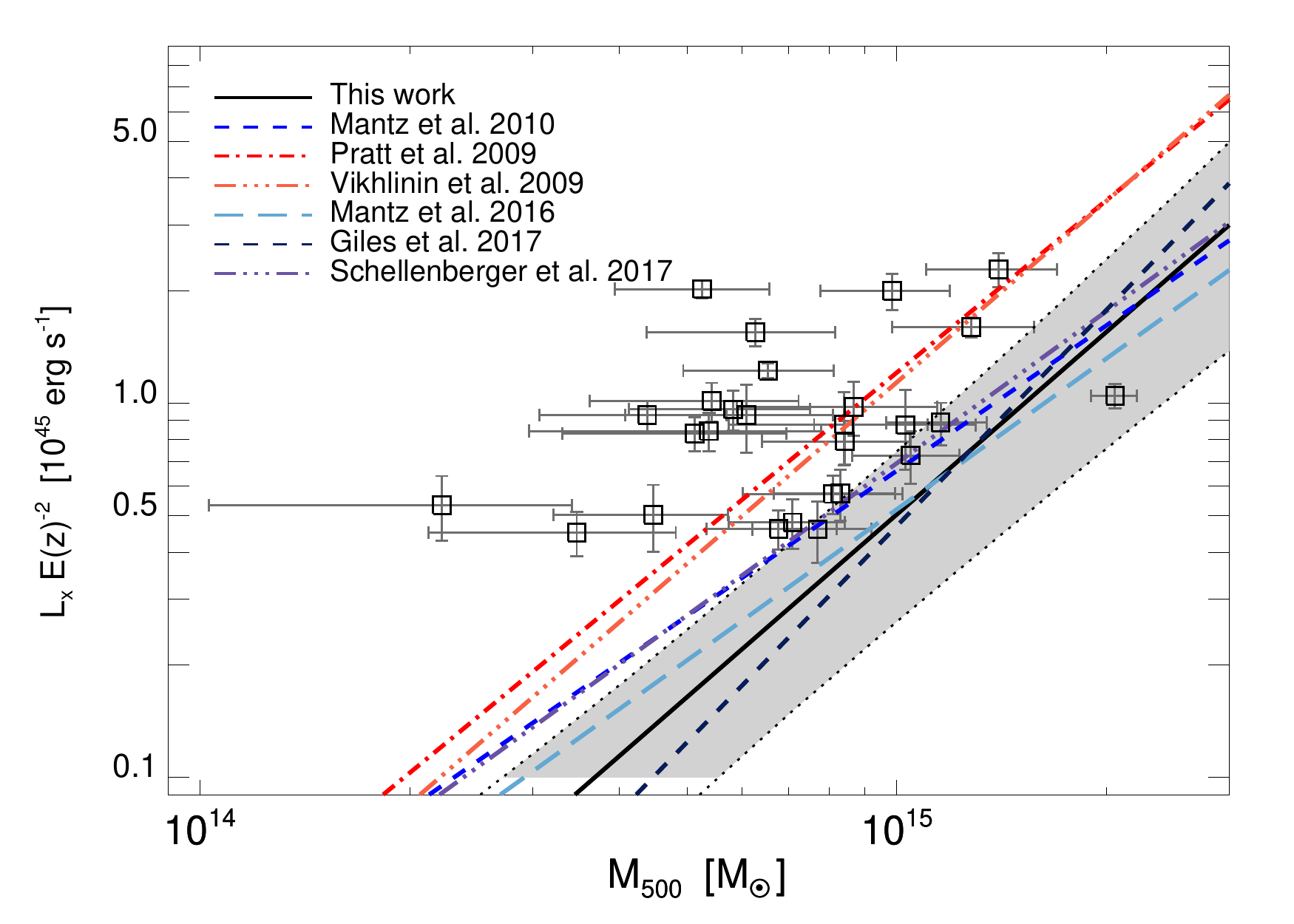} \caption{Luminosity-mass relation: The luminosities and our scaling relations are measured in the energy band $\text{0.1--2.4}$ keV. The scaling relations are represented for the luminosity-mass relation for the same energy band.
  The grey shaded region represents the 68 \% confidence level of our scaling relation. The measured values are generally up-scattered from the median relation. This is due to the Eddington and Malmquist biases in the sample which is corrected for in the scaling relation determination through our Bayesian analysis. Previous literature measurements of the luminosity-mass relation for the luminosity in the same energy band of $\text{0.1-2.4}$ keV are plotted for comparison with our constraints. 
  This is discussed in Section \ref{sec:compare}.}\label{fig:mllit}
 \end{figure}

\begin{figure*} \includegraphics[width=18cm]{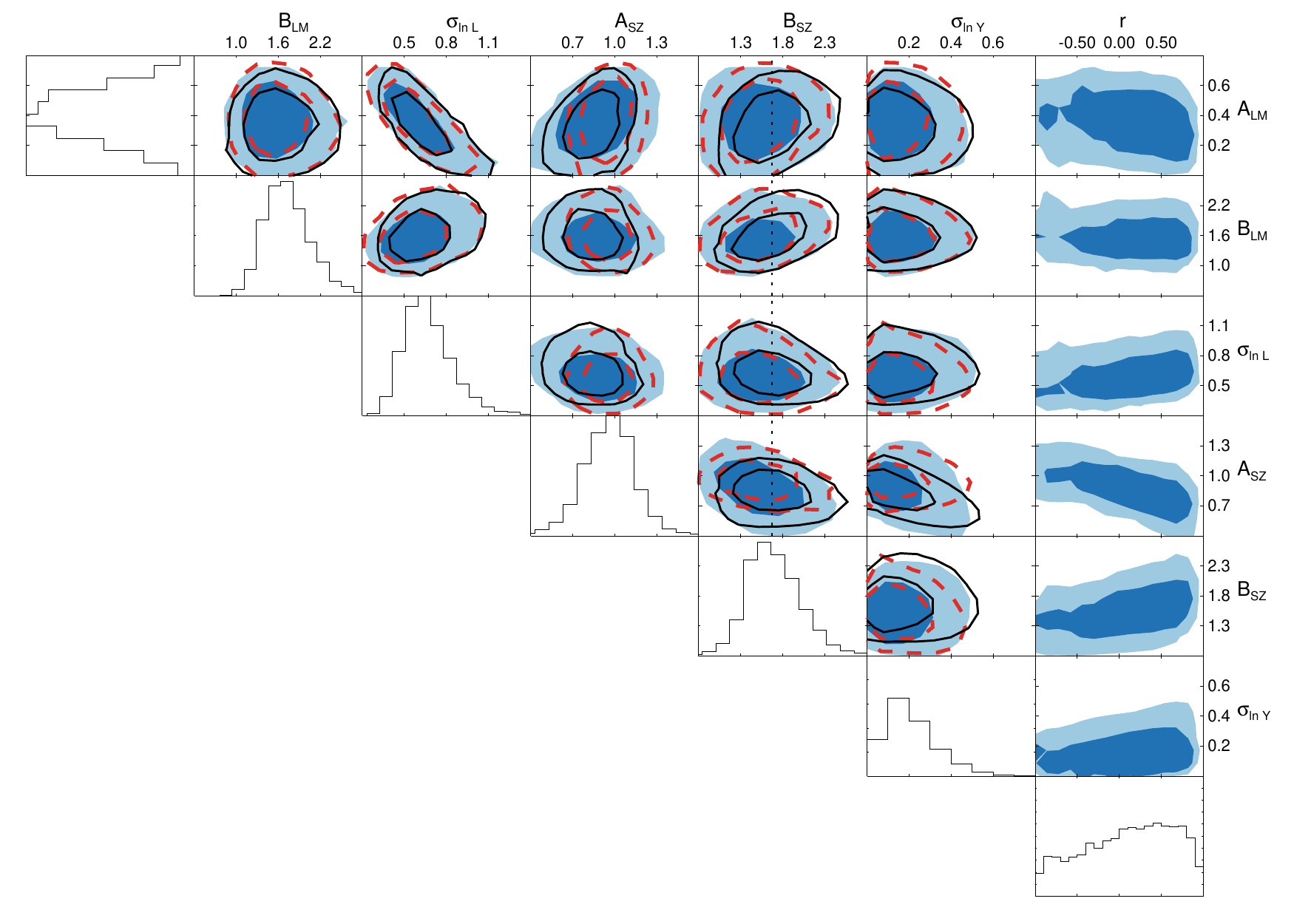}\caption{Same as Figure \ref{fig:ryml} but for $\sigma_{\rm WL|HM} =0.20M_{\rm HM}$.  The red contours are marginalized recovered confidence level for $r=0.0$, and black contours for $r=0.5$. The histograms show the marginalised distribution of parameters recovered from varying $r$.}
\label{fig:rymlbecker}\end{figure*}

\begin{figure*}
 \includegraphics{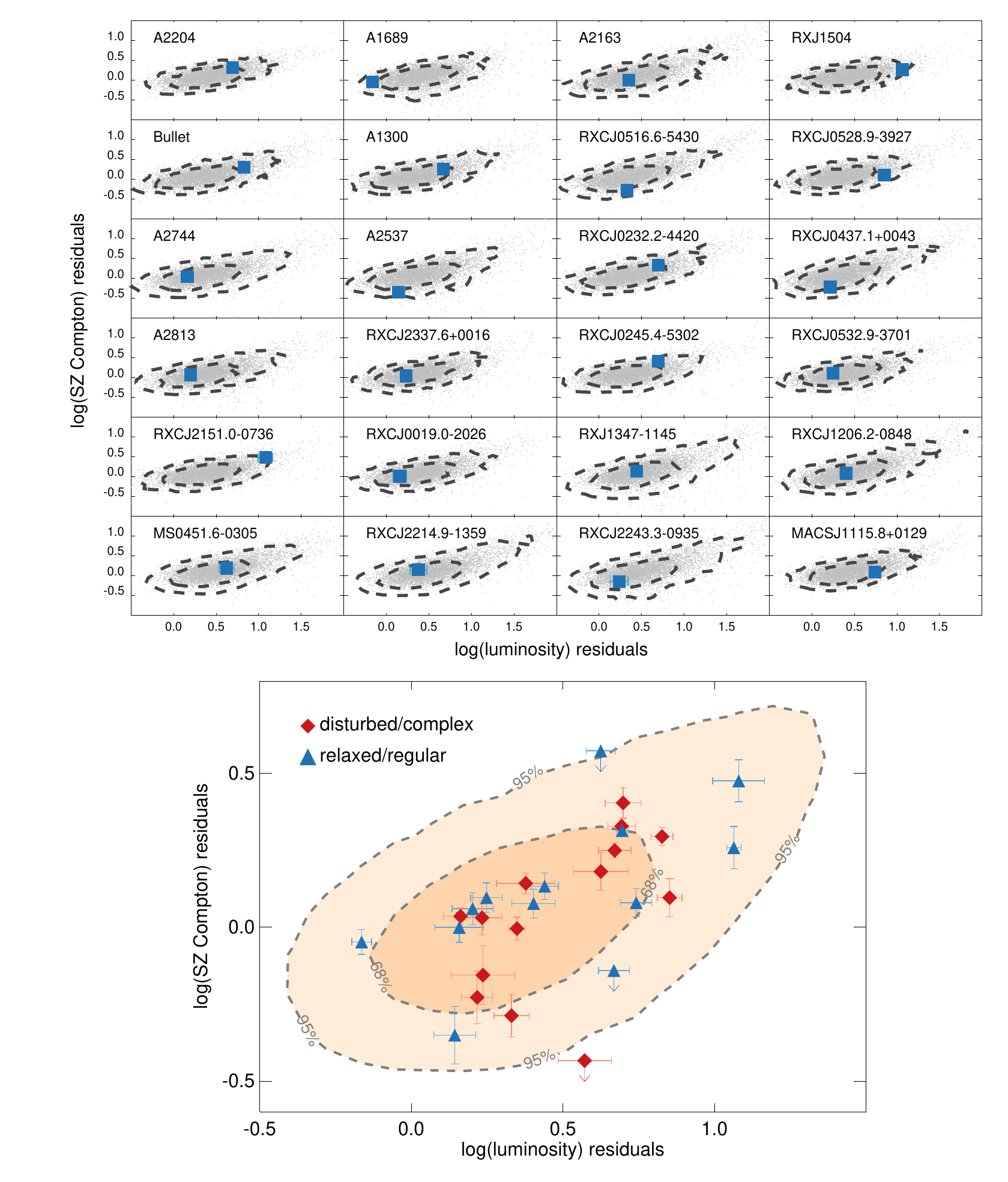}
  \caption{%
\textit{Top panel}: Luminosity and Comptonization residual data (blue square) after subtracting the median $Y_{\rm SZ}\text{--}M_{500}$ and $L_{\rm x}\text{--}M_{500}$ scaling relations for each of the 24 clusters that were detected with APEX-SZ. 
   The contours show the $68$\% and $95$\% confidence level prediction from random realisations of simulated residuals (grey points) for the median intrinsic covariance model.  
   \textit{Lower panel}: distribution of residuals of 24 clusters, showing a positive correlation. The Pearson correlation coefficient in residuals is 0.73.  The contours represent the $68$\% and $95$\% confidence levels obtained by combining the distributions shown in the top panel.
    The three non-detections in Compton-Y are also shown (downward arrows) with their upper limit set to 2$\sigma$. 
   }\label{fig:residuals}
\end{figure*}

 \begin{figure*}
  \includegraphics[keepaspectratio=1,width=\columnwidth]{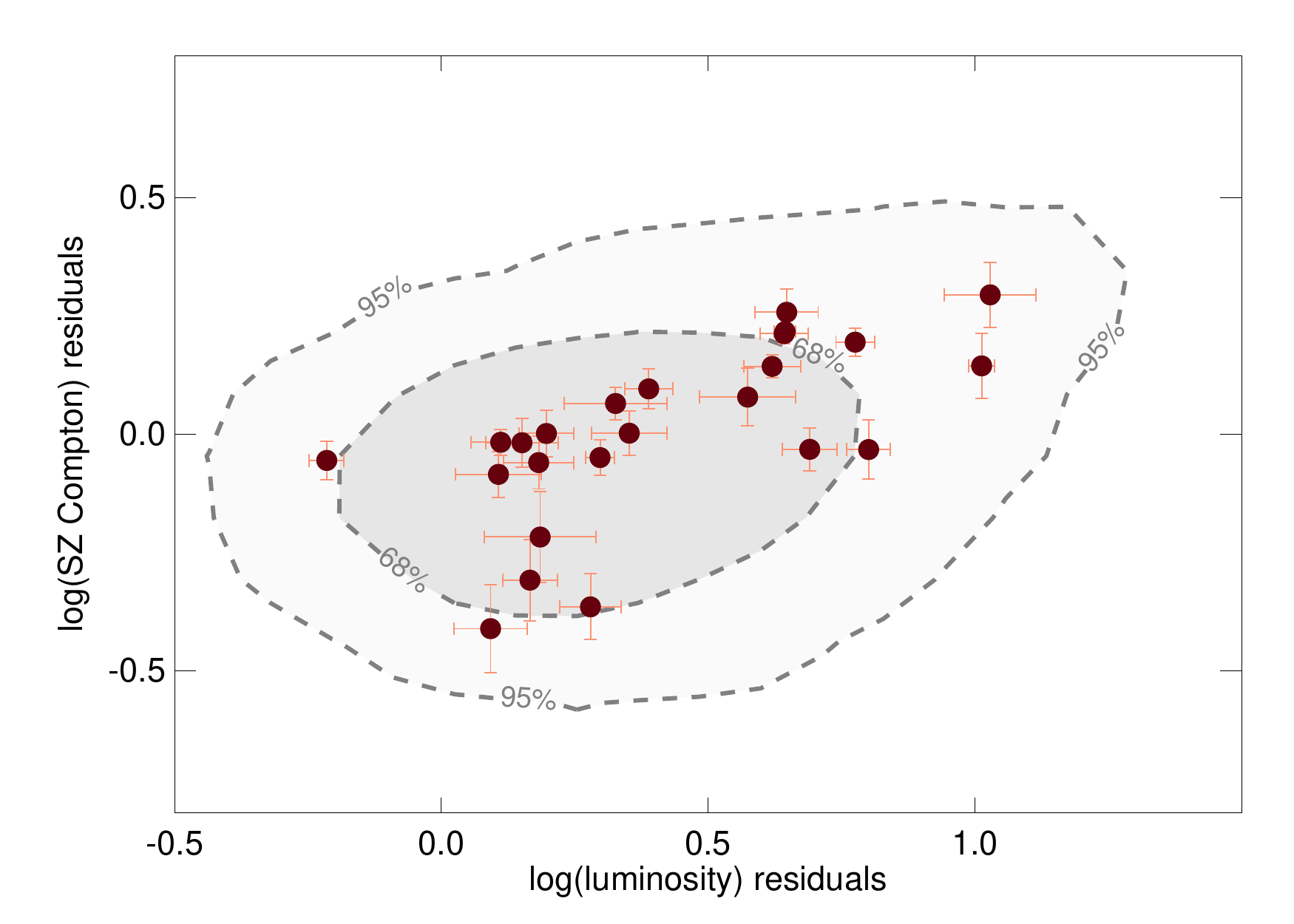}
  \includegraphics[keepaspectratio=1,width=\columnwidth]{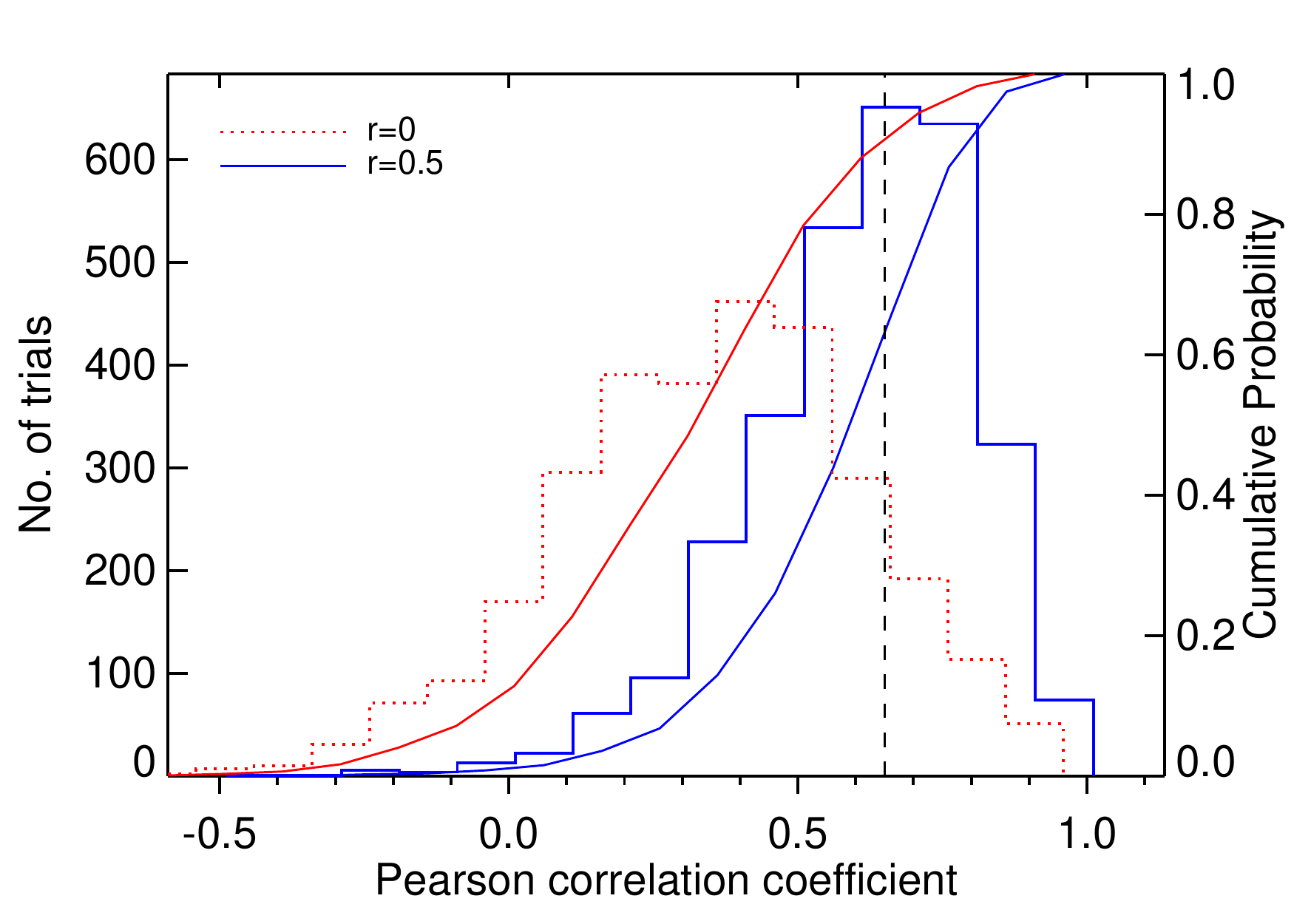}\caption{\textit{Left:} Same as Figure \ref{fig:residuals}, however, the residuals correspond to the median relation determined for the $r=0$ covariance model. \textit{Right:} Distributions of Pearson correlation coefficients in simulated realisations of 24 cluster residuals with un-correlated ($r=0$) and correlated ($r=0.5$) intrinsic scatters. The vertical dashed line indicates the measured Pearson correlation coefficient ($0.66$) in the residual data. 
 }\label{fig:residualsrzero}
\end{figure*}

\section{Robustness and limitations of the analysis}\label{sec:robust}
We now examine the robustness of the scaling relation analysis from the last section to potential modelling variations (Section \ref{sec:redshiftevolution}, \ref{sec:completeness} and \ref{sec:limitations}), systematic errors (Section \ref{sec:xraycenters} and \ref{sec:systematic}) and data selection choices (Section \ref{sec:outliers}).
\subsection{Redshift evolution of scaling relations}\label{sec:redshiftevolution}
We assumed a self-similar evolution in the $Y_{\rm SZ}\text{--}M_{500}$ relation for our analysis. In order to check the validity of this assumption for our measurements, 
we split the sample into three redshift ranges: $0.15$ to $0.22$, $0.27$ to $0.31$, $0.31$ to $0.55$, consisting of five, 15 and seven clusters respectively. We then fit the joint scaling relations with all parameters except for $A_{\rm SZ}$ fixed to the median values from Section \ref{sec:results:rfree}. The recovered medians and 68\% confidence levels of $A_{\rm SZ}$ are $0.80_{-0.12}^{+0.15}$, $0.81_{-0.08}^{+0.08}$, $0.90_{-0.13}^{+0.16}$ in the low, median and high redshift bins respectively. The normalisation is consistent within statistical errors in all three redshift ranges of the sample.
    We change the redshift evolution slope from self-similarity ($-2/3$) to $0$ based on the best fit value of the slope of the redshift evolution found in \cite{2015Sereno}, even though they do not find this deviation from self-similar evolution to be greater than 68\% confidence level. Assuming zero slope for the redshift evolution (i.e. no redshift evolution in the $Y_{\rm SZ}\text{--}M_{500}$ relation) increases the normalisation by $9\%$. The other parameters, including the correlation between the intrinsic scatters, are consistent with the results obtained in Section \ref{sec:results:rfree}.
 %

%
%
 
 \subsection{Treatment of completeness of the eDXL sample}\label{sec:completeness}
 
 We check the effect of varying completeness of the sample in the luminosity-redshift plane. Our model assumes high completeness for the sample, which enables us to apply an analytical integration of the normalisation of the likelihood model described in Section \ref{sec:method}. 
 {\it A posteriori}, we predict the cluster number count using our median $L_{\rm x}\text{--}M_{500}$ relation from our fiducial result, the eDXL sample selection function and 
 the same mass function used for our modelling. Our model predicts a total number of clusters to be 27. We re-compute the prediction of the cluster number counts, this time considering the completeness in the luminosity-redshift plane. 
 Using the same model as before, we predict a sample size of 24 clusters. This suggests an average completeness of $\approx 89$\%.
\subsection{Additional covariances in the scatter of mass observable}\label{sec:limitations}

Our scaling models treat the intrinsic scatter in lensing mass at fixed mass as being independent of scatter in the thermodynamic observables. 
Dark matter simulations \citep{2016Shirasaki, 2012Angulo} found a correlation in the range of 0.6--0.9 in intrinsic scatter of integrated Comptonization and weak-lensing mass. 
\cite{2017PennaLima} were unable to constrain this correlation for a sample with a size similar to the one used in this work, and given our lack of statistical power to constrain any more free parameters, we have ignored this correlation in the present work. 
Since our dominant source of bias is expected to come from the selection in luminosity, any correlation between Compton-$Y$ and lensing mass (intrinsic) scatters would be a second order effect. 
       
We note that the scatter in X-ray luminosities is sensitive to the physical processes near the core of the cluster.  
The scatter in weak-lensing mass, however, is more affected by averaging over projected structures over a larger area. Thus, we expect a weaker correlation between the scatters of $L_{\rm x}$ and weak-lensing masses. The correlation between the scatters of lensing mass with luminosity was found to be 0.41 from dark matter simulations \citep{2012Angulo}.

\subsection{Impact of mis-centring gNFW profile}\label{sec:xraycenters}
   In Section \ref{sec:results}, we analysed the scaling relations for measurements of $Y_{\rm SZ}$ with gNFW model centred on BCG. Alternatively, in Section \ref{sec:lum} we obtained X-ray centroids for eDXL clusters from ROSAT. Here, we re-compute the Compton-Y's  with these centroids as per procedure used in Section \ref{sec:inty} and fit the joint scaling relations with these measurements. 
   
   When using X-ray centroids rather than optical (BCG) centroids to determine the integrated Comptonizations, the resulting scaling relation parameters differ only marginally from the results of Section \ref{sec:results:rfree}. 
   The recovered medians and 68\% confidence levels are given in the last row of Table \ref{tab:selresults}. 
    The main difference is an increase in the intrinsic scatter of integrated Comptonization at fixed mass. This is indeed expected considering that the accuracy of the centroid estimation from ROSAT is limited. 
    We find that all deviations in the scaling parameters are within $1\sigma$ confidence with respect to the ones obtained in Section \ref{sec:results:rfree}. 
This is not surprising considering that 80\% of the eDXL clusters show less than half of the APEX-SZ beam offset in the two centres. 
The angular distance of the offset between optical and X-ray centroids is plotted for 27 eDXL clusters in Figure \ref{fig:centeroffset}. The cluster $\rm RXCJ1135.6-2019 $ shows the largest offset (approximately 2.5 arc minutes). The estimated $Y_{\rm sph,500}$ for this cluster at the X-ray centre produces a detection, but a non-detection at the optical centre as already discussed in Section \ref{sec:lum}. 
We provide further details on the impact of this cluster on the scaling relation in Section \ref{sec:outliers}. 

       \subsection{Systematic uncertainty in weak-lensing mass estimates }\label{sec:systematic}
       As discussed in Section \ref{sec:wlmass}, the weak-lensing mass estimates have a combined systematic uncertainty of approximately $8$\%. Other weak-lensing estimates \citep[e.g.,][]{2014Applegate} report similar level of systematic uncertainties, although the contributions and sources of systematics are not identical. 
      In our modelling, we have not accounted for this additional uncertainty in the lensing masses. 
       To the first order, we model the impact of this uncertainty inherent to the mass estimates on the normalisation of the scaling relations.  
      We consider a relative difference in the normalisation ($A_{\rm SZ}$), $\frac{\delta A_{\rm SZ}}{A_{\rm SZ}}$, for a fixed slope ($B_{\rm SZ}=1.51$). 
      The relative uncertainty comes out to be $+0.14$ and $-0.10$ for an $8\%$ underestimation and overestimation respectively.  These uncertainty levels are comparable to, but less than the confidence levels obtained for the normalisation parameter. 
       We note that this estimate ignores the effect of the mass function, however, this is expected to be a secondary effect. 

\subsection{Treatment of outliers}\label{sec:outliers}
 We explore the stability of our constraints to potential outliers. We identify a couple of such cluster measurements and drop them completely or replace them with alternate measurements from different modelling assumptions or literature. 
In the following series of tests, we find that the cluster mass measurement of A1689 has the strongest impact on the $L_{\rm x}\text{--}M_{500}$ relation at $\sim 1\sigma$ level. This, in turn, affects our $Y_{\rm SZ}\text{--}M_{500}$ relation at $\sim 1\sigma$. Since our cluster sample is X-ray selected and the variations in the constraints are at most $\sim 1\sigma$, we do not have a reason to exclude this cluster from our analysis. 
We describe these details below.  

In our sample, the weak-lensing mass estimate of $\rm A1689$ not only places it as the most massive cluster, it also has the most precise measurement in mass. From Figure \ref{fig:mllit}, we observe that $\rm A1689$ is the only cluster that is down-scattered in the $L_{\rm x}\text{--}M_{500}$ relation. We investigate how our scaling relation constraints are driven by this cluster. We drop it altogether from our Bayesian fitting and re-perform the joint analysis with 26 cluster measurements. The most notable change we observe a decrease of almost $0.7\sigma$ in the intrinsic scatter of the $L_{\rm x}\text{--}M_{500}$ relation, and this being anti-correlated with the $A_{\rm LM}$, increases the latter by $\sim 1\sigma$. 
We observe an increase of $\sim 0.8\sigma$ increase in the normalisation of the $Y_{\rm SZ}\text{--}M_{500}$ relation, and the slopes $B_{\rm SZ},~B_{\rm LM}$ are estimated to be steeper by $\sim 0.7\sigma$. 
The correlation parameter $r$ is lowered marginally by $0.4\sigma$, with over 77\% of the distribution still preferring a more positive or greater than zero correlation. In order to disentangle the effect of differences in the $L_{\rm x}\text{--}M_{500}$ relation impacting the changes in the $Y_{\rm SZ}\text{--}M_{500}$ relation, we fix the $L_{\rm x}\text{--}M_{500}$ relation and the $r$ parameter to the median values obtained from Section \ref{sec:results:rfree} and re-fit the $Y_{\rm SZ}\text{--}M_{500}$ relation by including and excluding the cluster in our analysis. We recover identical scaling relation parameters for the $Y_{\rm SZ}\text{--}M_{500}$ relation irrespective of whether $\rm A1689$ is included or excluded in the analysis.  
We infer that the steeper $B_{\rm SZ}$, and higher $A_{\rm SZ}$ are found mainly due to the impact the mass measurement of $\rm A1689$ has on the $L_{\rm x}\text{--}M_{500}$ relation. 

The weak-lensing mass estimates across the sample quoted in this paper have used the \cite{2013ApJ...766...32B} $c\text{--}M_{200}$ relation to break the degeneracy in concentration and mass. The intrinsic scatter in the relation has been ignored for the purpose of obtaining the mass estimates. In the NFW fitting with two free parameters without using the $c\text{--}M_{200}$ relation, we find the mass estimate to be 2$\sigma$ lower with $M_{500}=17.2\, _{-1.4}^{+1.4}\times 10^{14}M_{\sun}$. A detailed study of the cluster using strong lensing, weak-lensing as well as triaxiality information finds a cuspier core and a lower mass (\citealt{2015Umetsu}). There is general consensus that the cluster is elongated along the line of sight (\citealt{2012Sereno, 2013Sereno, 2013Limousin, 2015Umetsu}). We replace our lensing mass estimate for this cluster with the spherical mass estimate obtained from the detailed analysis in \cite{2015Umetsu} of $M_{500c}=$ $12.6\pm 1.9 \times 10^{14} M_{\sun}$ and re-measure the corresponding $Y_{\rm SZ,500}$ in the same aperture. 
We perform the joint fitting with the updated mass and integrated Comptonization for $\rm A1689$ and fit for all the seven free parameters. The recovered scaling relations are indifferent to dropping $\rm A1689$ from the fit with the exception that the intrinsic scatter in SZ is lowered to 28\% (decreases by 7\%). We find the slope, $B_{\rm SZ}$, to be $1.73_{-0.32}^{+0.34}$, which is steeper than found in Section \ref{sec:results:rfree} by $0.6\sigma$. We fix the $L_{\rm x}\text{--}M_{500}$ relation to our median estimate as done previously and find the difference in slope and normalisation reduce to $0.45\sigma$ and $0.3\sigma$ respectively. In conclusion, the  marginal changes in the $Y_{\rm SZ}\text{--}M_{500}$ relation mainly occur due to the $L_{\rm x}\text{--}M_{500}$ relation preferring higher normalisation, steeper slope and lower intrinsic scatter when $\rm A1689$ is not as massive.    

In addition, we investigate the non-detection in SZ for $\rm RXCJ1135.6-2019$. As discussed in Section \ref{sec:xraycenters}, the SZ measurement at the BCG centre is a non-detection versus a $3\sigma$ detection at the X-ray centre. 
We drop this cluster and re-fit the scaling relations by allowing the $r$ parameter to float. The parameter that shows the most significant change is $B_{\rm SZ}$ preferring a shallower value of $1.38_{-0.26}^{+0.24}$. This change is well within the confidence levels ($0.5\sigma$) of our fit results in Section \ref{sec:results:rfree}. We replace the $Y_{\rm SZ}$ measurement from BCG centre to X-ray centre and re-fit the relations and find the slope, $B_{\rm SZ}$, to be $1.42\, _{-0.23}^{+0.27}$ which remains shallower by $0.42 \sigma$ while rest of the scaling parameters are almost identical to the results obtained in Section \ref{sec:results:rfree}. The impact of this cluster measurement is marginal and at most on the slope of the $Y_{\rm SZ}\text{--}M_{500}$ relation. 

As a final step, we replace the mass of $\rm A1689$ by the value obtained in \cite{2015Umetsu}, the corresponding SZ measurement for this cluster within this aperture as done earlier and replace the SZ measurement of $\rm RXCJ1135.6-2019$ with the one obtained using the X-ray centroid. We jointly fit the seven free scaling parameters and find that the slope $B_{\rm SZ}$ is $1.59_{-0.28}^{+0.31}$ which effectively implies that the two outliers cancel each other's effects on the recovered slope. The other scaling parameters, namely, $A_{\rm SZ}$ and $A_{\rm LM}$ increase by $0.8\sigma$, $1 \sigma$ respectively. The intrinsic scatter $\sigma_{\ln L_{\rm x}}$  and $\sigma_{\ln Y_{\rm SZ}}$ are lowered by $1\sigma$ and $0.5\sigma$, respectively. The correlation parameter $r$ is lowered by $0.3\sigma$. The values of all scaling parameters except for $B_{\rm SZ}$ are similar to the ones obtained while fitting with lower mass measurement of $\rm A1689$ and leaving $\rm RXCJ1135.6-2019$ as a non-detection. 
Fixing the $L_{\rm x}\text{--}M_{500}$ relation to the median values in Section \ref{sec:results:rfree} reduces the $0.8\sigma$ difference in $A_{\rm SZ}$ to $0.3\sigma$ with respect to our fiducial result. In conclusion, the most dominant effect on the scaling relation comes from $\rm A1689$ which has been shown to mainly influence the $L_{\rm x}\text{--}M_{500}$ relation.  

\section{Discussion}\label{sec:discussion}
In this Section, we discuss the implications of our constraints on the scaling relations and the correlation coefficient of intrinsic scatters in $Y_{\rm SZ}$ and $L_{\rm x}$. First, we compare our recovered relations to previous work, which is summarised in Section \ref{sec:compare}. In Section \ref{sec:origin}, we discuss the origin of the correlation in intrinsic scatters and finally, we discuss the implications for future cosmological studies in Section \ref{sec:cosmology}.  
\subsection{Comparison to literature}\label{sec:compare}
    
\begin{figure} 
  \includegraphics[width=\columnwidth,keepaspectratio=1]{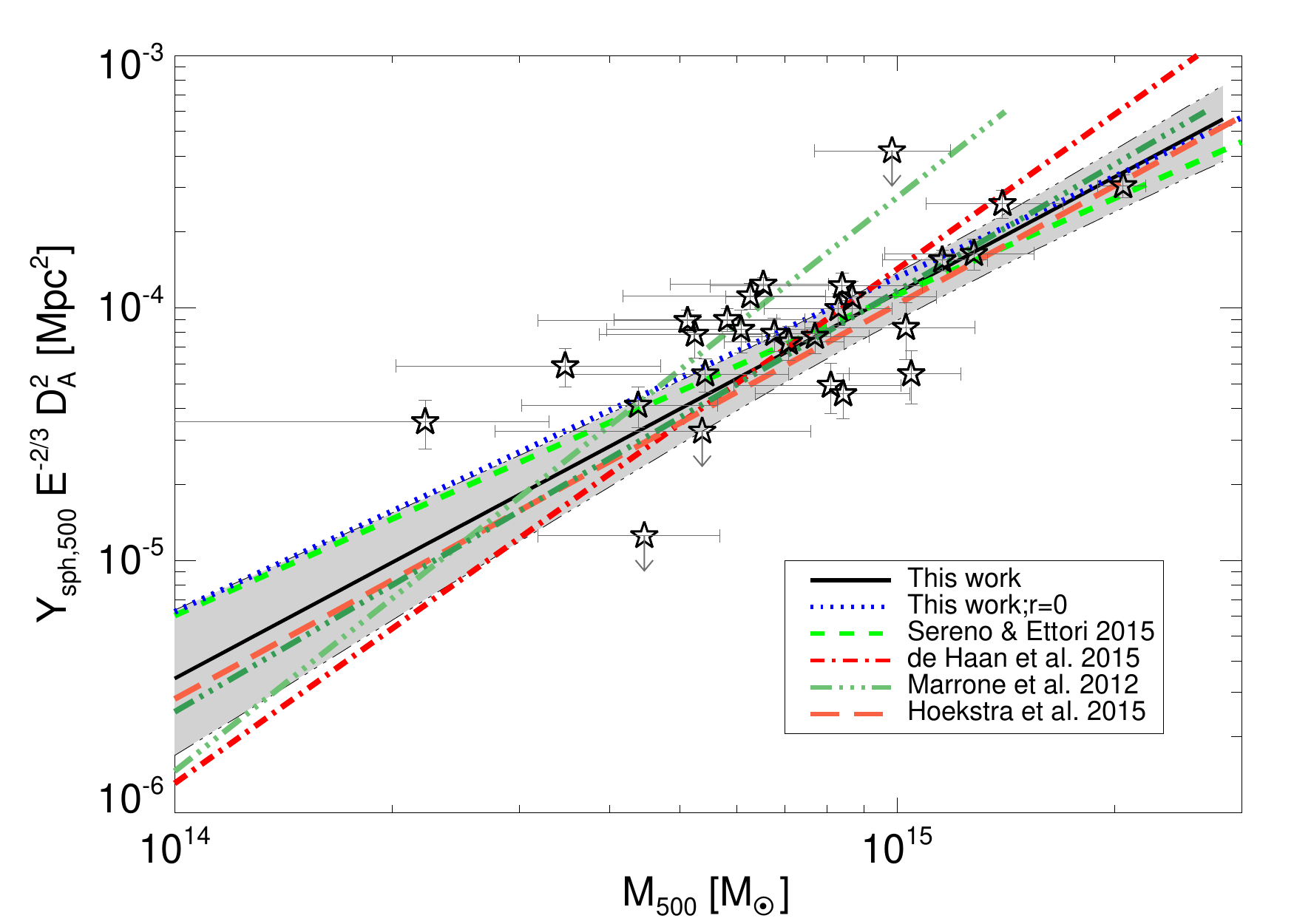} \caption{The median $Y_{\rm SZ}\text{--}M_{500}$ scaling relation is shown as the solid black line, with the 68\% confidence level indicated by the grey shaded region. 
   The constraints shown here were obtained by ignoring the weak-lensing intrinsic scatter. The scaling relations found in 
  other works that used weak-lensing mass to calibrate total mass within $R_{500}$ are overplotted. }
  \label{fig:compareWL}
\end{figure}

\begin{figure} 
  \includegraphics[width=\columnwidth,keepaspectratio=1]{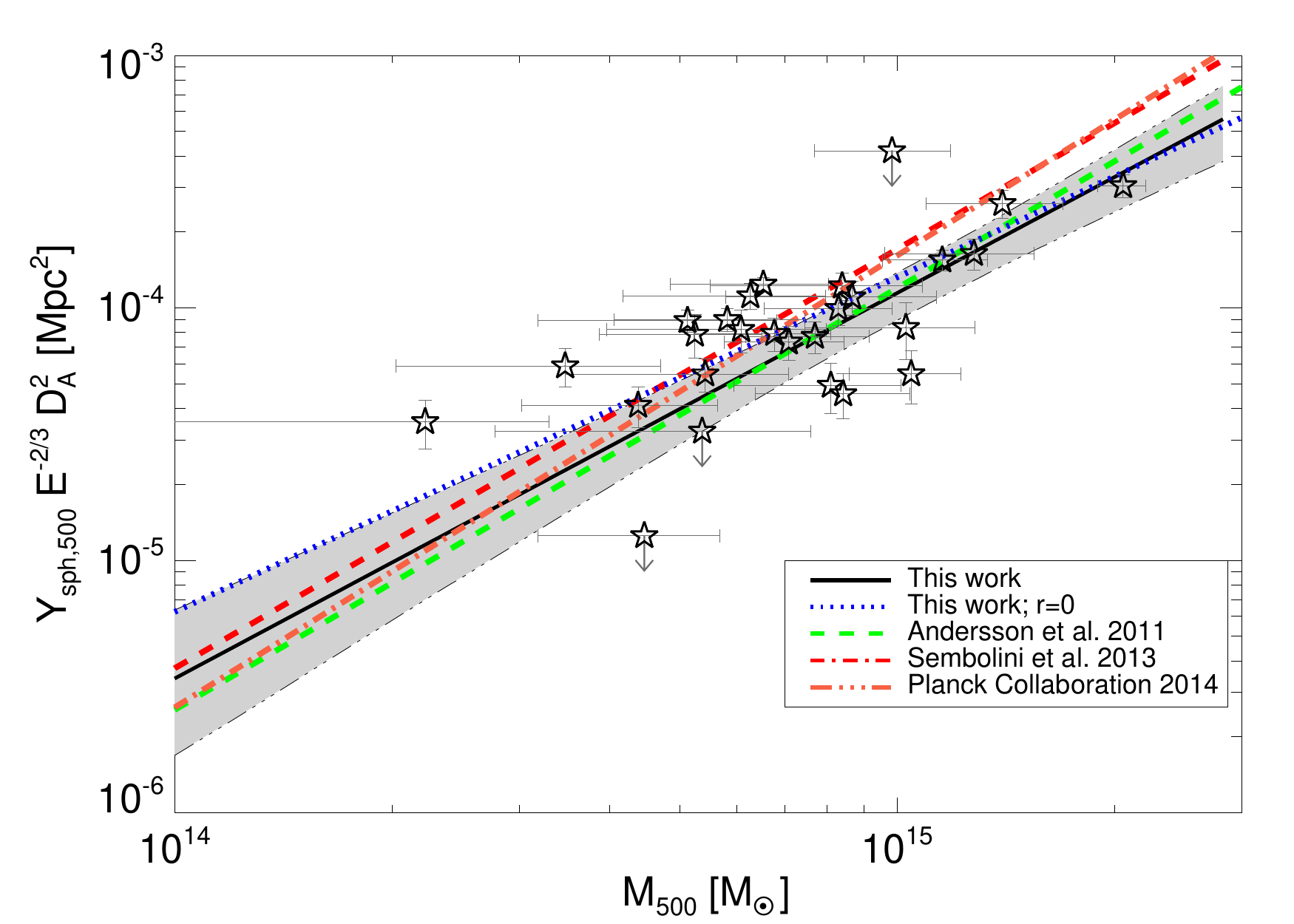}\caption{Same as Figure \ref{fig:compareWL} but comparing to previous works using mass proxies other than weak-lensing mass. The prediction from simulation work of Sembolini et al. (2013) is also shown for comparison.}\label{fig:compare}
\end{figure}

\begin{figure} 
\includegraphics[height=\columnwidth,keepaspectratio=true,angle=90]{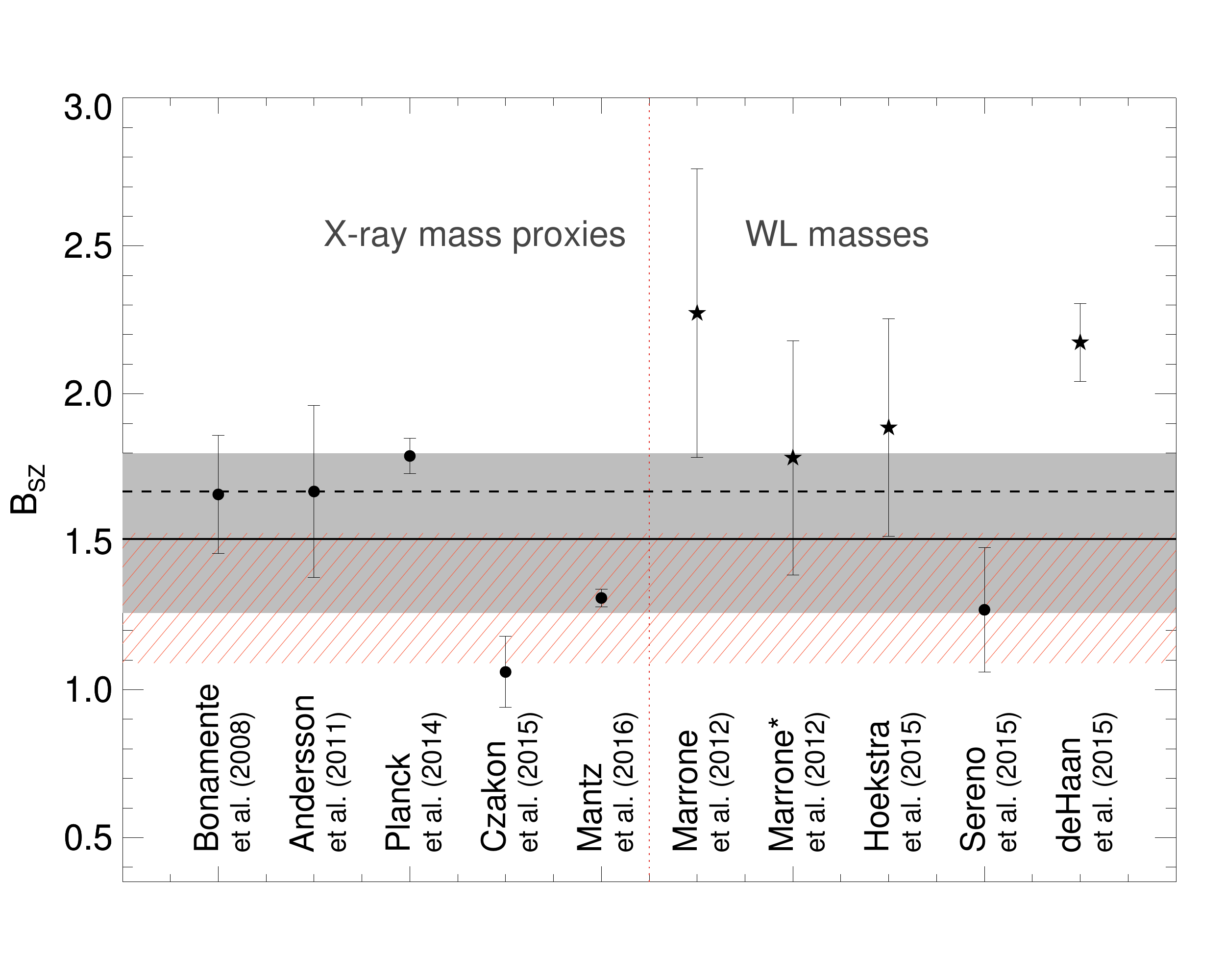}
\caption{Comparison of constraints on the slope of $Y_{\rm SZ,\Delta}-M_{\Delta}$ scaling relation between the literature and this work.
 The black solid line is our median slope value and the corresponding 1$\sigma$ confidence is represented as the grey shaded region. 
 The dashed line represents the self-similar slope value. 
 The 1$\sigma$ confidence on the slope constraints obtained with $r=0.0$ is indicated by the line filled region. We inverted the slope constraint for certain works which originally provided results for $1/B_{\rm SZ}$; these are represented as stars. The asterisk on Marrone et al. (2012) represents their result without A383.    }\label{fig:compare_slope}\end{figure}

We compare our median relation from Section \ref{sec:results:rfree} to other published work in Figures \ref{fig:compareWL} and \ref{fig:compare}. The scaling between $M_{500}$ and $Y_{\rm SZ,500}$ has been inferred in several previous publications using weak-lensing masses \citep[e.g.,][]{2012Marrone, 2012Hoekstra, 2015Hoekstra, 2015Sereno}  and using X-ray mass proxies \citep[e.g.,][]{2011Andersson, 2014Planck, 2016Mantzwtg}. The comparison with the former set of works is shown in Figure \ref{fig:compareWL} and with latter is shown in Figure \ref{fig:compare}. 
In both cases, there is general agreement in the scaling relation in the mass ranges of our eDXL sample with three exceptions \citep{2012Marrone, 2014Planck, 2016deHaanupdate}.
There are numerous differences other than the measurement of mass itself, such as, in the sample selection and follow-up statistics between this work and the above mentioned literature. 
We note that at lower mass end there is a mild disagreement with \cite{2012Marrone} and \cite{2016deHaanupdate}. At higher mass end there is some disagreement with \cite{2012Marrone}, \cite{2016deHaanupdate} and \cite{2014Planck}.  
The former two publications \citep[i.e.,][]{2012Marrone, 2016deHaanupdate} estimate a significantly steeper slope for the mass scaling of the integrated Comptonization than the self-similar value for the slope.   
The relation given in \cite{2012Marrone} used for comparison is the one including $\rm A383$ in their sample, and while dropping it they find a shallower slope that is consistent with self-similar slope and this scaling relation is more consistent within our uncertainties. In addition to this, their lensing masses from \cite{2010Okabe} have increased by $9\text{--}20$\% according to \cite{2016Okabe}. 
For comparison with \cite{2016deHaanupdate}, we use their $Y_{\rm x}\text{--}M_{500}$ relation, where $Y_{\rm x}$ is the X-ray equivalent of $Y_{\rm SZ}$. Their constraints on the $Y_{\rm x}\text{--}M_{500}$ relation are a byproduct of a full cosmological analysis. 
The disagreement with \cite{2014Planck} is due to their relation preferring a higher normalisation. This discrepancy could arise due to differences in the type of mass estimates used for the calibration and in the approaches adopted for mitigating selection biases.  In Figure \ref{fig:compare}, we also show the scaling relation prediction from numerical simulation of \cite{2013Sembolini} which indicates a higher normalisation than our constraints. 
A summary of slope values obtained from different literature and this work is presented in Figure \ref{fig:compare_slope}.
For comparison, we also include results using $R_{2500}$. We note that our recovered slope is in disagreement with \cite{2015Czakon}, who find a shallower slope for the $Y_{\rm SZ,2500}\text{--}M_{\rm x,2500}$ relation. Whereas, our biased estimate of slope when $r=0$ agrees well with the shallower estimate. 
   The ROSAT luminosities used in our work are
model dependent (Section \ref{sec:lum}), and suffer from low signal to noise and poor de-blending capabilities, in
particular for resolving out AGNs. Therefore, 
they are used in this work as a purely
phenomenological description for the
selection and not much importance
should be given to possible discrepancy with
other, more involved works. Nevertheless, we compare the $L_{\rm x}\text{--}M_{500}$ relation obtained in this work with other literature and they are shown in Figure \ref{fig:mllit}. 
Whenever possible, we use
scaling relations obtained directly from the
ROSAT luminosities in a similar band \citep[e.g., $L_{\rm BCS}\text{--}M$ relation given by][]{2017Giles}. 
   In addition, we compare with only those work that considered luminosity measurements without core-excision. To compare with \cite{2009Vikhlinin}, their $L_{\rm x}\text{--}M_{500}$ relation was modified to give the relation for luminosities in the same energy band as used in this work, i.e., $0.1\text{--}2.4$ keV. Similarly, the relation used for comparing with \cite{2009Pratt} was for the luminosity in the 0.1--2.4 keV energy band \citep[from Table B.2 in Appendix B of ][]{2009Pratt}. 
Our $L_{\rm x}\text{--}M_{500}$ relation is in good agreement with \cite{2010Mantz}, \cite{2016Mantzwtg}, \cite{2017Giles}, \cite{2017SchellenbergerI}. All of these work have accounted for their sample selection in a manner equivalent to ours. 
     We observe that there is a notable offset in our measured relation with respect to the best-fit of \citet{2009Vikhlinin} and \cite{2009Pratt}. 
     Possible sources of this offset could arise due to subtle differences between the approaches used by these authors and ours for accounting for the sample selection biases. The difference also could be due to different radius adopted for the luminosity measurements. In addition to this, the masses used in all of these publications for the calibration were primarily done with either hydrostatic masses or from $Y_{\rm x}$ with the exception of \cite{2016Mantzwtg}, who used some weak lensing masses in addition.  
However, the slope ($B_{\rm LM}$) is found to be consistent with all of the previous work \citep[e.g.,][]{2009Vikhlinin, 2009Pratt, 2010Mantz, 2016Mantzwtg, 2015Sereno, 2017Giles, 2017SchellenbergerI} within our $68\%$ confidence level. Our constraints on the intrinsic scatter of $L_{\rm x}$, $\sigma_{\ln L_{\rm x}}$, is $0.75\, _{-0.15}^{+0.21}$. 
This is consistent with the findings by \cite{2017Giles}, whereas it is higher than what was found by \cite{2010Mantz}, \cite{2016Mantzwtg}, \cite{2009Vikhlinin}, \cite{2017SchellenbergerI}. We note that the scatter is subject to how the luminosities are measured, such as removal of sub-structures, etc.

     \subsection{Origin of correlation between intrinsic scatters of $L_{\rm x}$ and $Y_{\rm SZ}$ at fixed mass}\label{sec:origin}
    Our constraint on the correlation parameter ($r$) suggests a positively correlated scatters of luminosity and SZ with 84\% of the marginalised distribution of this correlation lying above 0.12. In Section \ref{sec:robust:residuals}, we confirmed the consistency of our modelling by examining the residual data.  
    In Figure \ref{fig:residuals}, we indicate cluster morphologies as relaxed or disturbed. The classification was determined from using the centroid shift $w$ parameter in units of $R_{500}$ from weak-lensing estimate. The procedure for centroid shift calculation closely follows the method given in \cite{2013Weissmann}. We make a cut in $w$ at 0.0088. The clusters with centroid shifts less than this value are classified as relaxed or regular. The rest are determined to be disturbed. 
    From the distribution of the residuals, we do not find an indication of them being morphologically segregated.  
Since both SZ and luminosity probe the same ICM, they are expected to be correlated. Soft-band luminosity for massive clusters are essentially tracers of gas mass density and the Comptonization is sensitive to the product of gas mass density and temperature. 
The correlation between the scatters of luminosity and SZ observable could arise more due to fluctuations in gas mass fraction, as found in the pre-heating model of simulated clusters by \cite{2010Stanek}.  
For the pre-heating scenario they report the correlation in intrinsic scatters to be as strong as 0.88 between the gas mass fraction and the Comptonization and followed by a correlation of 0.78 between the scatters of integrated Comptonization and bolometric luminosity. In this case, the scatter in luminosity and SZ is largely driven by the scatter in the gas mass fraction. While \cite{2010Stanek} used bolometric luminosity that has stronger temperature dependence than the soft-band luminosity of the ROSAT survey for massive clusters, we can extrapolate that the scatter in soft-band luminosity is indeed dominated by gas mass fraction. 
Furthermore, more recent work by \cite{2016Truong} in their simulation with AGN contributions find a correlation of 0.71 between the intrinsic scatters of $Y_{\rm x}$ and gas mass and 0.66 between the scatters of bolometric luminosity and $Y_{\rm x}$. 

Our constraint on $r$ is consistent with these predictions from simulations and is likely to originate from the scatter of gas mass in the ICM. 

\subsection{Impact on cluster based cosmological studies}\label{sec:cosmology}

The mass-observable calibration plays a decisive role in cosmological studies of the cluster population.
Despite recent progress, it still dominates the error budget of current analyses \citep{2016Planckxxiv}.
We consider here the implications of our new estimate of the $Y_{\rm SZ} \text{--} M_{500}$ relation on future
SZ cluster surveys such as SPT-3G (\citealt{2014Benson}), AdvACT (\citealt{2016Henderson}) and the Simons Observatory\footnote[1]{\url{http://simonsobservatory.org/}}. 

In this context, we focus on a simplified model for the SPT-3G 2500 deg$^2$ survey. We take a fixed detection threshold 
of $Y_{\rm SZ,lim}=0.88\times10^{-5}\,\mathrm{Mpc^2}$, chosen to match the predicted SPT-3G number counts (5000 clusters) of \cite{2014Benson}. 
We ignore measurement errors for simplicity. 

We first estimate the effect of the statistical errors of our $Y_{\rm SZ}\text{--}M_{500}$ on the constraining power 
for such a survey. We note, however, that future survey calibration samples would be larger and we might expect better statistical constraining power. 
Here, we discuss the effect of current level of statistical uncertainties on cosmological analyses to illustrate the need for such improvements in the measurements of scaling relations. To do so, we compute the predicted number of clusters for the same cosmology but
using the scaling relation parameters of each point in our MCMC chain. The resulting distribution is
extremely broad with a 68\% confidence interval spanning the range 1000--7800. 
Although our data set is one of the largest available samples with detailed SZ measurements and reliable 
lensing masses, the combined error of all the free parameters in our modelling outweighs the 
statistical errors for such a survey (essentially Poisson errors given the large survey volume).
This shows that a tremendous calibration effort is required in order to really take advantage of the 
next generation of SZ surveys. A common procedure to circumvent this problem is to improve 
the precision of the mass calibration by resorting to constraints derived from X-ray observations 
\citep{2016deHaanupdate,2016Planckxxiv}. 
However, the propagation of X-ray information to the SZ observables also involves a number of 
modelling  assumptions, whose systematic effects are likely to dominate the error budget once larger cluster
samples are available.

Now, focusing on the average number of detected clusters, no strong claim can be made on revised 
expectations for the upcoming surveys since, within the uncertainties, our measurements are 
compatible with several of the previously available calibrations.
However, our work revealed the importance of some modelling assumptions on the final results, in particular 
the impact of the correlation between the SZ Comptonization and X-ray emission at fixed mass on scaling 
relations relying simultaneously on X-ray and SZ observables. In the case of our imaginary survey, the
predicted number of clusters would raise from $5000$ to $\sim21000$ if, instead of our median scaling 
relation, one used the median relation obtained without including the covariance in the scatter of $L_{\rm x}$ and 
$Y_{\rm SZ}$. We stress here that this shift purely originates from differences in the modelling assumptions 
and not from the statistical uncertainties inherent to our sample. The simulations described in 
Appendix~\ref{app:test} show, for instance, that even with much smaller statistical errors, similar 
deviations in the best fit or median values of the parameters are expected if the covariance between $L_{\rm x}$ and $Y_{\rm SZ}$ is 
neglected while modelling a correlated population. This demonstrates the necessity to always consider 
such a covariance term when X-ray and SZ observations are mixed. A detailed understanding of the nature 
of this covariance from both numerical simulations and other observations would definitely help in constraining the correlation and add relevant priors to the analysis of future surveys.

\section{Summary \& Conclusions}\label{sec:conclusions}
We study the statistical correlations between three galaxy cluster
observables: the integrated Comptonization, $Y_{\rm SZ}$, the X-ray
luminosity, $L_{\rm x}$, and the weak lensing mass, $M_{500}$.
Special attention is given to the sample selection bias and the
correlation between the intrinsic scatter of X-ray luminosity and SZ
Comptonization.

\begin{enumerate}
 \item We construct a complete sample of 30 clusters (eDXL) from the ROSAT X-ray survey catalogues. We obtain SZ effect measurements from the APEX-SZ experiment and lensing follow-up
 observations for 27 of these clusters. The global completeness of this sample is $\sim$ 90\%.

 \item We present and implement a Bayesian analysis method that allows
  to control the sample selection bias while fitting for the
  mass-observable scaling relations.  In the statistical formalism, we account for the impact of the measurement uncertainties of cluster properties,
the shape of the cluster mass function, intrinsic covariances of cluster observables at fixed mass and the selection function of the cluster sample. 

 \item We jointly constrain the $Y_{\rm SZ}\text{--}M_{500}$ and $L_{\rm
 x}\text{--}M_{500}$ scaling relations, while accounting for the correlation
 between the intrinsic scatter of $Y_{\rm SZ}$ and $L_{\rm x}$
 observables.
 The constraint on the correlation coefficient, $r=0.47_{-0.35}^{+0.24}$, is weak, but suggests a positive
 correlation, with 84\% of the marginalised distribution above
 $r=0.12$. The slope of the $Y_{\rm SZ}\text{--}M_{500}$ relation is found to be consistent with the self-similar expectation. At the current precision level, we find a general consistency in the relation with previous work. However, we note a marginally lower normalisation than given by \cite{2014Planck} at the pivot mass value of our sample. 
 Our $L_{\rm x}\text{--}M_{500}$ relation is measured using ROSAT luminosities for the sole purpose of accounting for the sample selection function. We find a slope for the $L_{\rm x}\text{--}M_{500}$ relation that is steeper than the self-similar slope value as noted by previous literature. Our constraints on the relation is found to be consistent with previous literature that account for sample selection in a manner equivalent to the one used in this work.
\item  We perform post-predictive checks that yield a consistent picture of the robustness of our modelling of the completeness of the sample, the measured correlation between the intrinsic scatters and the mass distribution. We vary our modelling assumptions on the redshift evolution of the $Y_{\rm SZ}\text{--}M_{500}$ relation, and the gNFW centroids for measuring $Y_{\rm SZ}$. We find our constraints to be stable within the confidence levels. 
In addition, we re-measure the scaling relation parameters by treating a couple of cluster measurements as possible outliers in our modelling. Among these, we find that removing the A1689 measurements altogether or replacing its mass by a lower estimate from literature dominantly affects the calibration of the $L_{\rm x}\text{--}M_{500}$ relation at $\sim 1\sigma$ and thereby also affects the $Y_{\rm SZ}\text{--}M_{500}$ relation at the same level. However, due to the selection in X-ray luminosities that yield a complete sample, we have no reason to remove this cluster from the analysis. 

 \item Using mock data emulating the eDXL mass proxy measurements, and
 assuming more precise measurements of the global observables, we show
 that ignoring the intrinsic covariance of cluster observables can lead to significant biases (depending on the strength of the correlation) in the measured
 mass-observable scaling relation, even when that particular
 observable plays no role in the sample selection.

\item In our eDXL/APEX-SZ sample, neglecting the correlation ($r$) of
 the intrinsic scatter between the integrated Comptonization and X-ray luminosity
 biases the normalisation (high by $1\text{--}2\sigma$) and slope (low by
 $1\sigma$) of the $Y_{\rm SZ}\text{--}M_{500}$ scaling relation.
 \end{enumerate} %

Cosmological studies using galaxy cluster scaling relations are
currently limited by the systematic uncertainties in those assumed
scaling laws. We show that, for future cluster-based cosmology
experiments, the biases induced in the measured scaling relations by not accounting for the
intrinsic covariance of cluster observables can dominate the error
budget of cosmological analyses.
It is prudent therefore to explicitly account for these covariance terms in the
modeling of the scaling relations for future cluster-cosmology
experiments.

\section*{Acknowledgements}
We thank Thomas Reiprich and Joe Mohr for useful discussions that has benefited this work. We acknowledge the help of Holger Israel in re-shaping part of this paper.
We thank Jens Erler and Kevin Harrington for proof-reading an early draft of this manuscript. We acknowledge the support and involvement of former graduate student Daniel Dahlin in the APEX-SZ Collaboration. We thank Reinhold Schaaf for providing support for the APEX-SZ data analysis software (BoA). We thank Hans B\"ohringer for sharing his knowledge that was helpful in the initial stages of target selection. 
AN was supported in the initial few months period through a stipend from the International Max Planck Research School (IMPRS) for Astronomy and Astrophysics at the Universities of Bonn and Cologne. AN was later supported by the Ph.D. scholarship from Deutscher Akademischer Austauchdienst (DAAD).
This work received support in part from the DFG Transregio program TR33 ``The Dark Universe''. FP acknowledges support by the German Aerospace Agency (DLR) with funds from the Ministry of Economy and Technology (BMWi) through grants 50 OR 1514 and 50 OR 1608.
 MK acknowledges the support of the Max Planck Gemeinschaft Faculty Fellowship program and the High Energy Group at MPE. MK also acknowledges the support of the DFG Cluster of Excellence ``Origin and Structure of the Universe''. 
CH acknowledges support from the Barbro Osher pro Suecia foundation and from the Swedish Research Council under grants 2006-3356 and 2009-4027. Work in the U.S. on the APEX-SZ instrument development and analysis was supported by the National Science Foundation under grants AST-0138348 and AST-0709497.






\bibliography{references} 
\bibliographystyle{mnras}


\appendix

\section{Bias due to correlations between the intrinsic scatters of cluster observables}\label{ch3:sec:samplebias}
To understand the types of biases to expect in a typical observable limited sample and how it propagates to the non-selecting observables, a simple demonstration of the sample selection biases with the help of a toy model is presented below. A similar discussion can be found in \cite{2011Allen}. Due to the subtleties involved in understanding of how the combination of cluster mass function, an observable limited selection and intrinsic covariance of cluster observables at fixed mass can distort the 
 distributions in the measurements plane of observables and produce bias in a sample, we describe this here.  



\begin{figure*}
 \includegraphics[scale=0.8]{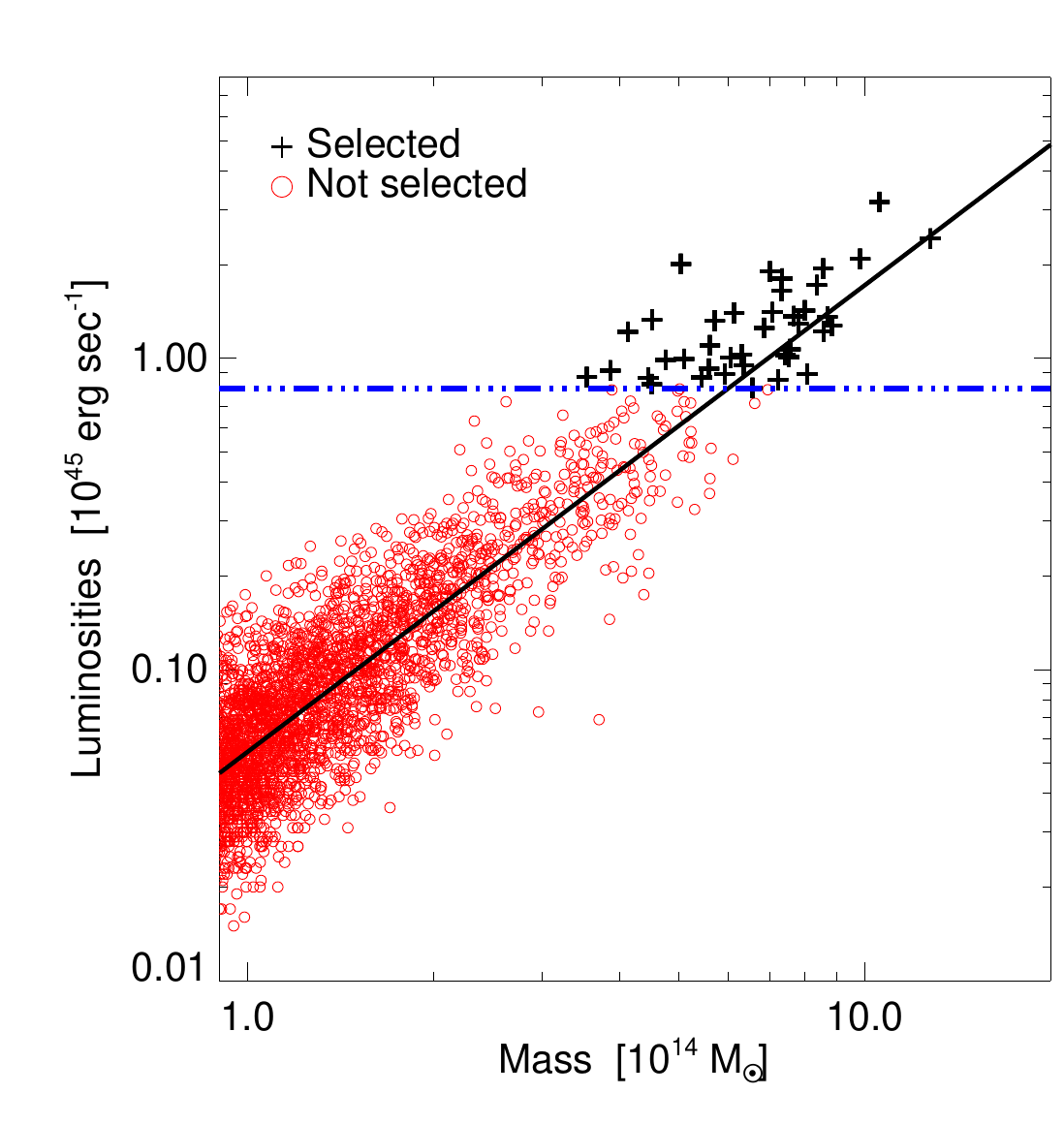}\\
 \includegraphics[width=16cm]{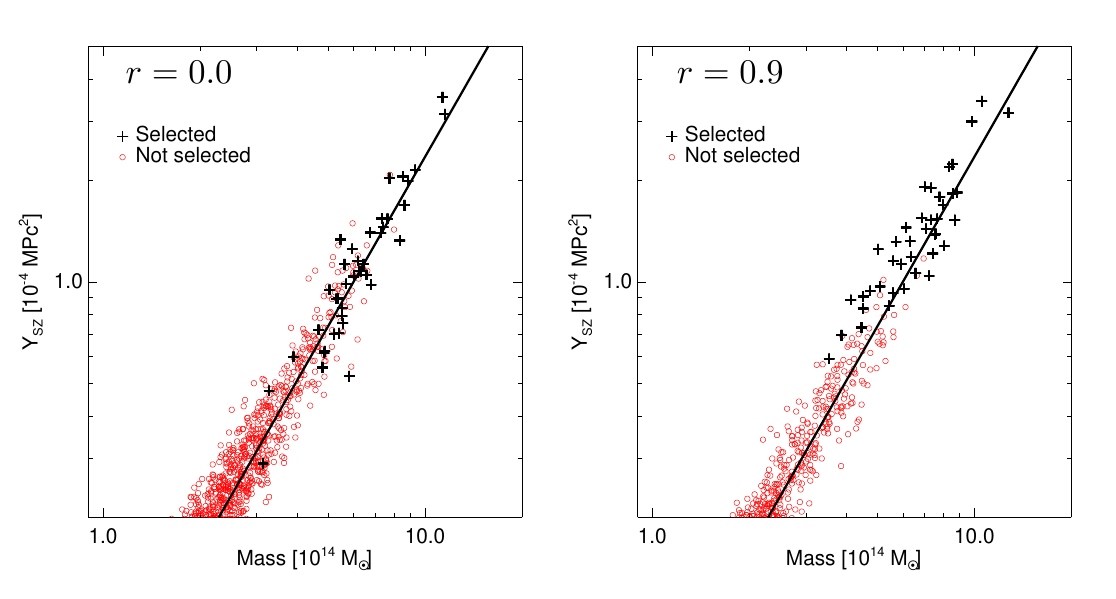}
 \caption{{\it Top}: A mock sample of clusters with X-ray luminosities vs. true mass generated using the Tinker mass function and a $L_{\rm x}\text{--}M$ scaling relation (black solid line) are shown. As predicted by the mass function, more massive objects are rarer than low mass systems. The intrinsic scatter of the $L-M$ relation mixes the population between mass bins and on applying a threshold cut in Luminosity (blue dashed line) this selects the intrinsically bright sources at low mass end. Together they produce a sample that is biased towards the low mass and intrinsically bright objects. {\it Bottom}: The sample generated above was used to also predict the sample in other cluster observables (e.g., $Y_{\rm SZ}$) when ({\it left}) luminosity and $Y_{\rm SZ}$ are uncorrelated in their scatters at fixed mass, and  ({\it right}) luminosity and $Y_{\rm SZ}$ are strongly correlated in their scatters at fixed mass.}
 \label{fig:biasdemo}
\end{figure*}
In Figure \ref{fig:biasdemo}, a toy model of cluster population is used to demonstrate how sample selection biases a sample such as the eDXL (Section \ref{sec:edxl}). 
It must be noted that this applies to any sample that is selected with a threshold cut in an observable. For simplicity, the observable is chosen to be the X-ray luminosity of clusters and the measurement scatters shall be ignored. After assuming a power-law relation of the luminosities with cluster mass, their one-to-one relation is distorted by cluster-to-cluster differences. 
We note here that the cluster mass function itself predicts more number of low-mass clusters than high mass systems. This is modelled by using the Tinker mass function \citep{2008Tinker}. 
The cluster-to-cluster differences is encapsulated by allowing for a log-normal intrinsic scatter in the cluster observables at fixed mass that distort the distribution in the observable-mass plane. Such a distribution of sources using the mass function, a scaling law and 40\% intrinsic scatter in X-ray luminosity is shown in the upper panel of Figure \ref{fig:biasdemo}. 
A selection threshold is chosen and applied to mimic the sample selection of a typical luminosity limited samples such as, LoCuSS \citep{2013Okabe, 2016Okabe}, REXCESS \citep{2009Pratt}, eDXL. 
If a scaling relation is measured naively to the black data points, the measured relation would be biased towards a relation that has a shallower slope and higher normalisation. 
Such a bias in scaling of mass-observable, where the sample was selected on the observable was identified by several authors and mitigation of such biases was recommended in measuring their scaling analysis \citep{2007Pacaud, 2006Stanek, 2009Vikhlinin, 2009Pratt, 2010Mantz}.

Now, we consider a correlation in the cluster observables on an individual cluster level as predicted from simulations \citep[][]{2010Stanek, 2012Angulo, 2016Truong}. 
For the toy model, this is considered to be 0.9 and cluster observable $Y_{\rm SZ}$ is also generated along with $L_{\rm x}$ using the \verb|mrandomn| routine in \verb|IDL|. 
Another set of generated sample was also considered with zero correlation between the intrinsic scatters. The two examples are shown in the lower panels of Figure \ref{fig:biasdemo}. 
In the zero correlation case, there are clusters that can be found to be both up-scattered and down-scattered from the assumed scaling relation. Whereas, in the strongly correlated scatters case, at the lower mass end of the selected sample, there is a bias towards up-scattered population of clusters in the follow-up observables. This biases the follow-up observables in similar fashion as the original selection observable. 
This demonstrates that even when the selection is not on the SZE observable, a selection in the X-ray luminosity and a correlation in these cluster observables at fixed mass can lead to a biased sample in SZE observable. 
In such an instance, ignoring the correlation would assume a relation for the SZE observable mass relation to be biased towards a higher normalisation and shallower slope scaling relation. 

\section{Derivation of the normalised likelihood for the eDXL sample}\label{app:lik}
Denoting measured variables with tilde, symbols $\tilde{M}_{\rm WL}$, $\tilde{Y}_{\rm SZ}$, and $\tilde{L}_{\rm x}$ represent the measured observables, namely, weak-lensing mass, SZ Compton parameter, and soft-band luminosities of a given cluster respectively.
$\sigma_{\tilde{M}_{\rm WL}}$, $\sigma_{\tilde{Y}_{\rm SZ}}$ and $\sigma_{\tilde{L}_{\rm x}}$ represent their uncertainties respectively. 

We assume that the ROSAT measurement scatter is log-normal, making it easier to compute analytically the normalisation of the likelihood.
For brevity, we drop the subscripts. We model the log-normal intrinsic scatters in the $L$ and $Y$ as correlated intrinsic scatters with a correlation coefficient parameter, $r$.
The selection function for eDXL cluster at a given redshift is \begin{equation}
P(\mathcal{I}=1| \tilde{L}, L^{\rm min})= 1 ~\mathrm{when }~ \tilde{L} \geq L^{\rm min}
\end{equation}
otherwise $P(\mathcal{I}=1|\tilde{L}, L^{\rm min})=0$. 
The full normalized likelihood is given as:
\begin{align}
\mathcal{L}=  \prod_i^{N_{\rm det}} \frac{\ P(\tilde{M}_i,\tilde{Y}_i,\tilde{L}_i|\theta)}{\int d\tilde{L}_i\ P( \mathcal{I}=1|\tilde{L}_i,\tilde{L}_{i}^{\rm min})\ P(\tilde{M}_i,\tilde{Y}_i,\tilde{L}_i|\theta) }\, .
\end{align}
Since measurements on luminosities are independent of other measured quantities, the contribution of each cluster in the sample to the normalisation can be reduced to $\int {\rm d}\tilde{L}\, P(\mathcal{I}=1| \tilde{L}, \tilde{L}^{\rm min})\, P(\tilde{L}|\theta)$ or $\int {\rm d}\ln\tilde{ L}\, P(\mathcal{I}=1|\ln \tilde{L}, \ln \tilde{L}^{\rm min}) \, P(\ln \tilde{L}|\theta)$, where the subscript $i$ has been ignored for simplicity.  
We consider,
\begin{multline}
 P(\ln\tilde{L}|\theta)= 
 \int {\rm d} \ln L\ P(\ln \tilde{L}|\ln L)\ P(\ln L|\theta)\\
~~~~~~~~~~~= \iint \mathrm{d}\ln L \ {\rm d}M \ P(\ln \tilde{L}|\ln L)\ P(\ln L|M,\theta)\ P(M)\\
 = \int {\rm d}M\ P(\ln \tilde{L}|\theta, M)\ P(M)\,. ~~~~~~~~~~~~~~~~~~~~
 \end{multline}

Assuming a log-normal distribution for measured luminosity distribution and the log-normal intrinsic scatter in the scaling law, the probability density, $P(\ln \tilde{L}|M,\theta)$, is given by
\begin{align}\label{app:eq:measlumprob}
  P(\ln \tilde{L}|M,\theta) =\frac{1}{\sqrt{2\pi(\sigma_{\ln \tilde{L}}^{2}+\sigma_{\ln L}^{2})}} \mathrm{exp}\biggl(-\frac{1}{2}\frac{(\ln \tilde{L}-\ln \hat{L}(M))^2}{\sigma_{\ln \tilde{L}}^{2}+\sigma_{\ln L}^{2}}\biggr).
\end{align}

To fully obtain the normalisation of the likelihood, we compute:
\begin{equation}
 \int_0^{\infty} dM \int_{-\infty}^{+\infty} d\ln \tilde{L} P(\mathcal{I}=1|\ln \tilde{L}, \ln \tilde{L}^{\rm min})P(\ln \tilde{L}|M,\theta) P(M)\, ,
 \end{equation}
 which reduces to $\int_0^{+\infty} dM \int_{\ln L^{\rm min}}^{+\infty} d\ln\tilde{L}\ P(\ln \tilde{L}|M, \theta) P(M)$.

Substituting for $P(\ln \tilde{L}|M,\theta)$ from Equation \eqref{app:eq:measlumprob} and integrating the above expression gives the final normalised likelihood as: 

\begin{align}\label{eq:sel}\mathcal{L} & = \prod_{i=1}^{N_{\rm det}} P(\tilde{M_i},\tilde{Y_i},\tilde{L_i}|\theta)  \nonumber \\
   &\qquad {}\times \frac{1 }  {\int_{0}^{+\infty}
      \frac{1}{2}\left[1-\mathrm{erf}\left(\frac{\ln \widetilde{L}^{\mathrm{min}}_{i}-\ln \hat{L}(M_{i}')}{\sqrt{2\left(\sigma_{\ln \tilde{L_{i}}}^{2}+\sigma_{\ln L_{i}}^{2}\right)}}\right)\right] P(M_{i}') dM_{i}' }\, .
    \end{align}

  The $\rm{erf}$ in the denominator arises due to the Heaviside step function used in the selection of luminosities. The denominator of the likelihood gives the probability of including the cluster in the sample.  The function $\hat{L}(M_{i})$ is given by the relation in Equation \eqref{eq:lm}.  The detailed description of probabilities related to $Y_{\rm SZ}$ and $M_{\rm WL}$ are given in appendix \ref{app:prob}. 
    The integrations over the nuisance parameters, namely, the true underlying values of the observables are computed via an MCMC by marginalising over the true observable variables. At each step of the MCMC, the normalization varies with the $L\text{--}M$ scaling parameters, which are left free in our fitting. Hence, the likelihood is re-normalised at every step in the MCMC. 

\section{Probabilities}\label{app:prob}
\subsection{Measurement probabilities}
\paragraph*{Measurement probability on weak-lensing and SZ observables:}{The conditional probability that $\tilde{M}_{\rm WL}$, $\tilde{Y}_{\rm SZ}$ are measured given true values, $M_{\rm WL}$  and $Y_{\rm SZ}$, is denoted by $P(\tilde{M}_{\rm WL},\tilde{Y}_{\rm SZ}|M_{\rm WL},Y_{\rm SZ})$.
We model the measurement probability densities on weak-lensing masses and the Compton-Y as bi-variate Gaussian probability density function: 
\begin{multline}\label{app:eq:measpdf}
P(\tilde{M}_{\rm WL},\tilde{Y}_{\rm SZ}|M_{\rm WL},Y_{\rm SZ})=\\ \left[\frac{1}{2\pi\sigma_{\tilde{M}_{\rm WL}}\sigma_{\tilde{Y}_{\rm SZ}} \sqrt{(1-\rho^2})}\right]F(\tilde{M}_{\rm WL},\tilde{Y}_{\rm SZ}|M_{\rm WL},Y_{\rm SZ})\, , \\
\mathrm{where} ~F(\tilde{M},\tilde{Y}|M,Y)=\\
\mathrm{exp}\left(-\frac{1}{2} \frac{1}{(1-\rho^2)} \left \lbrace \left(\frac{\tilde{M}-M}{\sigma_{\tilde{M}}}\right)^2 +\left(\frac{\tilde{Y}-Y}{\sigma_{\tilde{Y}} }\right)^2 -2\rho \frac{(\tilde{Y}-Y)(\tilde{M}-M)}{\sigma_{\tilde{M}} \sigma_{\tilde{Y}} } \right \rbrace\right)
\end{multline}
and $\rho$ is the correlation in the measurement uncertainties in the two observables.
}

\paragraph*{Measurement probability of luminosities}{$P(\tilde{L}_{{\rm x}}|L)$ is assumed to be log-normal with log-normal measurement uncertainty $\sigma_{\ln \tilde{L}} = \sigma_{\tilde{L}}/\tilde{L}_{x}$ . }
\subsection{Scaling model probabilities}\label{app:scal_prob}
Log-normal probability distribution of the $Y_{\rm SZ}$ and $L_{\rm x}$ at fixed mass is given by,
 
\begin{multline}
P(Y_{\rm SZ},L_{\rm x}|M,\theta_{y},\theta_{l},\sigma_{\ln L_{\rm x}},\sigma_{\ln Y_{\rm SZ}},r)=  
\frac{1}{2 \pi \sigma_{\ln L_{\rm x}}\sigma_{\ln Y}Y_{\rm SZ}L_{\rm x} \sqrt{1-r^2}}\\ 
\times \; \mathrm{exp} \biggl \lbrace -\frac{1}{2} \frac{1}{(1-r^2)} \biggl [ \biggl(\frac{\ln Y_{{\rm SZ}}-\ln \hat{Y}(M)}{\sigma_{\ln Y_{\rm SZ}}}\biggr)^2 + \biggl(\frac{\ln L_{\rm x}-\ln \hat{L}(M)}{\sigma_{\ln L_{\rm x}} }\biggr)^2 \\
-2r  \frac{(\ln Y_{\rm SZ} -\ln \hat{Y}(M))(\ln L_{\rm x}- \ln \hat{L}(M))}{\sigma_{\ln L_{\rm x}} \sigma_{\ln Y_{\rm SZ}} } \biggr ]\biggr \rbrace \, ,\\
\end{multline}
where $\hat{Y}( M)$ is the scaling power-law. 
\subsection{Intrinsic scatter in weak-lensing mass}\label{app:method:intscatter}
Here we calculate the marginalisation over the true weak-lensing mass $M_{\rm WL,500}$ for the scaling model described in Section \ref{sec:method:application:withintscatter}. 
We assume the true weak-lensing mass is unbiased and scatters from the halo mass $M_{\rm HM}$ with Gaussian distribution such that the dispersion is proportional to halo mass.  
The probability distribution of measured lensing masses is given by a Gaussian probability:
\begin{multline}\label{app:eq:pwl}
P(M_{\rm WL}|M_{\rm HM})=\\ \frac{1}{\sqrt{2\pi} \sigma_{M_{\rm WL|HM}}}\mathrm{exp}\left[-0.5 \left(\frac{M_{\rm WL}-M_{\rm HM}}{\sigma_{M_{\rm WL|HM}}}\right)^{2}\right],
\end{multline}
where $\sigma_{M_{\rm WL|HM}}=0.2M_{\rm HM}$. 
To calculate $P(\tilde{M}_{\rm WL}|M_{\rm HM})$, we compute the integral $\int d M_{\rm WL}\, P(\tilde{M}_{\rm WL}|M_{\rm WL})P(M_{\rm WL}|M_{\rm HM})$. Since in our case, the measurement probability of weak-lensing mass and the SZ observable measurements are correlated, we should take into account the cross-terms to compute the final integral. 
Consider the integral,
\begin{equation}\label{app:eq:wlscatmarginalisation}
\int \mathrm{d}M_{\rm WL}\ \mathrm{exp}\left[-0.5 \left(\frac{M_{\rm WL}-M_{\rm HM}}{\sigma_{M_{\rm WL|HM}}}\right)^{2}\right] F(\tilde{M}_{\rm WL},\tilde{Y}_{\rm SZ}|M_{\rm WL},Y_{\rm SZ})\, ,
\end{equation}
where  $F(\tilde{M}_{\rm WL},\tilde{Y}_{\rm SZ}|M_{\rm WL},Y_{\rm SZ})$ has the same meaning as given in equation \eqref{app:eq:measpdf}.
By defining,
\begin{align*}
A=\frac{1}{2}\left \lbrace \left [ \frac{1}{(1-\rho^2) \sigma_{\tilde{M}_{\rm WL}}^{2}}\right]+ \left [ \frac{1}{\sigma_{M_{\rm WL|HM}}^2}\right ] \right \rbrace\, ,
\end{align*}
\begin{align*}
B=\frac{1}{2} \left \lbrace \frac{1}{1-\rho^2} \left [ \frac{-2 {\tilde{M}_{\rm WL}}}{\sigma_{\tilde{M}_{\rm WL}}^{2}}+ \frac{2\rho (\tilde{Y}_{SZ}-Y_{\rm SZ})}{\sigma_{\tilde{M}_{\rm WL}} \sigma_{\tilde{Y}_{\rm SZ}}}\right ] + \left [ \frac{-2 M_{\rm HM}}{\sigma_{M_{\rm WL|HM}}^{2}} \right ] \right \rbrace\, 
\end{align*}
and  
\begin{multline*}
C= \frac{1}{2}\biggl \lbrace \biggl [ \biggl(\frac{\tilde{M}_{\rm WL}}{\sigma_{\tilde{M}_{\rm WL}}}\biggr)^{2}+
\biggl(\frac{\tilde{Y}_{\rm SZ}-Y_{\rm SZ}}{\sigma_{\tilde{Y}_{\rm SZ}}}\biggr)^{2}-\frac{2 \rho \tilde{M}_{\rm WL} (\tilde{Y}_{\rm SZ}-Y_{\rm SZ})}{\sigma_{\tilde{M}_{\rm WL}} \sigma_{\tilde{Y}_{\rm SZ}}}\biggr ] \\+ \biggl [ \frac{M^2_{\rm HM}}{\sigma_{M_{\rm WL|HM}}^{2}} \biggr ] \biggr \rbrace\, ,
\end{multline*}
the integral \eqref{app:eq:wlscatmarginalisation} reduces to
\begin{equation}
\int_{0
}^{+\infty} \mathrm{d}x \; \mathrm{e}^{-Ax^2-Bx-C}  = \frac{1}{2}\sqrt{\frac{\pi}{A}}  \mathrm{e}^{\frac{B^2}{4A} -C} \left[1- {\rm erf}\left(\frac{B}{2\sqrt{A}}\right)\right] \, ,
\end{equation}
where $M_{\rm WL}$ is the running $x$ variable in the above integral.
 
%
\section{Tests with simulated data}\label{app:test}
We analyse with the help of simulated mock data sets, the impact of ignoring correlation between scatters of response variables at a fixed true independent variable. We consider two response variables that are used synonymously with $L_{\rm x}$, $Y_{\rm SZ}$ for the sake of consistency. However, the response variables are inter-changeable when the selection is on some other observable such as temperature, $Y_{\rm SZ}$. 
The presented analysis can be extended to scaling relations with other thermodynamic properties as long as the selection is known and the independent variable is calibrated with mass of a cluster.  
\subsection{Mock data with correlated intrinsic scatters}\label{app:r}
\subsubsection{Generating mock samples} 
\begin{figure*}
 \includegraphics[keepaspectratio=true,width=15cm]{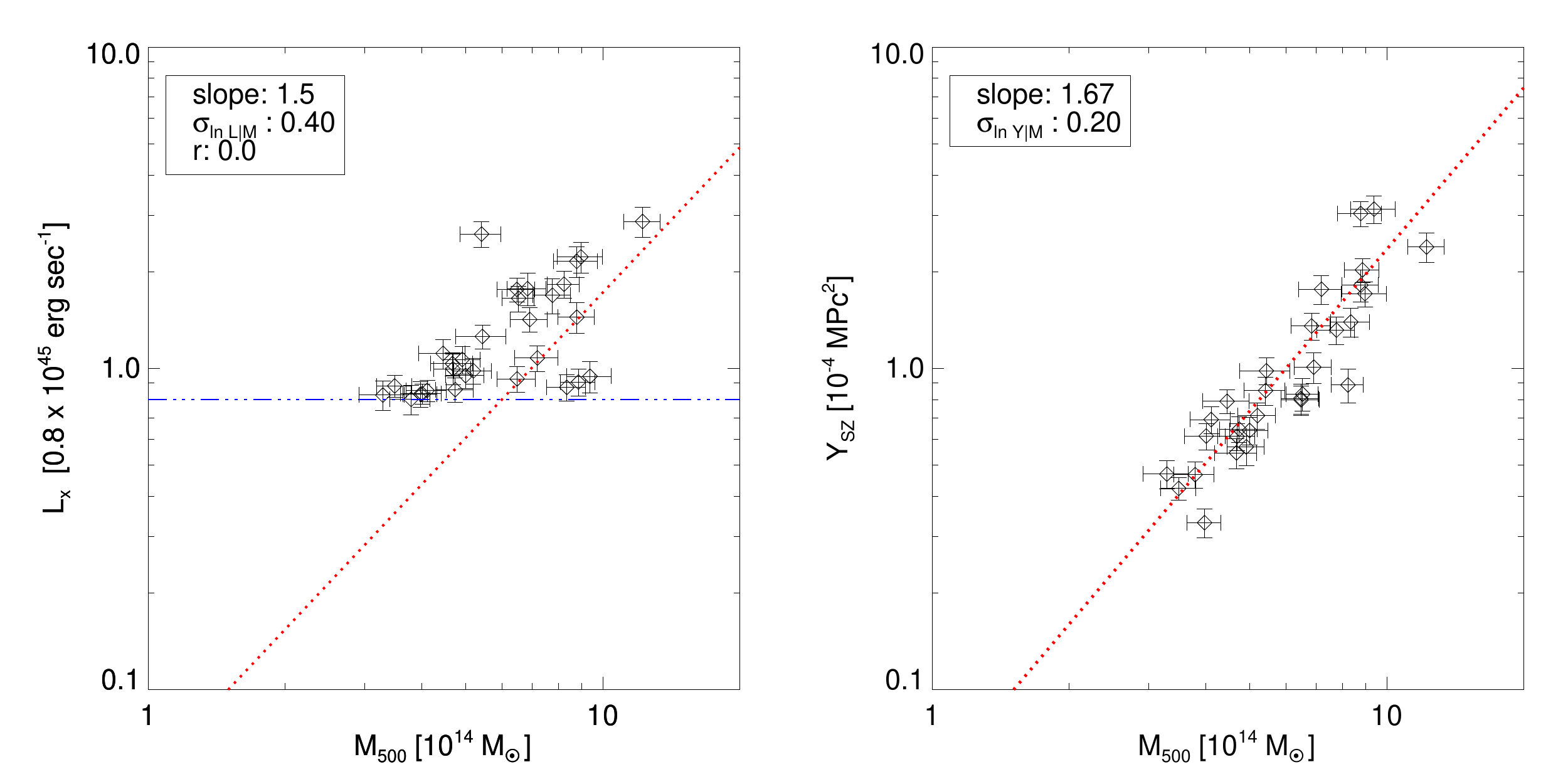}
\caption{An example set of simulated sample with their mock measurements at redshift $z=0.30$ is shown here. The sample was selected by applying threshold selection (blue horizontal line) in measured luminosities. 
The simulations of mock data included a correlated intrinsic scatters in the two response variables ($L_{\rm x}$, $Y_{\rm SZ}$) with the correlation coefficient set to $r=0.0$.
}\label{fig:r0}
\end{figure*}

\begin{figure*}
 \includegraphics[keepaspectratio=true,width=15cm]{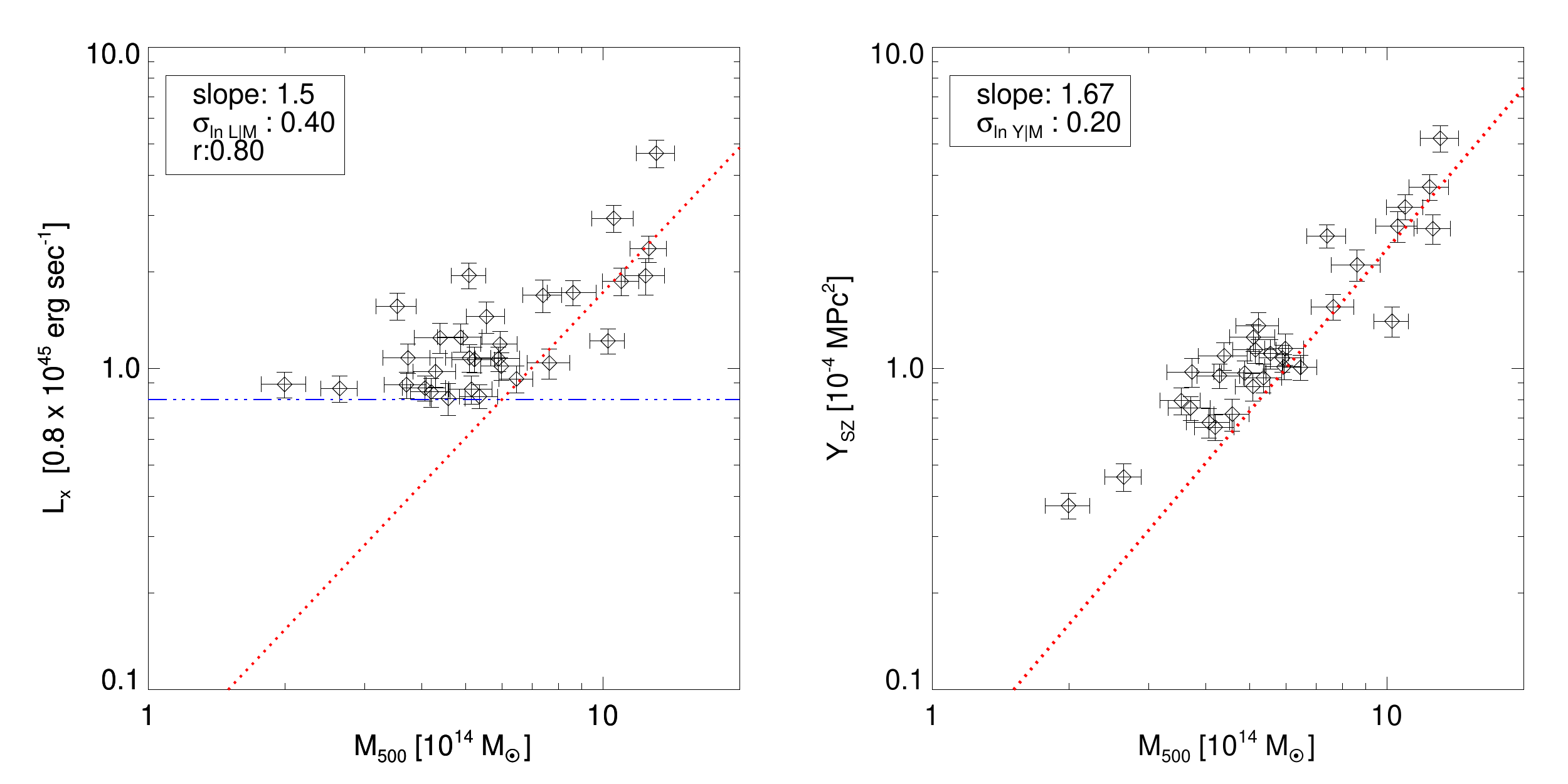}
\caption{An example set of simulated sample with their mock measurements at redshift $z=0.30$ is shown here. The sample was selected by applying threshold selection (blue horizontal line) in measured luminosities. 
The simulations of mock data included a correlated intrinsic scatters in the two response variables ($L_{\rm x}$, $Y_{\rm SZ}$) with the correlation coefficient set to $r=0.80$.}\label{fig:r0.8}
\end{figure*}

We use the  Tinker mass function \citep{2008Tinker} to generate a sample of galaxy clusters whose underlying distribution in terms total masses in $R_{500}$ is $\propto$ $p(M_{500},z)$. The mass function used here was for the cosmology where $\Omega_{m}=0.3$, $\Omega_{\Lambda}=0.7$, $H_{0}=70 \rm ~ km/s/Mpc$, $\sigma_{8}=0.82$,  $\Omega_{b} = 0.045$, and slope of the primordial power-spectrum,  $n_{pk} =-1.0$. 
Using a set of input scaling relation parameters for the $L_{\rm x}\text{--}M_{500}$ and the $Y_{\rm SZ}\text{--}M_{500}$ relations (Equations \eqref{eq:lm} and \eqref{eq:sz-mwl}), 
we generate true values of $Y_{\rm SZ}$ and $L_{\rm x}$ observables including a log-normal intrinsic covariance at fixed mass as per Equation \eqref{eq:matrix}. 
Subsequently, $Y_{\rm SZ}$, $M_{500}$ are further scattered with Gaussian measurement scatters assuming some relative percentage uncertainties. $L_{\rm x}$ is scattered with a log-normal measurement uncertainty. The mock cluster sample is then built based on the mock measured luminosities and an applied luminosity threshold. 
We consider three cases: 
\begin{enumerate} \item there are statistically small measurement uncertainties $\sim$ 10\% on all observables. 
\item the measurement uncertainties are larger as in the eDXL measurements.%
\item  the intrinsic scatters on response variables are larger and keeping the measurement uncertainties similar to realistic measurement uncertainties as in point (ii). We call this eDXL like samples due to the recovered scatters found in our analysis. The input scaling relation is shown in Table \ref{tab:rbias}.  \end{enumerate}
The mock sample size is kept fixed to 30 clusters at redshift 0.3 (median redshift of eDXL sample). 
Mock data were generated from the input scaling relations given in Table \ref{tab:rbias}. %
For these 30 clusters, the $Y_{\rm SZ}$ was simultaneously generated using a correlated intrinsic scatter in $Y_{\rm SZ}$ and $L_{\rm x}$ using the \verb|mrandomn| IDL routine. 
\paragraph{ Small measurement uncertainties }
{The sample with ``small'' errors were generated with 10\% measurement scatter. 
     We simulate the datasets consisting of 30 clusters with 10\% measurement scatter in each observable which is more precise than our APEX-SZ measurements. 
      We assume a set of input scaling relations parameters for the $L_{\rm x}\text{--}M_{500}$ and $Y_{\rm SZ}\text{--}M_{500}$. 
      An example set of mock dataset is shown in Figure \ref{fig:r0} for uncorrelated ($r=0$) intrinsic scatters at fixed mass. Figure \ref{fig:r0.8} shows mock data set of clusters for input correlation, $r=0.8$.}

\paragraph{Realistic measurement uncertainties}{The samples with ``real'' errors were generated by using the median of $\sigma_{\widetilde{M}_{\rm WL}}/\widetilde{M}_{\rm WL}$, $\sigma_{\widetilde{Y}_{\rm SZ}}/\widetilde{Y}_{\rm SZ}$,
$\sigma_{\widetilde{L}_{\rm x}}/\widetilde{L}_{\rm x}$ and keeping this ratio for the generated sample. For realistic measurement uncertainties, we use for the $L_{\rm x}\text{--}M_{500}$, $Y_{\rm SZ}\text{--}M_{500}$ relations, two sets of scaling parameters. The first set of input parameters is identical to those used in simulating mock samples with small measurement uncertainties. 
For the second set, we inject higher intrinsic scatters in $L_{\rm x}$ and $Y_{\rm SZ}$. The values of these scaling parameters were chosen based on the eDXL sample fit values from Section \ref{sec:results:rfree}.} 

Numerous realisations of mock samples for all of the above cases were generated for a range of input values of $r$. The analysis of these mock samples is presented in the next Section.  
\subsubsection{Analysis of scaling relations fit to  mock data}\label{app:test:method}
For each input relations and $r$ values given in Table \ref{tab:rbias}, we fit each mock sample with the model described in Section \ref{sec:method:application:nointscatter} and the normalised likelihood given in Appendix \ref{app:lik}. For all cases, we set $r=0$ while fitting mock data. This is to investigate the level of bias in scaling parameters if the $r$ parameter is ignored in the joint fit of the $L_{\rm x}\text{--}M_{500}$ and $Y_{\rm SZ}\text{--}M_{500}$ relations. The results are given in Table \ref{tab:rbias}. The means reported are an average of recovered modes of scaling relation parameters from fit to different mock samples (typically $\sim$ 35-60 samples). The uncertainty quoted on each parameter is the standard deviation of the sample distribution of recovered modes. This corresponds to an uncertainty in the scaling relation parameter from fitting a single set of mock cluster sample. Below, we define relative bias and its significance.   
  \begin{equation}\label{eq:relbias}
 \mathrm{relative ~bias} ~\times 100 = \frac{(\mathrm{recovered}-\mathrm{input})}{\mathrm{input}} ~~ \%
\end{equation}
\begin{equation}\label{eq:significance}
 \mathrm{Detection~ level ~of ~ the ~ bias} ~ = d = \frac{\mathrm{relative ~ bias}}{\frac{\Delta \mathrm{recovered}}{\mathrm{input}}}
\end{equation}
The relative bias and significance are quoted in Table \ref{tab:rbias}. The significance here is a measure of an average bias level w.r.t. the uncertainties on the recovered parameter for any cluster sample.
\begin{table*}\caption{The bias in recovered parameters of the scaling relations for the simulated cluster samples with different input correlation coefficient, $r$, values. 
The variable, $r$, refers to the correlation between the intrinsic scatters of $L_{\rm x}$ and $Y_{\rm SZ}$ at fixed mass. The bias in the SZ scaling parameters grows significant with strong correlation coefficient. The relative bias and significance quoted here were calculated using the equations \eqref{eq:relbias} and \eqref{eq:significance}}\label{tab:rbias}
 \begin{tabular}{c c c c c c c}
\hline
  $r$ &$A_{\rm LM}$&$B_{\rm LM}$&$\sigma_{\ln L}$&$A_{\rm SZ}$&$B_{\rm SZ}$&$\sigma_{\ln Y_{\rm SZ}}$\\
  input & 0.80	& 1.50	& 0.40	& 1.0	&	1.67& 0.20\\
  \hline
  \multicolumn{7}{c}{small measurement uncertainties}\\
  \hline
 0.0 &0.789 $\pm$ 0.137	&1.471 $\pm$    0.238	&0.400 $\pm$    0.096	&0.996  $\pm$  0.051	&1.676   $\pm$   0.159	&0.189 $\pm$ 0.061\\
  rel. bias  & $-$ 1.4 $\%$ & $-$2.0 $\%$ & 0.0 $\%$& $-$0.4 $\%$ & 0.4 $\%$&$-$5.3 $\%$\\
   $d$  & $-$ 0.08 $\sigma$ &$-$0.12 $\sigma$ &0.00 $\sigma$ &$-$0.09 $\sigma$ & 0.04$\sigma$&$-$ 0.17 $\sigma$\\
  \hline
 0.1& 0.832 $\pm$    0.117	&1.481$\pm$0.198	&0.392	$\pm$ 0.078&1.022 $\pm$ 0.057	&1.642$\pm$0.135	&0.171$\pm$0.064\\
  rel. bias  &  4.0 $\%$ & $-$1.3 $\%$ & $-$ 2.0 $\%$& 2.2 $\%$ & $-$ 1.6 $\%$&$-$ 14.5 $\%$\\
 $d$  & 0.27 $\sigma$ &-0.10 $\sigma$ &$-$ 0.10 $\sigma$ &0.39 $\sigma$ & $-$ 0.20 $\sigma$& $-$ 0.46 $\sigma$\\
  \hline
  0.5&0.822$\pm$0.110	&	1.535  $\pm$   0.213 &0.398	$\pm$ 0.081&1.118 $\pm$ 0.066	&1.476$\pm$ 0.145&0.173$\pm$0.052\\
  rel. bias  & 2.8 $\%$ & 2.4 $\%$ & $-$ 0.5 $\%$& 11.8 $\%$ & $-$ 11.6 $\%$& $-$ 13.5 $\%$\\
   $d$  & 0.20$\sigma$ & 0.16$\sigma$ & $-$0.03$\sigma$ & 1.79 $\sigma$ & $-$ 1.34$\sigma$& $-$  0.62$\sigma$\\
  \hline
  0.8&0.861$\pm$ 0.129	&1.525$\pm$ 0.210	& 0.350$\pm$   0.091	&1.161$\pm$	0.058&	1.466 $\pm$0.110&0.121 $\pm$ 0.056\\
   rel. bias  & 7.7 $\%$ & 1.7 $\%$ & $-$ 12.5 $\%$& 16.1 $\%$ &$-$ 12.2 $\%$& $-$ 39.3 $\%$\\
 $d$ & 0.48 $\sigma$ & 0.12$\sigma$ & $-$ 0.55$\sigma$ & 2.76 $\sigma$ & $-$ 1.91$\sigma$& $-$ 1.40 $\sigma$\\
 \hline 
 \multicolumn{7}{c}{realistic measurement uncertainties }\\
 \hline
 0.1& 0.926 $\pm$    0.190	&1.516$\pm$0.240	&0.360	$\pm$ 0.123&1.014 $\pm$ 0.118	&1.568$\pm$0.398	&0.22$\pm$0.110\\
  rel. bias  &  15.7 $\%$ & 1.1 $\%$ & $-$ 10.6 $\%$& 1.4 $\%$ & $-$ 6.1 $\%$& 12.7 $\%$\\
 $d$  & 0.66 $\sigma$ &0.07 $\sigma$ &$-$ 0.34 $\sigma$ &0.12 $\sigma$ & $-$ 0.26 $\sigma$&  0.23 $\sigma$\\
 0.6 &1.014$\pm$0.133	&	1.438  $\pm$   0.298 &0.310	$\pm$ 0.076&1.128 $\pm$ 0.132	&1.418$\pm$ 0.331&0.203$\pm$0.088\\
  rel. bias  & 26.8 $\%$ & $-$ 4.1 $\%$ & $-$ 23.2 $\%$& 12.8 $\%$ & $-$ 15.1 $\%$&  1.6 $\%$\\
   $d$  & 1.61$\sigma$ & $-$ 0.21$\sigma$ &$-$ 1.23$\sigma$ & 0.97 $\sigma$ & $-$ 0.76$\sigma$&  0.04$\sigma$\\
   \hline
   \multicolumn{7}{c}{ eDXL-like samples}\\
      input & 0.30	& 1.50	& 0.60	& 1.0	&	1.67& 0.40\\
      \hline
   $r=0.3$ & 0.37 $\pm$ 0.12 & 1.61 $\pm$ 0.29 & 0.54 $\pm$ 0.11 & 1.11 $\pm$ 0.14 & 1.70 $\pm$ 0.42 & 0.29 $\pm$ 0.17\\
  rel. bias & 23.3\% & 7.3\%&$-$0.1& 11.0\%&1.8\%&$-$27.5\%\\
 $d$  &0.58$\sigma$&0.38$\sigma$&$-$0.55$\sigma$&0.79$\sigma$&0.07$\sigma$&$-$0.65$\sigma$\\
   $r =$ 0.6 &0.440 $\pm$    0.150	&1.428 $\pm$ 0.297	&0.453	$\pm$ 0.167&1.396 $\pm$ 0.180	&1.318 $\pm$ 0.370	&0.267 $\pm$ 0.134\\
  rel. bias  & 46.7 $\%$ & $-$ 4.8 $\%$ & $-$ 24.5 $\%$& 39.6 $\%$ &$-$ 21.1 $\%$& $-$ 49.6 $\%$\\
   $d$  & 0.94$\sigma$ & $-$ 0.24$\sigma$ &$-$ 0.88$\sigma$ & 2.20 $\sigma$ & $-$ 0.95$\sigma$&  $-$ 0.99$\sigma$\\
  \hline
   \end{tabular}
\end{table*}

We focus our analysis so as to examine the impact of the intrinsic covariance on the robustness of measurements of the $Y_{\rm SZ}\text{--}M$ relation. 
The impact on the $L_{\rm x}\text{--}M$ relation here is of little relevance. To measure the $L_{\rm x}\text{--}M$ relation itself, the formalism used here remains the same but does not require the additional cluster observables (which play no role in sample selection) scaling relations.
The recovered values of the $L_{\rm x}\text{--}M$ relation in our multi-observable mass scaling show some deviations from input values due to degeneracies of the $L_{\rm x}\text{--}M$ relation parameters with $Y_{\rm SZ}\text{--}M$ scaling parameters. 
The degeneracies are mainly observed due to our restrictive incorrect prior on the correlation parameter in our mock fits. 
Releasing the fixed prior would mitigate the impact on the $L_{\rm x }\text{--}M$ and $Y_{\rm SZ}\text{--}M$ relations. 
Therefore, the analysis presented with these mock data serves the main purpose of examining the impact of intrinsic covariance on measuring 
scaling relations of cluster observables that play no role in the sample selection (e.g., $Y_{\rm SZ}$) with mass. 

\subsection{Mock data with intrinsic scatter in the lensing mass} \label{app:scat}
  The procedure for the mock data generation is identical to the previous section with couple of exceptions. We introduce scattered lensing masses ($M_{\rm WL}$) from true halo mass ($M_{\rm HM,500}$). We introduce a 20\% log-normal scatter between the halo mass ($M_{\rm HM}$) and the weak-lensing mass. All observables are generated in similar manner as in previous Section using the scaling relation models defined in Section \ref{sec:method:application:withintscatter}. We also use the redshift distribution as well as the redshift dependent luminosity thresholds of the eDXL sample. We simulate with small and real measurement uncertainties as done before. 
\paragraph{Small measurement uncertainties}{We use mock samples with 10\% measurement uncertainties to test for bias. The mock samples include zero correlation between intrinsic scatters of $L_{\rm x}$ and $Y_{\rm SZ}$. We use the identical method used in Appendix \ref{app:test:method} and fit the scaling relations with $r=0$. The mean recovered parameters from numerous mock samples and the standard deviations are quoted in Table \ref{tab:scat}. We observe significant bias in the slope and intrinsic scatter parameters of the $Y_{\rm SZ}-M_{500}$ scaling relations. On the $L_{\rm x}-M_{500}$ relation the bias in scaling parameters are lower than their confidence level obtained for single mock data set.
}
\paragraph{Realistic measurement uncertainties}{%

We fit the scaling relations parameters to the mock samples with realistic measurement uncertainties using the same method as in Appendix \ref{app:test:method}. The mock samples considered include zero correlation between intrinsic scatters of $L_{\rm x}$ and $Y_{\rm SZ}$.
As done in previous section, we fit these mock samples by ignoring the $r$ parameter. The average of recovered modes and the standard deviation of the sample distribution of recovered modes are summarised in Table \ref{tab:scat}. 
 We inject $r=0.6$ for the mock samples along with the 20\% intrinsic scatter in lensing mass. We fit these mock data while ignoring $r$ and the scatter in lensing mass. At $r$ equals 0.6, the total bias in the normalisation  is significant at a level of 2.7$\sigma$ with respect to uncertainties expected for a single mock data set. The slope parameter of the $Y_{\rm SZ}-M_{500}$ relation is biased low by $\sim 1 \sigma$. We re-fit the mock samples with $r=0.6$ fixed in the fitting routine.  The results are shown in the final row of Table \ref{tab:scat}. We observe that the bias in the normalisation drops to $0.5\sigma$ which is similar to the bias obtained in the mock samples that were generated with $r=0$. 
 The bias in the slope value reduces from $0.34$ to $0.17$. 
\begin{table*}\caption{Scatter between $m_{\rm lens}$ and $m_{\rm halo}$ assumed to be 0.20 (20 \%). The redshift dependent luminosity cuts were applied to the set of mock data used here.}\label{tab:scat}
 \begin{tabular}{c c c c c c c}
 \hline
    $r$ &$A_{\rm LM}$&$B_{\rm LM}$&$\sigma_{\ln L}$&$A_{\rm SZ}$&$B_{\rm SZ}$&$\sigma_{\ln Y_{\rm SZ}}$\\
   Input &	0.3&	1.5&0.6	&1.0	&1.67 	&0.4\\
   \hline
   \multicolumn{7}{c}{small measurement uncertainties}\\
   \hline
    $r$ $=$ 0 &0.235 $\pm$    0.123	&1.476$\pm$0.238	&0.676	$\pm$ 0.151&1.118 $\pm$ 0.133	&1.356 $\pm$ 0.225	&0.511 $\pm$ 0.079\\
   rel. bias  & $-$ 21.6 $\%$ & $-$ 1.6 $\%$ &  12.6 $\%$& 11.8 $\%$ &$-$ 18.8 $\%$&  27.7 $\%$\\
   $d$  & $-$ 0.53$\sigma$ & $-$ 0.10$\sigma$ &$+$ 0.50$\sigma$ & 0.89 $\sigma$ & $-$ 1.40$\sigma$&  1.41$\sigma$\\
   \hline
\multicolumn{7}{c}{realistic measurement uncertainties}\\
   \hline
   $r$ $=$ 0 &0.33 $\pm$    0.15	&1.44 $\pm$ 0.22	&0.60	$\pm$ 0.12&1.09 $\pm$ 0.16	&1.55 $\pm$ 0.32	&0.44 $\pm$ 0.12\\
   rel. bias  & 10.0 $\%$ & $-$ 4.0 $\%$ & $-$ 0.0 $\%$& 9.0 $\%$ &$-$ 7.0 $\%$&  10.0 $\%$\\
   $d$  & 0.2$\sigma$ & $-$ 0.3$\sigma$ &$-$ 0.0$\sigma$ & 0.6 $\sigma$ & $-$ 0.4$\sigma$&  0.3$\sigma$\\
   \hline
  $r$ $=$ 0.6 &0.434 $\pm$    0.213	&1.406 $\pm$ 0.254	&0.501	$\pm$ 0.164&1.590 $\pm$ 0.217	&1.326 $\pm$ 0.322	&0.342 $\pm$ 0.128\\
  rel. bias  & 44.7$\%$ & $-$ 6.3$\%$ & $-$ 16.5$\%$& 59.0$\%$ &$-$ 20.6$\%$& $-$ 14.5$\%$\\
   $d$  & 0.63$\sigma$ & $-$ 0.37$\sigma$ &$-$ 0.60$\sigma$ & 2.72 $\sigma$ & $-$ 1.07$\sigma$&  $-$ 0.45$\sigma$\\
    $r$ $=$ 0.6 (fixed) &0.34 $\pm$    0.15	&1.43 $\pm$ 0.22	&0.60	$\pm$ 0.13&1.10 $\pm$ 0.21	&1.49 $\pm$ 0.21	&0.45 $\pm$ 0.16\\
   $d$  & 0.27$\sigma$ & $-$ 0.32$\sigma$ &$-$ 0.00$\sigma$ & +0.50 $\sigma$ & $-$ 0.86$\sigma$&  $+$ 0.31$\sigma$\\
   \hline
 \end{tabular}

\end{table*}
 
\begin{figure*}
 \includegraphics[width=14cm,keepaspectratio=true]{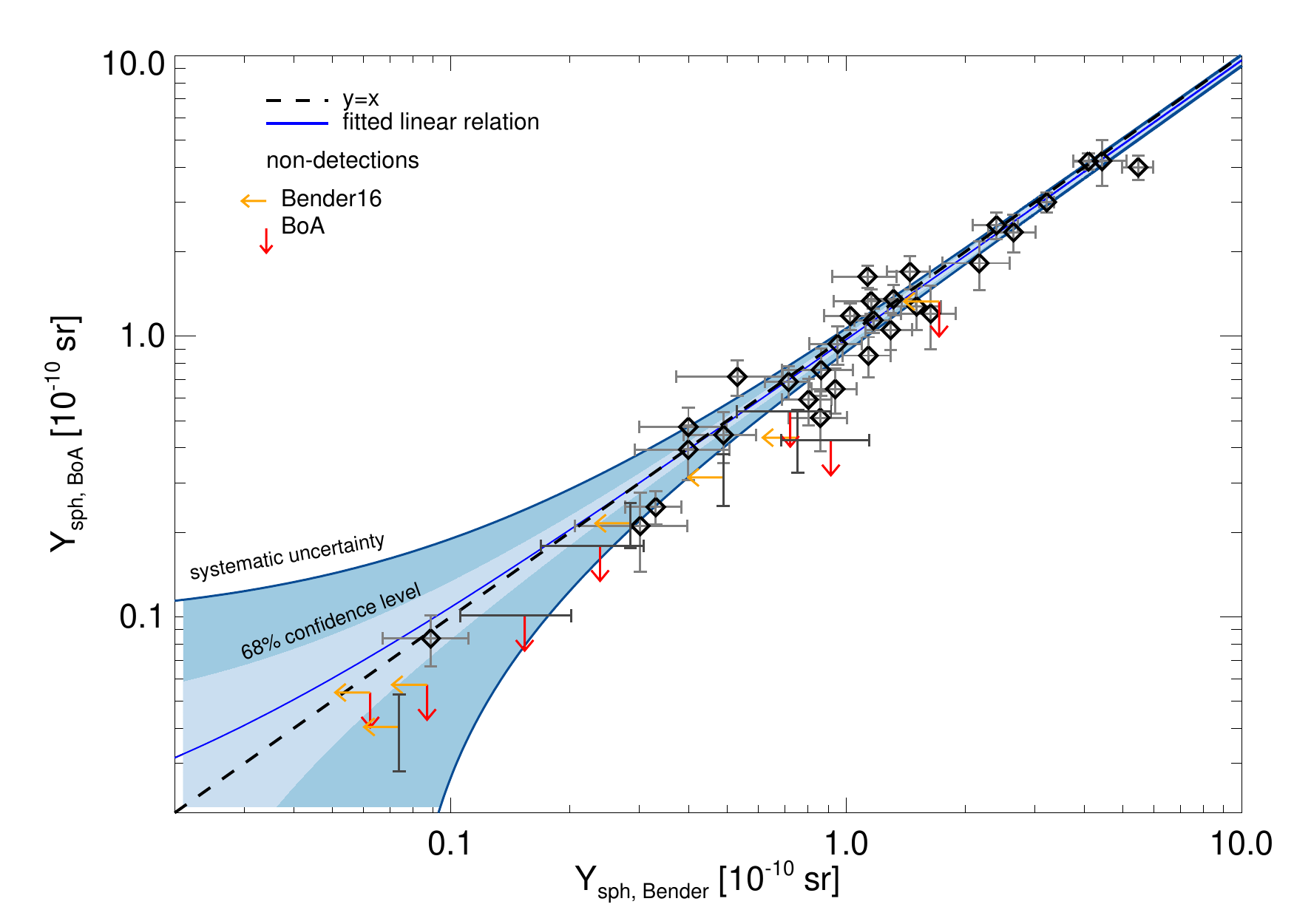}\caption{$Y_{\rm sph, 500}$ measured from two pipelines (BoA and MATLAB (Bender et al. 2016)) are shown here along with their uncertainties and upper limits wherever appropriate. The correlation between the two pipelines using the Kelly method shows that the measurements are statistically consistent for the 41 clusters in the sample. The solid line is the best-fit relation and the dashed line is the one-to-one relation. }\label{fig:boavsmatlab}
\end{figure*}

\section{Comparison with Bender et al. }\label{app:bender}
Here, we compare the measured $Y_{\rm sph,500}$ from the previously published ones from APEX-SZ in \cite{2016Bender} and the measured ones with the independent pipeline BoA. The gNFW model parametrizations, cluster centroids and $R_{500}$'s were kept identical to \cite{2016Bender}. We re-fit the model to the BoA reduced maps using the radial binning method described in this paper. The re-measured $Y_{500,\rm BoA}$ and the literature values are plotted in the Figure \ref{fig:boavsmatlab}. 

Across 41 clusters in the APEX-SZ targets that were used in the previous study, in general, there is a good agreement in the measurements.
We use the \cite{2007Kelly} to quantify the linear relation (Equation \ref{app:eq:bender}) between the two sets of measurements. 
We define,
\begin{equation}\label{app:eq:bender}
 \frac{Y_{\rm BoA}}{10^{-10} ~{\rm sr}} = \alpha + \beta \frac{Y_{\rm Bender}}{10^{-10} {\rm sr}} + \sigma_{\rm int}.
\end{equation}

We obtain constraints on $\alpha =0.012 \pm 0.027$, $\sigma_{\rm int}= 0.056 \pm 0.031$ and $\beta = 0.96 \pm 0.04$.
We note that these measurements are expected to be highly correlated as they are measured from the same data sets. However, the analyses were performed independently using different pipelines. The analysis described in Section \ref{sec:obs} was homogeneously applied to all clusters, in contrast to the cluster signal-to-noise optimisation adopted in the previous work by \cite{2016Bender}.
 In the current work, we use the point source transfer function to model the filtered data, whereas \cite{2016Bender} used a fixed cluster transfer function. One-to-one comparison of the two pipelines is, however, beyond the scope of this work.  
 
%
\bsp	
\label{lastpage}
\end{document}